\documentclass[twocolumn,numberedappendix,twocolappendix,appendixfloats]{openjournal_dhayaa}

\usepackage{amsmath, bm}
\usepackage{fontawesome5, multirow}
\usepackage{color}
\usepackage{xcolor}
\usepackage{ulem}

\usepackage{natbib}
\setcitestyle{aysep={}}

\usepackage[colorlinks=true
  ,urlcolor=blue
  ,anchorcolor=blue
  ,citecolor=blue
  ,filecolor=blue
  ,linkcolor=blue
  ,menucolor=blue
  ,linktocpage=true
  ,pdfproducer=medialab
  ,pdfa=true
]{hyperref}

\usepackage{graphicx}	
\usepackage{amsmath}	

\usepackage{fontawesome5}
\usepackage{color}
\usepackage{xcolor}
\usepackage{xspace}

\usepackage{enumitem}
\setlist{nosep}

\newcommand{\eg}{{\sl e.g.}, }       
\newcommand{\ie}{{\sl i.e.}, }

\newcommand{\mpc}{\ensuremath{\, {\rm Mpc}}}

\newcommand{\chat}{\hat{c}}
\newcommand{\selhat}{\hat{s}}
\newcommand{\bhat}{\hat{b}}
\newcommand{\mathR}{\mathcal{R}}
\newcommand{\mathD}{\mathcal{D}}
\newcommand{\mathB}{\mathcal{B}}
\newcommand{\mathW}{\mathcal{W}}

\newcommand{\decade}{DECADE\xspace}
\newcommand{\Balrog}{\textsc{Balrog}\xspace}
\newcommand{\Cosmos}{\textsc{Cosmos}\xspace}
\newcommand{\PAUS}{\textsc{Paus}\xspace}
\newcommand{\CTRT}{\textsc{C3R2}\xspace}
\newcommand{\Boss}{\textsc{BOSS}\xspace}
\newcommand{\eBoss}{e\textsc{BOSS}\xspace}

\definecolor{orcidlogocol}{HTML}{A6CE39}
\definecolor{purple}{RGB}{128, 0, 128}
\definecolor{kelly}{RGB}{76, 187, 23}

\newcommand{\OrcidID}[1]{ \href[urlcolor = red]{https://orcid.org/#1}{\textcolor{lightgray}{\faOrcid}}}
\newcommand{\OrcidIDName}[2]{\href{https://orcid.org/#1}{#2}}

\defcitealias{y3-shapecatalog}{\textsc{GS21}}
\defcitealias{Myles:2021:DESY3}{\textsc{MA21}}
\defcitealias{Sanchez:2020:NoiseSOM}{\textsc{S20}}
\defcitealias{Sanchez:2023:highzY3}{\textsc{S23}}
\defcitealias{Gatti:2022:WzY3}{\textsc{GG22}}
\defcitealias{paper1}{\textsc{Paper I}}
\defcitealias{paper2}{\textsc{Paper II}}
\defcitealias{paper3}{\textsc{Paper III}}
\defcitealias{paper4}{\textsc{Paper IV}}

\newcommand*{\vcenteredhbox}[1]{\begingroup
\setbox0=\hbox{#1}\parbox{\wd0}{\box0}\endgroup}

\begin{document}
{\hfill FERMILAB-PUB-25-0064-LDRD-PPD}

\title{The \decade cosmic shear project II: photometric redshift calibration of the source galaxy sample} 
\shortauthors{Anbajagane et. al}
\shorttitle{The \decade cosmic shear project II: photometric redshift calibration}

\author{\OrcidIDName{0000-0003-3312-909X}{D. Anbajagane} (\vcenteredhbox{\includegraphics[height=1.2\fontcharht\font`\B]{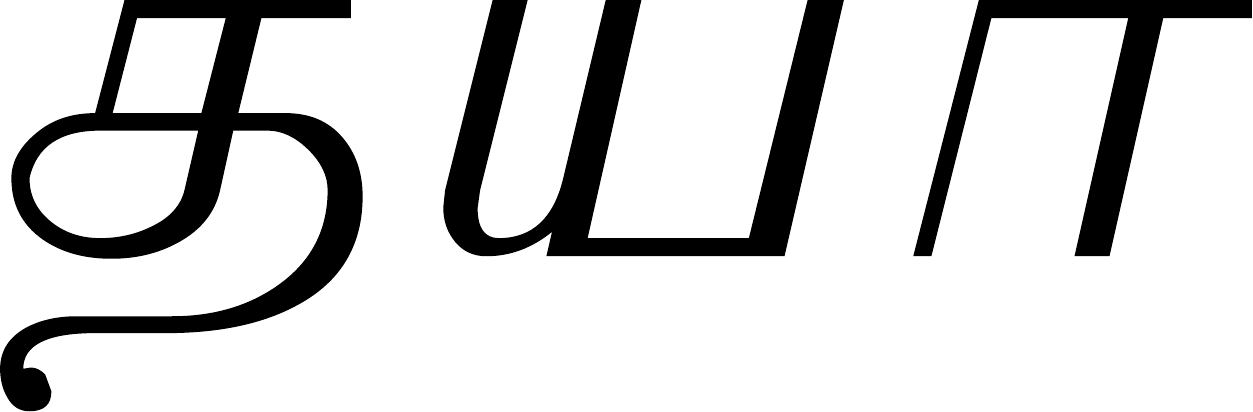}})$^\star$}
\affiliation{Department of Astronomy and Astrophysics, University of Chicago, Chicago, IL 60637, USA}
\affiliation{Kavli Institute for Cosmological Physics, University of Chicago, Chicago, IL 60637, USA}
\email{$^{\star}$dhayaa@uchicago.edu}

\author{\OrcidIDName{0000-0001-8505-1269}{A.~Alarcon}}
\affiliation{Institute of Space Sciences (ICE, CSIC),  Campus UAB, Carrer de Can Magrans, s/n,  08193 Barcelona, Spain}

\author{\OrcidIDName{0000-0002-5279-0230}{R.~Teixeira}}
\affiliation{Department of Astronomy and Astrophysics, University of Chicago, Chicago, IL 60637, USA}
\affiliation{Department of Physics, Duke University Durham, NC 27708, USA}

\author{\OrcidIDName{0000-0002-7887-0896}{C.~Chang}}
\affiliation{Department of Astronomy and Astrophysics, University of Chicago, Chicago, IL 60637, USA}
\affiliation{Kavli Institute for Cosmological Physics, University of Chicago, Chicago, IL 60637, USA}

\author{\OrcidIDName{0000-0002-6002-4288}{L.~F.~Secco}}
\affiliation{Kavli Institute for Cosmological Physics, University of Chicago, Chicago, IL 60637, USA}

\author{\OrcidIDName{0000-0003-0478-0473}{C.~Y.~Tan}}
\affiliation{Department of Physics, University of Chicago, Chicago, IL 60637, USA}
\affiliation{Kavli Institute for Cosmological Physics, University of Chicago, Chicago, IL 60637, USA}

\author{\OrcidIDName{0000-0001-8251-933X}{A.~Drlica-Wagner}}
\affiliation{Fermi National Accelerator Laboratory, P. O. Box 500, Batavia, IL 60510, USA}
\affiliation{Department of Astronomy and Astrophysics, University of Chicago, Chicago, IL 60637, USA}
\affiliation{Kavli Institute for Cosmological Physics, University of Chicago, Chicago, IL 60637, USA}

\author{\OrcidIDName{0000-0002-6904-359X}{M.~Adamow}}
\affiliation{Center for Astrophysical Surveys, National Center for Supercomputing Applications, 1205 West Clark St., Urbana, IL 61801, USA}
\affiliation{Department of Astronomy, University of Illinois at Urbana-Champaign, 1002 W. Green Street, Urbana, IL 61801, USA}

\author{\OrcidIDName{0000-0002-4588-6517}{R.~A.~Gruendl}}
\affiliation{Center for Astrophysical Surveys, National Center for Supercomputing Applications, 1205 West Clark St., Urbana, IL 61801, USA}
\affiliation{Department of Astronomy, University of Illinois at Urbana-Champaign, 1002 W. Green Street, Urbana, IL 61801, USA}

\author{\OrcidIDName{0000-0002-3730-1750}{G.~Giannini}}
\affiliation{Kavli Institute for Cosmological Physics, University of Chicago, Chicago, IL 60637, USA}

\author{\OrcidIDName{0000-0001-7774-2246}{M.~R.~Becker}}
\affiliation{Argonne National Laboratory, 9700 South Cass Avenue, Lemont, IL 60439, USA}

\author{\OrcidIDName{0000-0001-6957-1627}{P.~S.~Ferguson}}
\affiliation{DIRAC Institute, Department of Astronomy, University of Washington, 3910 15th Ave NE, Seattle, WA, 98195, USA}

\author{\OrcidIDName{0009-0005-1143-495X}{N.~Chicoine}}
\affiliation{Department of Astronomy and Astrophysics, University of Chicago, Chicago, IL 60637, USA}
\affiliation{Department of Physics and Astronomy, University of Pittsburgh, 3941 O’Hara Street, Pittsburgh, PA 15260}

\author{\OrcidIDName{0000-0002-7523-582X}{Z.~Zhang}}
\affiliation{Department of Astronomy and Astrophysics, University of Chicago, Chicago, IL 60637, USA}
\affiliation{Department of Physics, Stanford University, 382 Via Pueblo Mall, Stanford, CA 94305, USA}
\affiliation{SLAC National Accelerator Laboratory, Menlo Park, CA 94025, USA}

\author{\OrcidIDName{0000-0003-4394-7491}{K.~Herron}}
\affiliation{Department of Physics and Astronomy, Dartmouth College, Hanover, NH 03755, USA}

\author{\OrcidIDName{0000-0003-2911-2025}{D.~Suson}}
\affiliation{Department of Chemistry and Physics, Purdue University Northwest 2200, 169th Ave, Hammond, IN 46323}

\author{\OrcidIDName{0000-0002-3173-2592}{A.~N.~Alsina}}
\affiliation{Instituto de Física Gleb Wataghin, Universidade Estadual de Campinas, 13083-859, Campinas, SP, Brazil}

\author{\OrcidIDName{0000-0002-6445-0559}{A.~Amon}}
\affiliation{Department of Astrophysical Sciences, Princeton University, Peyton Hall, Princeton, NJ 08544, USA}

\author{\OrcidIDName{0000-0003-4383-2969}{C.~R.~Bom}}
\affiliation{Centro Brasileiro de Pesquisas F\'isicas, Rua Dr. Xavier Sigaud 150, 22290-180 Rio de Janeiro, RJ, Brazil}

\author{\OrcidIDName{0000-0002-3690-105X}{J.~A.~Carballo-Bello}}
\affiliation{Instituto de Alta Investigaci\'on, Universidad de Tarapac\'a, Casilla 7D, Arica, Chile}

\author{\OrcidIDName{0000-0003-1697-7062}{W.~Cerny}}
\affiliation{Department of Astronomy, Yale University, New Haven, CT 06520, USA}

\author{\OrcidIDName{0000-0002-5636-233X}{A.~Choi}}
\affiliation{NASA Goddard Space Flight Center, 8800 Greenbelt Rd, Greenbelt, MD 20771, USA}

\author{\OrcidIDName{0000-0003-1680-1884}{Y.~Choi}}
\affiliation{NSF National Optical-Infrared Astronomy Research Laboratory}

\author{\OrcidIDName{0000-0003-4480-0096}{C.~Doux}}
\affiliation{Université Grenoble Alpes, CNRS, LPSC-IN2P3, 38000 Grenoble, France}

\author{\OrcidIDName{0000-0002-1407-4700}{K.~Eckert}}
\affiliation{Department of Physics and Astronomy, University of Pennsylvania, Philadelphia, PA 19104, USA}

\author{\OrcidIDName{0000-0001-6134-8797}{M.~Gatti}}
\affiliation{Kavli Institute for Cosmological Physics, University of Chicago, Chicago, IL 60637, USA}

\author{\OrcidIDName{0000-0003-3270-7644}{D.~Gruen}}
\affiliation{University Observatory, Faculty of Physics, Ludwig-Maximilians-Universität, Scheinerstr. 1, 81679 Munich, Germany}
\affiliation{Excellence Cluster ORIGINS, Boltzmannstr. 2, 85748 Garching, Germany}

\author{\OrcidIDName{0000-0001-9994-1115}{W.~G.~Hartley}}
\affiliation{Department of Astronomy, University of Geneva, ch. d’Ecogia 16, 1290 Versoix, Switzerland}

\author{\OrcidIDName{0000-0001-6718-2978}{K.~Herner}}
\affiliation{Fermi National Accelerator Laboratory, P. O. Box 500, Batavia, IL 60510, USA}

\author{\OrcidIDName{0000-0002-9378-3424}{E.~M.~Huff}}
\affiliation{Jet Propulsion Laboratory, California Institute of Technology, 4800 Oak Grove Dr., Pasadena, CA 91109, USA}

\author{\OrcidIDName{0000-0001-5160-4486}{D.~J.~James}}
\affiliation{Applied Materials Inc., 35 Dory Road, Gloucester, MA 01930}
\affiliation{ASTRAVEO LLC, PO Box 1668, Gloucester, MA 01931}

\author{\OrcidIDName{0000-0003-2511-0946}{N.~Kuropatkin}}
\affiliation{Fermi National Accelerator Laboratory, P. O. Box 500, Batavia, IL 60510, USA}

\author{\OrcidIDName{0000-0002-9144-7726}{C.~E.~Mart\'inez-V\'azquez}}
\affiliation{International Gemini Observatory/NSF NOIRLab, 670 N. A'ohoku Place, Hilo, Hawai'i, 96720, USA}

\author{\OrcidIDName{0000-0002-8093-7471}{P.~Massana}}
\affiliation{NSF's NOIRLab, Casilla 603, La Serena, Chile}

\author{\OrcidIDName{0000-0003-3519-4004}{S.~Mau}}
\affiliation{Department of Physics, Stanford University, 382 Via Pueblo Mall, Stanford, CA 94305, USA}
\affiliation{Kavli Institute for Particle Astrophysics \& Cosmology, P.O.\ Box 2450, Stanford University, Stanford, CA 94305, USA}

\author{\OrcidIDName{0000-0002-4475-3456}{J.~McCullough}}
\affiliation{Department of Astrophysical Sciences, Peyton Hall, Princeton University, Princeton, NJ USA 08544}

\author{\OrcidIDName{0000-0003-0105-9576}{G.~E.~Medina}}
\affiliation{Dunlap Institute for Astronomy \& Astrophysics, University of Toronto, 50 St George Street, Toronto, ON M5S 3H4, Canada}
\affiliation{David A. Dunlap Department of Astronomy \& Astrophysics, University of Toronto, 50 St George Street, Toronto ON M5S 3H4, Canada}

\author{\OrcidIDName{0000-0001-9649-4815}{B.~Mutlu-Pakdil}}
\affiliation{Department of Physics and Astronomy, Dartmouth College, Hanover, NH 03755, USA}

\author{\OrcidIDName{0000-0001-6145-5859}{J.~Myles}}
\affiliation{Department of Astrophysical Sciences, Princeton University, Peyton Hall, Princeton, NJ 08544, USA}

\author{\OrcidIDName{0000-0001-9438-5228}{M. ~ Navabi}}
\affiliation{Department of Physics, University of Surrey, Guildford GU2 7XH, UK}

\author{\OrcidIDName{0000-0002-8282-469X}{N.~E.~D.~Noël}}
\affiliation{Department of Physics, University of Surrey, Guildford GU2 7XH, UK}

\author{\OrcidIDName{0000-0002-6021-8760}{A.~B.~Pace}}
\affiliation{Department of Astronomy, University of Virginia, 530 McCormick Road, Charlottesville, VA 22904, USA}

\author{\OrcidIDName{0000-0002-7354-3802}{M.~Raveri}}
\affiliation{Department of Physics and INFN, University of Genova, Genova, Italy}

\author{\OrcidIDName{0000-0001-5805-5766}{A.~H.~Riley}}
\affiliation{Institute for Computational Cosmology, Department of Physics, Durham University, South Road, Durham DH1 3LE, UK}

\author{\OrcidIDName{0000-0002-1594-1466}{J.~D.~Sakowska}}
\affiliation{Department of Physics, University of Surrey, Guildford GU2 7XH, UK}

\author{\OrcidIDName{0000-0003-3054-7907}{D.~Sanchez-Cid}}
\affiliation{Physik-Institut, University of Zurich, Winterthurerstrasse 190, CH-8057 Zurich, Switzerland}
\affiliation{Centro de Investigaciones Energéticas, Medioambientales y Tecnológicas (CIEMAT), Madrid, Spain}

\author{\OrcidIDName{0000-0003-4102-380X}{D.~J.~Sand}}
\affiliation{Steward Observatory, University of Arizona, 933 North Cherry Avenue, Tucson, AZ 85721-0065, USA}

\author{\OrcidIDName{0000-0003-3402-6164}{L.~Santana-Silva}}
\affiliation{Centro Brasileiro de Pesquisas F\'isicas, Rua Dr. Xavier Sigaud 150, 22290-180 Rio de Janeiro, RJ, Brazil}

\author{\OrcidIDName{0000-0002-1831-1953}{I.~Sevilla-Noarbe}}
\affiliation{Centro de Investigaciones Energ\'eticas, Medioambientales y Tecnol\'ogicas (CIEMAT), Madrid, Spain}

\author{\OrcidIDName{0000-0001-6082-8529}{M.~Soares-Santos}}
\affiliation{Physik-Institut, University of Zurich, Winterthurerstrasse 190, CH-8057 Zurich, Switzerland}

\author{\OrcidIDName{0000-0003-1479-3059}{G.~S.~Stringfellow}}
\affiliation{Center for Astrophysics and Space Astronomy, University of Colorado, 389 UCB, Boulder, CO 80309-0389, USA}

\author{\OrcidIDName{0000-0003-4341-6172}{A.~K.~Vivas}}
\affiliation{Cerro Tololo Inter-American Observatory/NSF NOIRLab, Casilla 603, La Serena, Chile}

\author{\OrcidIDName{0000-0003-1585-997X}{M.~Yamamoto}}
\affiliation{Department of Astrophysical Sciences, Princeton University, Peyton Hall, Princeton, NJ 08544, USA}

\begin{abstract}
We present the photometric redshift characterization and calibration for the Dark Energy Camera All Data Everywhere (DECADE) weak lensing dataset: a catalog of 107 million galaxies observed by the Dark Energy Camera (DECam) in the northern Galactic cap. The redshifts are estimated from a combination of wide-field photometry, deep-field photometry with associated redshift estimates, and a transfer function between the wide field and deep field that is estimated using a source injection catalog. We construct four tomographic bins for the galaxy catalog, and estimate the redshift distribution, $n(z)$, within each one using the self-organizing map photo-z (SOMPZ) methodology. Our estimates include the contributions from sample variance, zeropoint calibration uncertainties, and redshift biases, as quantified for the deep-field dataset. The total uncertainties on the mean redshifts are $\sigma_{\langle z \rangle} \approx 0.01$. The SOMPZ estimates are then compared to those from the clustering redshift method, obtained by cross-correlating our source galaxies with galaxies in spectroscopic surveys. The two estimates are consistent within $\Delta \langle z \rangle < 0.1$, where the comparison precision is limited by uncertainties in the clustering redshift estimates.
\end{abstract}


\section{Introduction}

Over the past two decades, weak gravitational lensing (WL) has emerged as a leading technique for constraining the cosmological parameters pertaining to the content and distribution of matter in our Universe. WL manifests as a bending of light from distant ``source galaxies'' by the matter distribution between the source and the observer \citep[see][for a review of weak gravitational lensing]{Bartelmann2001}. Thus, WL probes the large-scale structure (LSS) of our Universe and any processes that impact this structure; this includes a range of cosmological signatures such as those from modified gravity \citep[\eg][]{Schmidt:2008:MG_WL}, primordial correlations \citep[\eg][]{Anbajagane2023Inflation, Goldstein:2024:inflation, Primordial1, Primordial2} etc.

When using WL for constraining cosmological models, two necessary measurements are the location and the shape of the galaxies. From these two measurements, one can estimate the spatial correlations in the alignments of galaxy pairs on the sky, which is commonly referred to as cosmic shear and has resulted in some of our most precise measurements of cosmology \citep{Asgari2021, Amon2022, Secco2022, Dalal2023, Li2023b}. In the early development of the cosmic shear method, most efforts focused on developing accurate and precise shear estimation in the regime of noisy, complicated data \citep{Kitching2012, Kitching2013, Mandelbaum2015}. Recently, however, the focus has also shifted towards characterizing the ``location'' of the galaxy sample -- in particular, the location in the radial direction, which is simply the redshift distribution of the source galaxies \citep{Hildebrandt2020, Wright2020, Myles:2021:DESY3, Gatti:2022:WzY3, vandenBusch2022, Rau2023}.

Weak lensing is typically performed using wide field,
photometric surveys that consist of a handful of broad-band photometric filters. This means a single galaxy has estimates of its flux over several, broad wavelength windows. In spectroscopic surveys, a redshift can be precisely estimated by measuring the wavelength of an observed emission/absorption line from a galaxy and comparing it to the rest-frame wavelength of the same line \citep[\eg][]{Dawson:2013:BOSS, Dawson:2016:eBOSS}. For photometric surveys, the coarseness in the wavelength coverage makes it challenging to achieve redshift accuracy that is comparable to those of spectroscopic surveys. The standard weak lensing analysis therefore is performed in a ``tomographic'' fashion where galaxies are divided in rough tomographic bins of width $\Delta z \sim 0.3$, and within each bin, the galaxies are projected into a two-dimensional map on the sky \citep[\eg][]{Asgari2021, Amon2022, Secco2022, Dalal2023, Li2023b}. These measurements are therefore not sensitive to clustering in the radial direction, but can still probe the evolution of structure across redshift by comparing the spatial clustering across tomographic bins \citep{Hu:1999:WL_Tomography}. For such a tomographic analysis, one need not estimate redshifts for each individual galaxy and instead, one only requires the redshift distribution, $n(z)$, of the galaxy ensemble in each tomographic bin.

Depending on the characteristics of the data, state-of-the-art weak lensing surveys have adopted different methodologies to estimate the tomographic redshift distributions. The Dark Energy Survey \citep[DES,][]{Flaugher2005} has undergone several phases of increasingly sophisticated techniques: the early analyses in DES Year 1 \citep[Y1,][]{Hoyle2018} used the template-fitting method, Bayesian Photometric redshifts \citep[BPZ,][]{Benitez2000}, to generate redshifts for each individual source galaxy and then derive the $n(z)$ of the ensemble. This was then calibrated (alongside uncertainty quantification on the $n(z)$) using the 30-band photometric redshifts of galaxies in the Cosmic Evolution Survey \citep[\Cosmos,][]{Laigle2016} and also via clustering redshift measurements \citep{Schneider:2006:WZ,Newman:2008:WZ,Menard2013}, which use spatial cross-correlations between the target source-galaxy sample and another reference galaxy sample with known, high-quality redshifts \citep{Davis2017, Gatti:2018:WZ}. 

In the more recent Year 3 (Y3) analyses, DES employed a redshift method that uses two self-organizing maps \citep[SOMs][]{Kohonen:1982:SOM, Kohonen:2001:SOM, Carrasco:2014:SOM, Masters:2015:SOM}. The method is referred to as self-organizing maps photo-z \citep[SOMPZ,][]{Buchs2019} and is the technique used in this paper. It employs a principled, Bayesian framework to transfer redshift information from a ``deep-field'' sample --- which has precise photometry and high-quality redshifts --- to the actual source-galaxy sample measured in the wide-field data.

The Kilo-Degree Survey \citep[KiDS,][]{deJong2015} has an in-built advantage over DES as its survey footprint overlaps with the Visible and Infrared Survey Telescope for Astronomy (VISTA) Kilo-Degree Infrared Galaxy Survey \citep[VIKING,][]{Edge:2013:VIKING}, so all the galaxies have 4-band photometry in the optical \textit{and} have 5-band photometry in the near-infrared. The earlier analyses in KiDS+VIKINGS-450 \citep{Hildebrandt2020} used a direct re-weighting technique to derive the $n(z)$ as well as its uncertainty -- this can be seen as an early attempt of a SOM-like method. In more recent analyses, KiDS also use a SOM for mapping between redshift and photometry \citep{Wright2020, Wright:2020:KidsSOMS, vandenBusch2022}. 

The Hyper Suprime-Cam Subaru Strategic Program \citep[HSC-SSP,][]{Aihara2018a} is unique compared to DES and KiDS as their data is much deeper (and therefore their galaxy sample probes higher redshifts), making the redshift calibration especially challenging. In the Year 1 cosmic shear analysis, a large number of template-fitting and machine learning photometric redshift estimation codes were used \citep{Tanaka2018, Hikage2019} and the uncertainty was taken to be the difference between all the methods. In the Year 3 analysis, a combination of photometry and clustering-based methods were used \citep{Rau2023}. Due to the high redshift of the source-galaxy sample, the clustering-based method cannot provide constraints for a part of the highest redshift data, as in that regime there is no statistically relevant reference sample with high-quality redshift estimates. As a result the highest redshift bin(s) had a larger calibration uncertainty that was marginalized over in the cosmology analysis \citep{Dalal2023, Li2023b}.

The main goal of this paper is to characterize the redshift distribution of a new weak lensing catalog -- the DECam All Data Everywhere (\decade) shear catalog (\href{\#cite.paper1}{Anbajagane \& Chang et al. \citeyear{paper1}}).  The catalog consists of 107 million galaxies observed by the Dark Energy Camera (DECam) in the northern Galactic cap, from imaging data that combines many public standard and survey observing programs. The catalog covers a region of $5,\!412 \deg^2$ at a median limiting magnitude of $r = 23.6$, $i = 23.2$, $z  = 22.6$ (estimated at ${\rm S/N} = 10$ in a $2\arcsec$ aperture). Compared to the other weak lensing catalogs mentioned above, the DECADE shape catalog has a slightly shallower magnitude limit and slightly larger sky coverage compared to DES Y3. It is also derived from imaging data that has significant spatial inhomogeneity in the data quality/depth given the images were collected under a multitude of different community observing programs. 

Our efforts in this work adopt many of the methodologies developed in the DES Y3 analysis \citep[][henceforth \citetalias{Myles:2021:DESY3}]{Myles:2021:DESY3} as well as the Y3 high-redshift analysis \citep[][henceforth \citetalias{Sanchez:2023:highzY3}]{Sanchez:2023:highzY3}. Given the nature of the DECADE data, this analysis also serves as a stress-test of the redshift calibration methods on less ideal data with significant, spatially varying depth. The results could have implications for upcoming surveys like the Vera C. Rubin Observatory Legacy Survey of Space and Time (LSST), in informing the level of data quality variations that can be accommodated for weak lensing analyses.

This paper is the second in a series of four papers describing the DECADE cosmic shear analysis. The first paper (\href{\#cite.paper1}{Anbajagane \& Chang et al. \citeyear{paper1}}, hereafter \citetalias{paper1}) describes the shape measurement method, the final cosmology sample, and the robustness tests and image simulation pipeline from which we quantify the uncertainty in our cosmology sample. This paper (hereafter \textsc{Paper II}) describes the tomographic bins and redshift distributions for our cosmology sample, calibrated using a combination of SOMPZ, clustering redshifts, and synthetic source injection, together with a series of validation tests. The third paper, (\href{\#cite.paper3}{Anbajagane \& Chang et al. \citeyear{paper3}}, hereafter \citetalias{paper3}) defines the modelling choices of the cosmology analysis and quantifies the robustness of our cosmology constraints (including the redshift estimation) against the significant inhomogeneity in our imaging data. Finally, the fourth paper (\href{\#cite.paper4}{Anbajagane \& Chang et al. \citeyear{paper4}}, hereafter \citetalias{paper4}) performs the cosmological inference with our cosmic shear measurements. 

This paper is organized as follows. In Section~\ref{sec:data} we introduce the different datasets used in this work.  Section~\ref{sec:Methods} describes the SOMPZ methodology used in deriving the redshift distributions in each tomographic bin, our procedure for estimating uncertainties in these distributions, and also the clustering redshift method used as a cross-check. A key element of the SOMPZ method is the survey transfer function, and Section~\ref{sec:Balrog} our estimate of this, made from a synthetic source injection pipeline that is built for the \decade data. We present the final redshift distribution and uncertainties in Section~\ref{sec:Results_Nz}, including validation tests using the clustering redshift data. We conclude in Section~\ref{sec:conclusions}.

\section{Data}
\label{sec:data}

When using the SOMPZ method for estimating photometric redshifts, a number of different datasets are needed. The primary requirements for the method are the wide-field, source-galaxy sample that we wish to estimate redshifts for (Section~\ref{sec:sec:wide}), a deep-field sample that contains low-noise photometric measurements (Section~\ref{sec:sec:deep}), and a redshift sample that is a subset of the deep-field sample that has high-precision redshift estimates in addition to the low-noise photometry (Section~\ref{sec:sec:Zsample}). We also need a probabilistic transfer function for connecting deep-field photometric measurements with those from the wide-field data (Section~\ref{sec:sec:Balrog}). Finally, this work uses clustering redshift measurements (see Section~\ref{sec:sec:Wz}) to validate our SOMPZ-based redshift distributions. This requires a reference catalog that contains spectroscopic information and has spatial overlap with our source galaxy sample (Section~\ref{sec:sec:reference}). We now describe each of these samples further below.
    
\subsection{Wide sample}\label{sec:sec:wide}

The wide-field sample contains all galaxies that are part of the DECADE weak lensing catalog, which is described in detail in \citetalias{paper1}. In particular, the sample is obtained after performing the exact selections described in that work. This includes both selections on the objects properties, as well as selections based on the survey foregrounds/footprint. 

In this work, the wide-field sample is characterized primarily by the flux and flux errors in the $riz$ bands. These are obtained from the \textsc{Metacalibration} estimator \citep{Sheldon2017, Huff2017}, which fits galaxy images with an elliptical Gaussian profile. \textsc{Metacalibration} provides five variants for the fluxes: the fiducial estimate on the actual object image and then four variant estimates measured on images that have been sheared by a small amount in different directions. We only use the unsheared/fiducial flux estimate for results presented in this work. However, we will use the four variant estimates to generate four variant tomographic binning assignments. The bin for a given galaxy, which is detailed further below in Section~\ref{sec:sec:Tomobins}, is assigned using the galaxy flux and flux errors. We perform the assignment procedure five times per galaxy: once for each of the five variant fits, since each variant has slightly different flux and flux errors. The assignment is a sample selection, and therefore must be computed for all variant fluxes in order to estimate any redshift binning-based selection ``response'' in the shear estimates \citep{Sheldon2020}. See \citetalias{paper1} for more details on the calibration of the \decade shape catalog. We also note that our redshift estimates account for this selection response through the use of the response factor, $R$, in Equation \eqref{eqn:shear_weight} below.

While the \decade survey has significantly more variations in its image quality (\eg depth) relative to that of other weak lensing surveys such as DES, the actual source galaxy number density variations across the sky are fairly similar between \decade and DES Y3. This is shown in more detail in Figure 1 of \citetalias{paper3}. Similar to DES Y3, our source galaxies are cut on $m_i < 23.5$ to limit the sample to a magnitude regime with better redshift calibration. This cut also significantly alleviates the impact of depth variations on galaxy number density. As a result, the final \decade galaxy sample has similar number density variations as the DES Y3 sample.

\subsection{Deep sample}\label{sec:sec:deep}

We use the same deep-field galaxy sample as used in \citetalias{Myles:2021:DESY3}. In particular, this is the catalog presented in \citet{Hartley:2022:Y3Deepfields}, based on the DES Year 3 deep-field data. It contains the photometry of galaxies in the $ugriz$ bands, measured in DECam images, as well photometry in the $JHK_s$ bands measured in images taken by the VISTA infrared camera \citep{Dalton2006} in the VISTA Deep Extragalactic Observations \citep[VIDEO,][]{Jarvis2013} and UltraVISTA \citep{McCracken2012} public surveys. The object model fits are obtained using the \textsc{Fitvd} algorithm.\footnote{\url{https://github.com/esheldon/fitvd}} The photometry are dereddened to correct for interstellar extinction and include additional photometric calibrations described in \citet{Hartley:2022:Y3Deepfields}. This sample is used to obtain a more precise mapping between galaxy photometry and redshift than is possible when using the wide-field galaxy data alone, since the latter catalog contains noisy flux measurements from only three photometric bands while the deep-field catalog has low-noise measurements from eight bands.

We place a number of quality cuts on this sample. Primarily, the sample is required to have a measurement in all eight bands. In addition, we also place the color cut,
\begin{equation}\label{eqn:df_colorcuts}
    \prod_{i = 1} (m_{i - 1} - m_{i} > -1); \,\, i \in [u, g, r, i, z, J, H, K_s],
\end{equation}
where $m_i$ is the magnitude in a given band, and $m_{i - 1}$ is the same in the next bluest band. Equation \eqref{eqn:df_colorcuts} is defined as a product over boolean arrays and is thus equivalent to a logical ``AND'' operation. The above cut removes objects that exhibit extreme blue colors in any consecutive colors, defined as colors computed between photometric bands with bandpasses 
 that have adjacent wavelength coverage. Such objects are more likely to be affected by image artifacts in the pixel-level processing of the data. This color cut only removes 1\% of our final deep-field galaxy sample, and is included as a conservative cut to produce a more pristine sample. Running the SOMPZ pipeline without using this cut results in shifts to the mean redshifts, $\Delta \langle z \rangle$, that are within 2\% to 10\% of the calibration uncertainties derived in this work (Table \ref{tab:z_calibration}).

In addition, it was noted in \citetalias{Myles:2021:DESY3} that the SOMPZ calibration can be optimized by only using deep-field galaxies that have a reasonable likelihood of being detected in the wide-field dataset. In their work, this cut is performed using \Balrog, a synthetic source injection pipeline that can accurately predict the detection probabilities of different galaxies \citep{Suchyta:2016, Everett:2022, Anbajagane:2025:Y6Balrog}. See Section~\ref{sec:Balrog} for details on our implementation of this method for the \decade data. In this work, we follow \citetalias{Myles:2021:DESY3} and \citetalias{Sanchez:2023:highzY3} in only using deep-field galaxies that are detected at least once in the wide-field data. After all cuts mentioned above, we find 178,301 galaxies remaining in our deep-field sample.

\subsection{Redshift sample} \label{sec:sec:Zsample}

The redshift sample is a subset of the deep sample (Section~\ref{sec:sec:deep}), containing galaxies that have high-quality redshift estimates. The redshifts are obtained from a number of different datasets: (i) \Cosmos \citep{Laigle2016, Weaver:2022:Cosmos}, (ii) \PAUS plus \Cosmos \citep{Alarcon:2021:PausCosmos}, (iii) \CTRT \citep{Masters:2017:C3R2_DR1, Masters:2019:C3R2_DR2, Stanford:2021:C3R2_DR3}, (iv) VVDS \citep{LeFevre:2005:VVDS}, (v) z\Cosmos \citep{Lilly:2007:zCOSMOS}. The first two are multi-band photometric redshift estimates (with 30 and 66 bands, respectively) that have much higher precision than what can be achieved with the eight-band $ugrizJHK_s$ measurements of the deep-field sample. The remaining three surveys are spectroscopic datasets. For photometric datasets, we use a point-estimate for the redshift, taken as the median redshift from the galaxy's $p(z)$ as estimated by a given survey. 

This redshift sample is similar to that used in \citetalias{Myles:2021:DESY3} and \citetalias{Sanchez:2023:highzY3}, with two notable exceptions: First, our \CTRT data includes the third data release, whereas the DES Y3 works only had the first two data releases available for their analyses. This third data release increases the number of available spectra by 20\%. Next, we use an updated \Cosmos catalog. The two DES works mentioned above \citepalias{Myles:2021:DESY3, Sanchez:2023:highzY3} use the \Cosmos 2015 release from \citet{Laigle2016}, whereas in this work we use the updated, \Cosmos 2020 catalog from \citet{Weaver:2022:Cosmos}. This updated catalog uses two different image processing algorithms for photometric estimates, denoted as \textsc{Classic} and \textsc{Farmer}, and also two template fitting algorithms to obtain redshifts, denoted \textsc{Eazy} and \textsc{LePhare}. \citet{Laigle2016} used the \textsc{Classic} processing with the \textsc{LePhare} model. In this work, we use all four catalog variants from \citet{Weaver:2022:Cosmos} --- through a simple Monte-Carlo sampling as described in Section~\ref{sec:sec:Zuncert} --- to marginalize over differences in the redshift estimates corresponding to changes in the processing pipeline. Our tests of the redshift bias in the different \Cosmos samples indicate that the \textsc{Eazy} redshift estimate from the \textsc{Classic} image processing is more accurate (compared to the other versions of \Cosmos 2020) for the magnitude range relevant for our wide-field sample; see Figure~\ref{fig:Zsample_bias} for the estimated bias of this sample, and also Figure 13 in \citet{Weaver:2022:Cosmos} which shows that for $i$-band magnitudes brighter than $m < 24$ the \textsc{Eazy} model-derived redshifts have lower uncertainty and bias than those of the \textsc{LePhare} model. However, we perform a conservative analysis here and fold in the variation of redshift estimates across all four samples; see Section~\ref{sec:sec:Zuncert} for more details.

\begin{figure}
    \centering
    \includegraphics[width=\columnwidth]{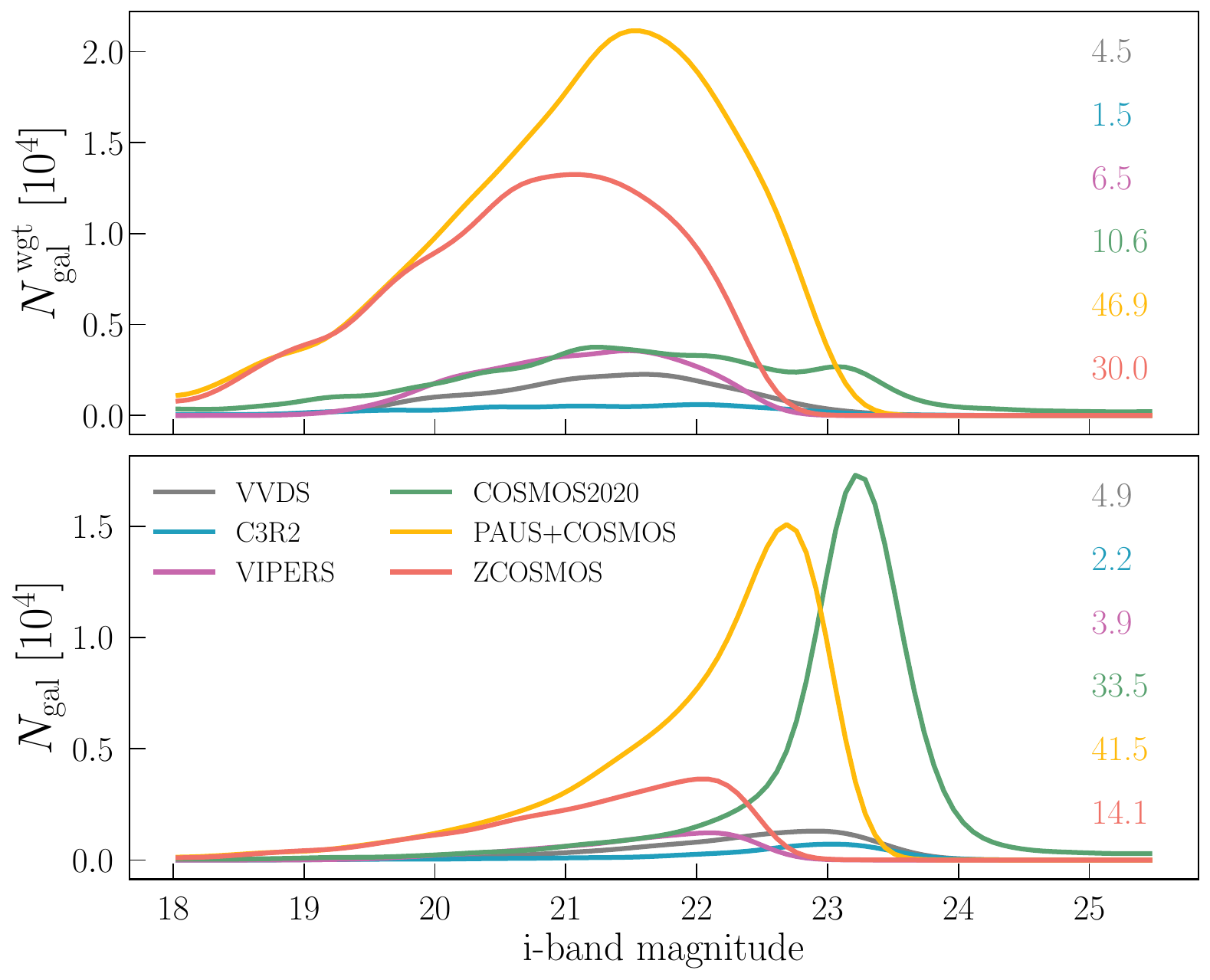}
    \caption{The weighted (top) and raw (bottom) counts of galaxies in the redshift sample, split by the source of their redshift information. The numbers on the right are the percentage contribution of a given sample to the total redshift sample. \Cosmos and \PAUS are photometric estimates (from 30 bands, and 66 bands, respectively) while the rest are spectroscopic samples. The distributions were smoothed with a narrow Gaussian kernel for visualization purposes. Our fiducial redshift estimates (which use weights) are primarily informed by the spectroscopic samples and the \PAUS plus \Cosmos sample.}
    \label{fig:Zsample}
\end{figure}

Of the fiducial deep-field sample defined above in Section~\ref{sec:sec:deep}, there are 51,541 objects (roughly 30\%) with a high quality redshift estimate. Figure~\ref{fig:Zsample} shows the magnitude distribution of the redshift sample, both in the weighted (top) and unweighted (bottom) case. The former represents the effective information taken from each redshift sample after accounting for the probability of detection and the lensing weights (described in Section~\ref{sec:sec:SOMPZ}), whereas the latter is simply the raw number counts. While \Cosmos is the dominant contributor to the raw counts, employing the weighting (which generally favors brighter, lower-redshift objects over fainter, higher-redshift ones\footnote{This is because fainter objects have a lower detection probability (see Figure~\ref{fig:BalrogCompleteness}), and they are generally low signal-to-noise objects with small sizes, and therefore have a lower lensing weight; see Figure 4 in \citetalias{paper1}.}) changes the dominant sample to a combination of the spectroscopic samples and the \PAUS plus \Cosmos sample, which contribute $\approx 45\%$ each. This is similar to the results of \citetalias{Myles:2021:DESY3} (see their ``SPC'' sample in Table 1).

\subsection{\Balrog sample}\label{sec:sec:Balrog}

The \Balrog sample refers to a catalog of synthetic injections in the \decade imaging dataset. This is obtained by injecting deep-field galaxies (Section~\ref{sec:sec:deep}) into images from which the wide-field galaxies are derived and then postprocessing these augmented images to derive synthetic galaxy catalogs. The simulated data therefore contain all the complexities in the images, such as the point-spread function, sky background, mask, crowding, artifacts, etc. See Section~\ref{sec:sec:Balrog} for a detailed description of the method. Through this method we can connect each deep-field galaxy with the distribution of properties it can exhibit in the (noisier) wide-field data, \ie we can estimate the distribution $P(X_{\rm wide} | Y_{\rm deep})$, where $X$ and $Y$ can be different types of measurements from completely different algorithms. For this reason we will refer to this dataset as the ``transfer function.'' Naturally, the \Balrog sample has quantities from both the deep-field galaxies, like the $ugrizJHK_s$ fluxes, and the wide-field galaxies, such as all \textsc{Metacalibration} measurements.

DES has its own synthetic source injection pipeline, denoted as \Balrog \citep{Suchyta:2016, Everett:2022, Anbajagane:2025:Y6Balrog}. We adopt the same nomenclature for our source injection pipeline within \decade but note that our pipeline was built as part of the \decade cosmic shear project and is a modified version of our image simulations pipeline in \citetalias{paper1}. See Section \ref{sec:Balrog} for more details.

\subsection{Clustering redshifts reference catalog} \label{sec:sec:reference}

Finally, we use a reference catalog of spectroscopically measured galaxies from the \Boss \citep[Baryon Oscillation Spectroscopic Survey,][]{Dawson:2013:BOSS} and \eBoss \citep[extended \Boss,][]{Dawson:2016:eBOSS} surveys. Specifically, we use the CMASS, ELG, LOWZ, LRG, and QSO samples.\footnote{The acronyms correspond to the various samples selection: constant mass-selected, emission line galaxies, low redshift, luminous red galaxies, and quasars. See the cited references for more details} These galaxies are used to estimate the redshift distribution of our source-galaxy sample using the clustering redshift method (Section~\ref{sec:sec:Wz}), which cross-correlates the sky positions of two galaxy samples and uses the correlation amplitudes to estimate the redshift distribution. See Section~\ref{sec:sec:Wz} for more details on the method, which broadly follows that of the DES Y3 calibration efforts \citep[][henceforth \citetalias{Gatti:2022:WzY3}]{Gatti:2022:WzY3}. We do not use any weights for the reference catalog.

\begin{figure}
    \centering
    \includegraphics[width=\columnwidth]{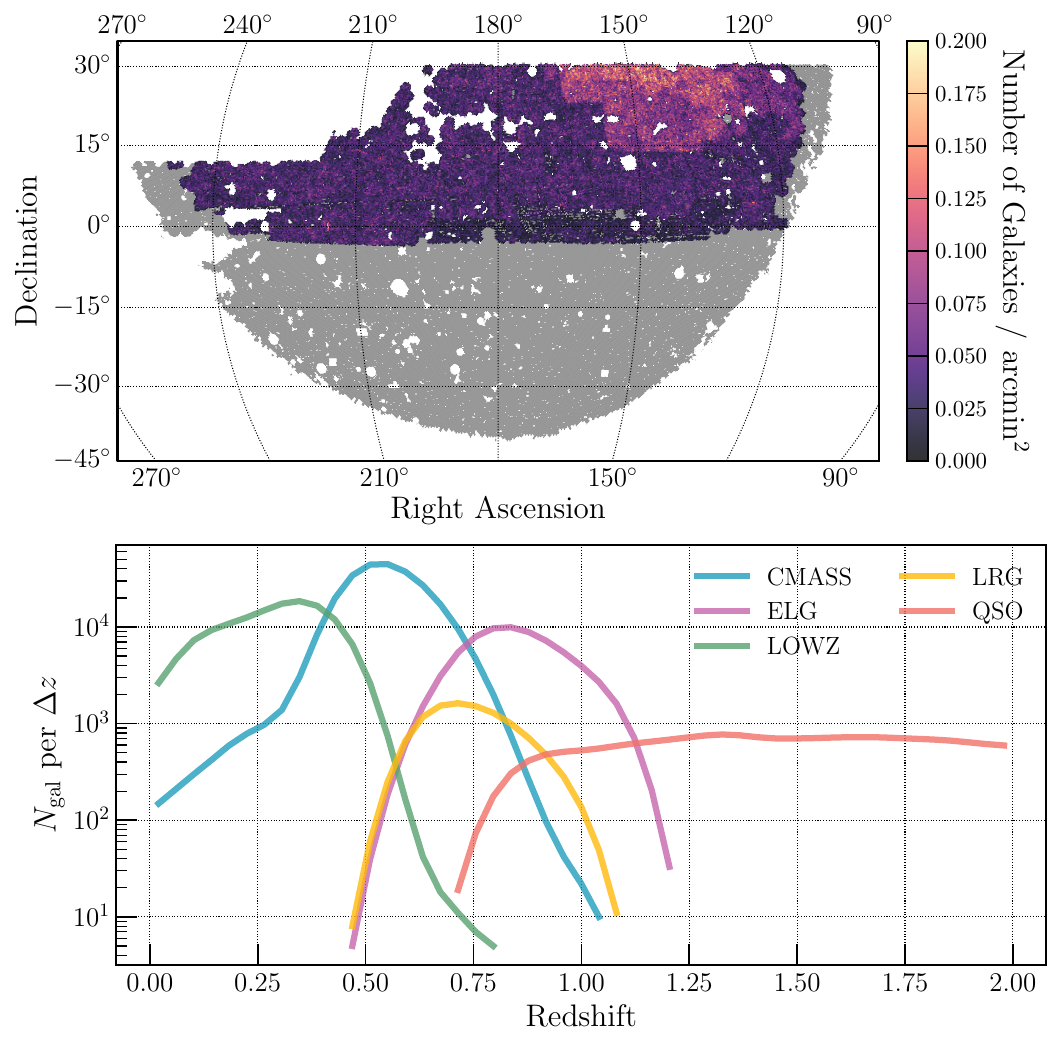}
    \caption{The distribution of the reference sample, across the sky (top) and across redshift (bottom). The reference sample has an overlap of around $3,\!000 \deg^2$ with the source-galaxy sample. We show the \decade footprint in gray for comparison. The inhomogeneity in number counts of the reference sample is due to the different samples (denoted in legend of bottom panel) spanning different areas on the sky. The number counts in the bottom panel have been smoothed by a narrow Gaussian kernel for visualization purposes.}
    \label{fig:ref_sample}
\end{figure}

Figure~\ref{fig:ref_sample} shows the distribution of the reference sample across the sky and across redshift. The sample spans $\approx\! 3,\!000 \deg^2$ of the DECADE footprint. This is nearly three times the area overlap compared to the overlap found in the DES analysis \citepalias[][see their Table 1]{Gatti:2022:WzY3}; given our footprint spans higher declination than DES, we have more overlap with the spectroscopic surveys. As a consequence, we are able to use more galaxies in our clustering redshift estimates and reduce the sample variance of the measurement. In practice, however, our sample is less optimal than that used in DES Y3. As we will discuss in Section~\ref{sec:sec:Wz}, the clustering redshift estimates use measurements on small-scales, where the limiting uncertainty is shot noise --- which depends on sample number density --- and not sample variance. Our reference sample has a lower number density than the one used in \citetalias{Gatti:2022:WzY3}, as the latter contains sky regions that have been observed to much greater depth by the spectroscopic surveys. As a result, our clustering redshift measurements are noisier than those of \citetalias{Gatti:2022:WzY3}, particularly for higher redshifts.

\section{Methods}\label{sec:Methods}

The primary method of our redshift calibration is the SOMPZ technique, as is the case in DES Y3 \citepalias[\eg][]{Myles:2021:DESY3, Sanchez:2023:highzY3}. This technique is described in Section~\ref{sec:sec:SOMPZ}. Similar to DES Y3, we add to this method three sources of uncertainty, associated with sample variance (Section~\ref{sec:sec:3sdir}), photometric zeropoint calibrations (Section~\ref{sec:sec:ZPuncert}), and redshift bias (Section~\ref{sec:sec:Zuncert}). We also describe the clustering redshift method, which we use to cross-check the SOMPZ-based $n(z)$, in Section~\ref{sec:sec:Wz}.

\subsection{Self-Organizing Maps Photometric Redshifts (SOMPZ)}\label{sec:sec:SOMPZ}

As mentioned above, SOMs are a type of unsupervised learning method used to represent high-dimensional data in a low-dimensional, discrete space. This makes them ideal for clustering and pattern recognition in data. We follow previous large-scale structure studies, such as DES \citep[\citetalias{Myles:2021:DESY3, Sanchez:2023:highzY3};][]{Giannini2022, Campos:2024:PZ} and KiDS \citep{Wright2020, Wright:2020:KidsSOMS, vandenBusch2022}, in using SOMs to estimate photometric redshifts for our galaxy sample.

In the SOMPZ methodology, galaxies are classified into photometric \textit{phenotypes}, where a phenotype refers to some subcategory of the galaxy sample. The subcategory can be defined using any combination of properties from the sample. For our work, we will use the galaxy photometry (the flux and flux errors) since galaxy colors are correlated with redshift \citep[\eg][]{Wang:1998:redshiftcolor, Masters:2015:SOM}. We note that the SOM used in \citetalias{Myles:2021:DESY3} only utilized the galaxy fluxes and not their errors. In this work, we use the updated SOM algorithm from \citet[][see their Appendix A]{Sanchez:2020:NoiseSOM} that builds the SOM using both fluxes and flux errors. This algorithm is better suited for working with noisier/fainter objects. The updated SOM framework from \citet{Sanchez:2020:NoiseSOM} has already been used in the DES analysis of \citetalias{Sanchez:2023:highzY3} to study high-redshift galaxies ($0.8 < z < 3$) and has also been validated for weak lensing purposes in anticipation of DES Y6 \citep{Campos:2024:PZ}. When building/training the SOM on a given sample, we use at most two million galaxies from the sample; if the sample is larger (as is the case with the wide-field galaxies) then we randomly subsample it. The training step defines the different phenotypes from the data and generates the SOM weights. The trained SOM can then be used to classify all galaxies from the sample into the SOM-based phenotypes.

The SOMPZ redshift estimates are obtained by learning a probabilitsic relationship between wide-field photometry and redshift. This is done in three distinct steps, which use two different SOMs and three different samples. The samples are the wide-field sample (Section~\ref{sec:sec:wide}), deep-field sample (Section~\ref{sec:sec:deep}) --- which also includes the redshift sample (Section~\ref{sec:sec:Zsample}) --- and the \Balrog sample (Section~\ref{sec:sec:Balrog}). The two SOMs are (i) a wide SOM, which is trained using the $riz$ wide-field, \textsc{Metacalibration}-based photometry and has 32$\times$32 cells, and; (ii) the deep SOM, which is trained using the $ugrizJHK_s$ deep-field photometry and has 48$\times$48 cells. 

The SOM size, denoted by the number of cells, is a hyper-parameter in the model and can take any integer value. Previous works have used different sizes for each SOM: \citetalias{Myles:2021:DESY3} used 32$\times$32 and 64$\times$64 for the wide SOM and deep SOM, respectively, while the subsequent work of \citetalias{Sanchez:2023:highzY3} used 22$\times$22 cells for the wide SOM and 48$\times$48 cells for the deep SOM. The change in the latter work, relative to the former, was because they had a more stringent magnitude selection ($22<m_i<23.5$), and therefore fewer deep-field galaxies passed the selection cuts. A general heuristic is that the SOM size should scale with the volume of the photometric space \citepalias[][see their Section 3.1]{Sanchez:2023:highzY3}. Our work finds similar behavior as \citetalias{Sanchez:2023:highzY3} for the deep-field sample, as the slightly shallower depth of \decade compared to DES Y3 means the number of detectable objects is smaller. However, our wide-field, source-galaxy sample has a similar size to DES Y3. Thus, we set the deep-field SOM size based on \citetalias{Sanchez:2023:highzY3} and the wide-field SOM size based on \citetalias{Myles:2021:DESY3}.

\begin{table}
    \centering
    \begin{tabular}{c|c}
    Notation & Description \\
    \hline
    $c$   & cell index in deep SOM \\
    $\chat$   & cell index in wide SOM \\
    $\selhat$ & Selection cuts of wide-field sample \\
    $\bhat$ & Tomographic binning selection \\
    \hline
    \end{tabular}
    \caption{Table describing the notation used throughout this work.}
    \label{tab:notation}
\end{table}

The SOMPZ method estimates the final redshift distribution of the wide-field sample through a Bayesian approach, as a composition of conditional probability distributions. Namely, we write:
\begin{equation}\label{eqn:pz_chat}
    p(z|\chat, \selhat) = \sum_c p(z|c, \chat, \selhat)p(c|\chat, \selhat),
\end{equation}
where $c$ and $\chat$ refer to a deep and wide SOM cell, respectively, and $\selhat$ refers to the selection criteria for the wide catalog (see Section~\ref{sec:sec:wide} above or Section 3.2 of \citetalias{paper1}). Table~\ref{tab:notation} also provides a brief overview of the notation in this work. Given Equation~\eqref{eqn:pz_chat}, we can define the redshift distribution in a given tomographic bin as,
\begin{equation}\label{eqn:pz_bhat}
    p(z|\bhat, \selhat) = \sum_{\chat \in \bhat}\sum_c p(z|c, \chat, \selhat)p(c|\chat, \selhat)p(\chat |  \bhat, \selhat),
\end{equation}
where $\sum_{\chat \in \bhat}$ indicates a summation over only wide cells, $\chat$, that are assigned to a given tomographic bin $\bhat$. Section~\ref{sec:sec:Tomobins} details the procedure through which SOM cells (and therefore galaxies) are assigned to tomographic bins.

While Equation \eqref{eqn:pz_bhat} is the exact expression for probabilistically assigning a redshift distribution to a sample, it requires an estimate of $p(z|c, \chat, \selhat)$. In practice, this probability distribution is poorly measured given the possible unique pairs of $(c, \chat)$ numbers $32^2 \times 48^2=$ 2.4 million, which is of the same order as the available number of galaxies in our \Balrog sample. Thus, an alternative is to condition on $\bhat$ instead of $\chat$. This step, called ``bin-conditionalization'', modifies Equation \eqref{eqn:pz_bhat} as
\begin{equation}\label{eqn:pz_bhat_bc}
    p(z|\bhat, \selhat) \approx \sum_{\chat \in \bhat}\sum_c p(z|c, \bhat, \selhat)p(c|\chat, \selhat)p(\chat |  \bhat, \selhat).
\end{equation}
This swap in the conditioning variable is allowed under the (reasonable) assumption that the redshift of a galaxy in deep SOM cell $c$ is only weakly dependent on its noisy, wide-field photometry. This is the approach used in \citetalias{Myles:2021:DESY3} and is verified further in \citetalias{Sanchez:2023:highzY3} (see their Appendix B3).

Each term in Equation~\eqref{eqn:pz_chat} and Equation~\eqref{eqn:pz_bhat} can be interpreted as follows:

\begin{enumerate}
    \item $p(z|c, \bhat, \selhat)$ is the probability distribution for the redshift of the galaxies assigned to a given deep SOM cell $c$, a given tomographic bin $\bhat$, and that also pass the sample selection cuts $\selhat$.  \vspace{10pt}
    \item $p(c|\chat, \selhat)$ is the probability of a galaxy being assigned to deep SOM cell $c$, given it is also assigned to the wide SOM cell $\chat$, and also given it passes the wide-field selection cuts. This term is the transfer function, and is obtained through our \Balrog sample (Section~\ref{sec:sec:Balrog}).\vspace{10pt}
    \item $p(\chat |\bhat, \selhat)$ is the probability of a galaxy from the wide-field catalog being classified into cell $\chat$, given it passes selection criteria $\selhat$ and is assigned to bin $\bhat$.\vspace{10pt}
\end{enumerate}

Throughout this work, all galaxies are weighted by a modified shear weight,
\begin{equation}\label{eqn:shear_weight}
    w_z = R w_\gamma,
\end{equation}
where $w_z$ is the weight used for the redshift calibration work, $R$ is the shear response measured in \textsc{Metacalibration}, and $w_\gamma$ are the shear weights defined in \citetalias{paper1} (see their Section 3.3). These shear weights are used in all analyses done with the shear catalog \citepalias{paper1, paper4} and help increase the signal-to-noise of the measurements. The factor of $R$ in Equation \eqref{eqn:shear_weight} is then required to correct for shear-dependent measurement biases; see Section 2.2 and Equation (10) in \citet{Maccrann2022ImSim} for a more detailed description of the interaction between the lensing response, $R$, and the effective source-galaxy redshift distribution. The $R$ used in the weight above is not the individual galaxy's estimate of the response and is instead the average response computed on a grid of signal-to-noise and size-ratio. This follows the same procedure used to obtain the weights $w_\gamma$, as detailed in \citetalias{paper1} (see their Figure 4). In practice, the weights are propagated through our pipeline using the formalism detailed in Appendix D of \citetalias{Myles:2021:DESY3}, which pertains to the uncertainty sampling technique described below in Section~\ref{sec:sec:3sdir}.

The redshift distributions, $p(z|\bhat, \selhat)$, are also postprocessed in one additional way. Namely, we force the distribution to follow $n(z \rightarrow 0) \rightarrow 0$ (``ramping''). This is a well-motivated modification given our prior knowledge that the source-galaxy number density approaches zero as the redshifts approaches $z \rightarrow 0$. In practice, we follow \citetalias{Myles:2021:DESY3} (see their Appendix B4) and transform the distribution as $n(z) \rightarrow n(z) f(z)$ with a function $f(z)$ that increases linearly across $0 < z < 0.05$:
\begin{align}
    f(z) = 
    \begin{cases}
        z/0.05 & 0 < z \leq 0.05\\
        1 & z > 0.05.\\
    \end{cases}
\end{align}
The SOMPZ method does not automatically fold-in this prior (or an analogous one) as the algorithm is purely statistical in its formalism, with no cosmological context/information about the allowed distribution of sources over redshift. This aspect of the method will be relevant later as well, when we discuss comparisons with clustering-based estimates of the $n(z)$. In practice, the inclusion of ramping changes the mean redshift of the $n(z)$ by $\Delta \langle z \rangle \lesssim 10^{-3}$ in the first tomographic bin, and $\Delta \langle z \rangle \lesssim 10^{-4}$ in the rest. These shifts are less than $4\%$ of the final calibration uncertainties (Table \ref{tab:z_calibration}).

\subsection{Construction of tomographic bins}\label{sec:sec:Tomobins}

The cosmological power of weak lensing datasets can be amplified by splitting the data into different tomographic redshift bins, as this increases the sensitivity to redshift-dependent signals \citep{Hu:1999:WL_Tomography}. In principle, there is complete freedom in how we assign galaxies to bins: a good binning choice increases the constraining power of the data, but a bad binning choice will not generate biases in the analysis and only degrade constraining power (depending on the analysis being performed). Our binning procedure follows \citetalias{Myles:2021:DESY3} and is performed as follows:

\begin{enumerate}
    \item For each wide cell, $\chat$, find all galaxies from the \Balrog sample that are assigned to that cell.\vspace{5pt}
    \item Select \Balrog galaxies in cell $\chat$ whose deep-field counterpart is part of the redshift sample. Then compute the median redshift of the galaxies in this cell.\vspace{5pt}
    \item Rank the cells by their median redshift. Add the ranked cells one-by-one to the lowest redshift bin until the probability $\sum_{\chat \in A} p(\chat | \selhat) \approx 0.25$. Then assign cells to the next lowest redshift bin, until the probability reaches $0.25$ again. This is repeated to construct four tomographic bins with nearly equal number counts of galaxies.
\end{enumerate}

The above procedure results in each $\chat$ being assigned to one unique bin (out of the four tomographic bins). Galaxies are assigned to each bin based on the $\chat$ cell they correspond to. The tomographic binning of the sample is then propagated to the rest of our analyses, including to the shear calibration presented in \citetalias{paper1}. 

For 39 out of the 1024 cells of the wide SOM, all \Balrog galaxies in that $\chat$ phenotype have no redshift information. A SOM structure is such that similar phenotypes cluster around the same part of the grid. Therefore, we can use neighboring cells to provide an approximate estimate of the median redshift in the 39 empty cells. We do so using the \texttt{CloughTocher2DInterpolator}\footnote{We note that this interpolator is used to interpolate noise images for our shear estimation procedure \citep{Sheldon2017, Huff2017} and is therefore a natural choice for interpolating across somewhat noisy quantities in the SOM grid.} \citep{Alfeld:1984:Interp} as implemented in \textsc{Scipy} \citep{Virtanen2020Scipy}. We treat the SOM map of $\langle z \rangle$, shown in Figure \ref{fig:WideSOM}, as an image and pass it to the interpolator.\footnote{The exact ``distance'' between SOM cells is given by the SOM $U$-matrix \citep{Kohonen:1982:SOM, Kohonen:2001:SOM}. The chosen interpolator scheme, however, treats the cells as pixels in an image and considers the Euclidean distances between these pixels during interpolation. Any inaccuracies in this step do not result in biased $n(z)$ samples as they only affect the tomographic bin choices and for only 1\% of the sample.} We find 32 out of the 39 cells are assigned to the 4th tomographic bin, with the remaining cells split between the other three bins. We reiterate that any reasonable interpolation procedure is an equivalently valid choice as different binning methods can only change the optimality of the sample and cannot induce biases. Only 1\% of our source galaxy sample is contained in these 39 cells.

\begin{figure*}
    \centering
    \includegraphics[width = 2\columnwidth]{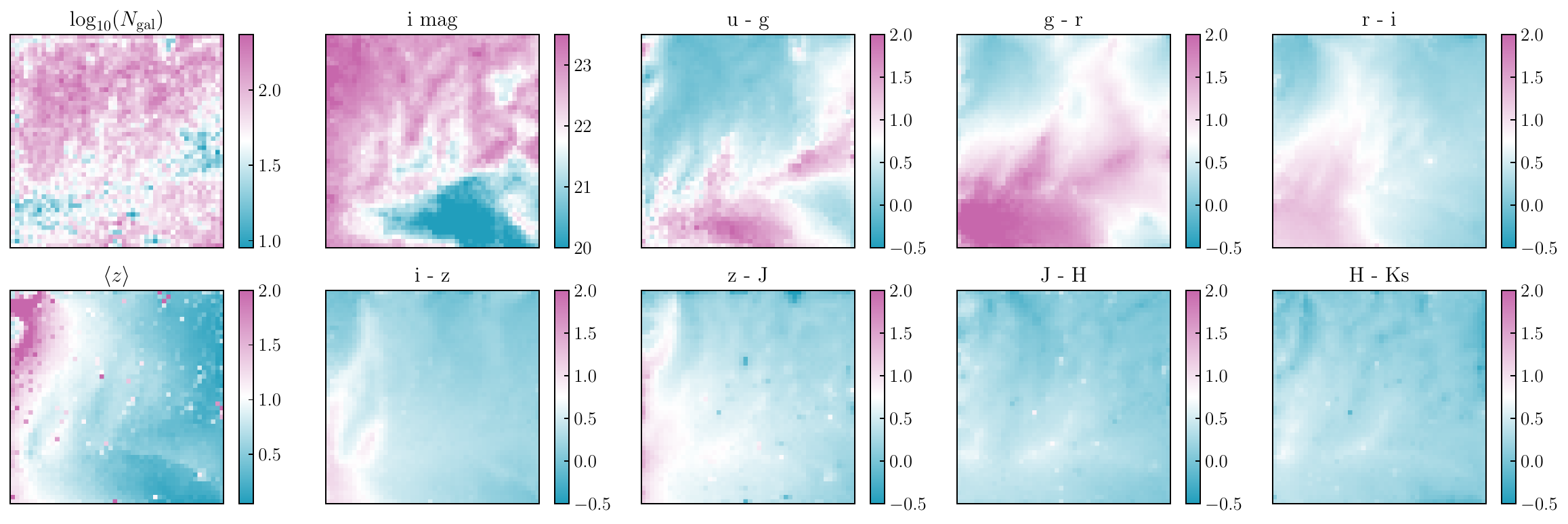}
    \caption{The properties of the deep SOM, which has $48\times48$ cells. We show the number counts of galaxies per cell, the average $i$-band magnitude, the average colors, and the average redshift per cell. There is clear spatial structure in the 2D maps of different properties, as is expected from using a SOM to classify galaxy photometry into phenotypes.}
    \label{fig:DeepSOM}
\end{figure*}

\begin{figure*}
    \centering
    \includegraphics[width = 2\columnwidth]{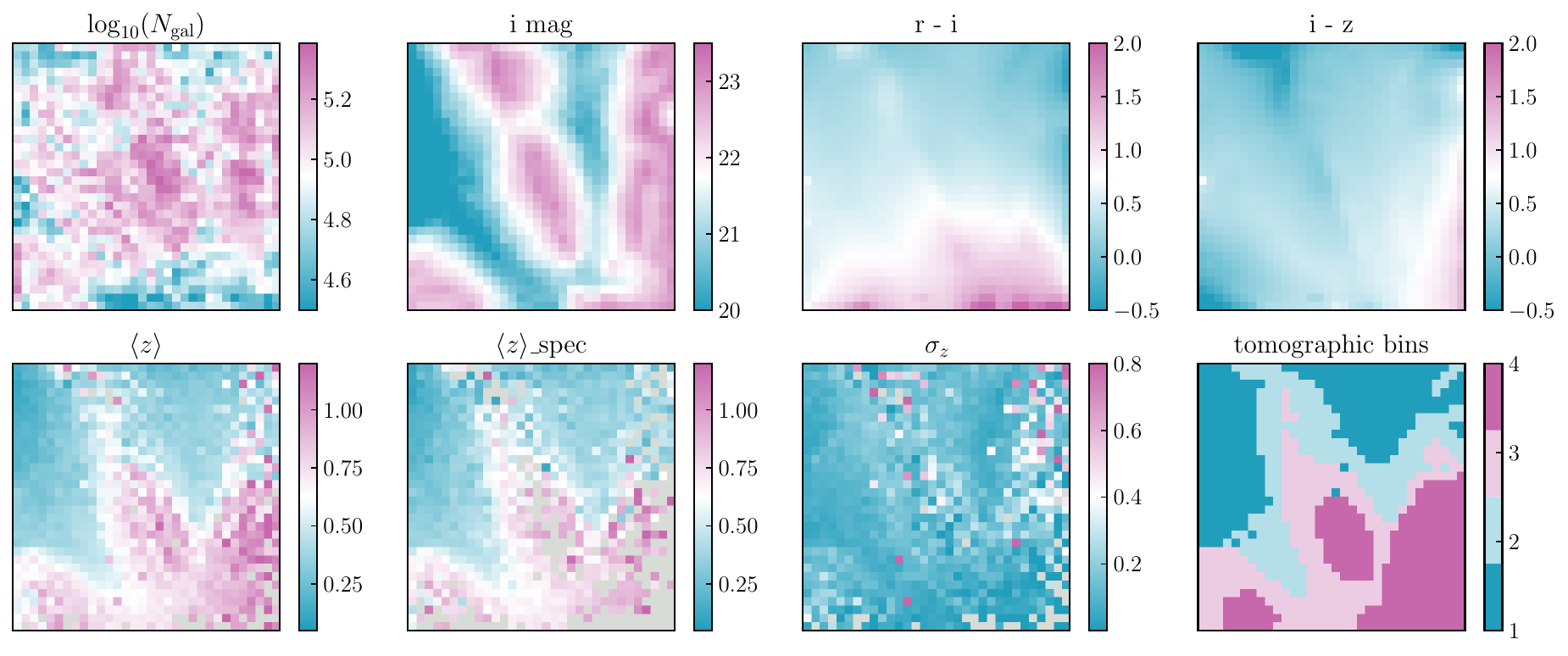}
    \caption{The properties of galaxies in the wide SOM, which has $32\times32$ cells. We show the number counts of galaxies per cell, the average $i$-band magnitude, the average colors, the average redshift per cell, the dispersion in the redshift per cell, and the tomographic bin assignment per wide cell. We also show one version of the average redshift estimated using only spectroscopic redshifts. Both versions of $\langle z\rangle$ are obtained by computing the mean redshift of \Balrog sample galaxies in a given $\chat$ cell. The cells with faint magnitudes are  frequently assigned to tomographic bins 3 and 4.}
    \label{fig:WideSOM}
\end{figure*}

Figures~\ref{fig:DeepSOM} and \ref{fig:WideSOM} show the colors, magnitudes, redshifts, etc. of the deep SOM and wide SOM, respectively. Both SOMs show clear structures in the magnitude/color-space that are also correlated with the average redshift in a given cell.

\subsection{Three-step Dirichlet sampling (3sDir)} \label{sec:sec:3sdir}

The SOMPZ method describes the probabilitic mapping of galaxy photometry to a redshift, and produces a distribution $n(z)$ for the wide-field galaxy sample. However, this method on its own does not quantify the uncertainties on this redshift distribution. Such quantification is a key requirement for performing robust cosmological analyses with weak lensing datasets. The primary component of the uncertainty is the sample variance and shot noise of the deep-field dataset. The deep-field catalogs used in our work (Section~\ref{sec:sec:deep}) correspond to four patches of the sky of $\approx\! 1.5\deg^2$ each, and therefore contribute significant sample variance towards any inference that uses their measurements. In addition, the finite number of galaxies in our final deep sample ($N_{\rm gal}^{\rm \,deep} = 178,301$) will contribute a shot noise term as well.

The uncertainty contributions from sample variance and shot noise are added to our pipeline using the Three-step Dirichlet method, abbreviated by ``3sDir'' henceforth, as described in Appendix D of \citetalias{Myles:2021:DESY3}. For the sake of brevity, we do not replicate their detailed discussion in this work and only reproduce the main, salient points of the method.

We start by rewriting Equation \eqref{eqn:pz_bhat} as
\begin{equation}\label{eqn:3sdir}
    p(z|\bhat, \selhat) \approx \sum_{\chat \in \bhat}\sum_c \frac{f^\mathR_{zc}}{f^\mathR_c} f_{c}^\mathD \frac{f_{c\chat}^\mathB}{f_c^\mathB f_{\chat}^\mathB} f_{\chat}^\mathW,
\end{equation}
where the coefficients $f$ represent the joint probability of the quantities listed in the subscript. For example $f_{zc}$ is interpreted as $p(z, c)$. The superscripts correspond to the sample the coefficients are estimated for: $\mathR$ is the redshift sample, $\mathB$ is the \Balrog sample, $\mathD$ is the deep-field sample, and $\mathW$ is the wide-field sample. The coefficients can be accurately represented by the number counts of galaxies, \ie $f_{X} = N_{X}$ where $X$ is some set of selections on one or more of $z, c, \chat$. The probability of the coefficients given the observed counts, $N_X$, is
\begin{align}\label{eqn:Probf}
    p(\boldsymbol{f} | N_X) = \,\, & p(N_X | \boldsymbol{f}) p(\boldsymbol{f})\nonumber\\
    & = \bigg[\prod_X (f_X)^{N_X}\bigg]p(\boldsymbol{f}).
\end{align}
Here, $\boldsymbol{f} = {f_X}$ is the vector of all values of $f$ in the different bins corresponding to selection $X$.

In \citetalias{Myles:2021:DESY3}, the prior $p(\boldsymbol{f})$ was chosen to be a Dirichlet distribution, which has advantageous analytic properties such as enforcing $f > 0$ and $\sum f = 1$, and is also the natural distribution for the likelihood of binned counts. Under this choice, the posterior distribution in Equation \eqref{eqn:Probf} can be rewritten as,
\begin{align}\label{eqn:Probf_Dir}
    p(\boldsymbol{f} | N_X) = \,\, & Dir(\boldsymbol{f}; \boldsymbol{\alpha} = \{N_X + \epsilon\})\nonumber\\
    & \propto \delta\bigg(1 - \sum_X f_X\bigg)\bigg[\prod_X (f_X)^{N_X - 1 + \epsilon}\bigg],
\end{align}
where $\boldsymbol{\alpha}$ is the vector of coefficients corresponding to the vector $\boldsymbol{f}$. The delta function enforces $\sum_X f_X = 1$, and we add a small positive number $\epsilon$ to the counts to ensure the distribution cannot get zero or negative counts as input; some bins, for example in a grid of $z$ and $c$, will have no galaxies and therefore be empty. Each $f$ coefficient in Equation \eqref{eqn:3sdir} can be modelled using Equation \eqref{eqn:Probf_Dir}. We can now use our measurements, which are the counts $N_X$, to sample different possible combinations of the coefficients $f_X$ and generate different $n(z)$ samples. This ensemble of samples will account for the shot noise, due to finite sample sizes.

In addition, we will account for sample variance of the galaxy counts in the deep field.\footnote{The wide-field dataset contributes minimal sample variance to the $n(z)$ estimates given that data covers an area 1000 times larger than the deep field. It also contributes minimal shot noise given its significantly larger sample size.} This variance will also vary as a function of redshift, which must be accounted for. We follow \citetalias{Myles:2021:DESY3} to include sample variance in the existing method by simply rescaling the coefficients of the Dirichlet distribution,
\begin{equation}\label{eqn:alpha_CV}
    \alpha_i \rightarrow \alpha_i / \lambda_i,
\end{equation}
where the index $i$ runs over all values in the vector $\boldsymbol{f}$. The mean and variance of $f_i$ follow $\langle f_i \rangle = \alpha_i / \sum_i \alpha_i$ and ${\rm Var}(f_i) = \alpha_i / (\sum_i \alpha_i)^2$, respectively. As a result, the transformation of Equation \eqref{eqn:alpha_CV} does not change the mean but scales the variance as $\text{Var}(f_i) \rightarrow \lambda_i\text{Var}(f_i)$, for coefficients with $\alpha_i \ll \sum_i \alpha_i$. Under this prescription, $\lambda_i$ is estimated as the the variance from both sample variance and shot noise divided by the variance from just shot noise. \citetalias{Sanchez:2020:NoiseSOM} have validated the accuracy of the entire sample variance formalism in 3sDir using simulations.

Note that different cells, $c$, that have overlapping redshift distributions will have correlated counts due to sample variance. We must therefore ensure the correct correlations are induced into the data during sampling. The work of \citetalias{Sanchez:2020:NoiseSOM} handles this step by grouping highly correlated cells, $c$, into ``super''-phenotypes, $T$. These different super-phenotypes are disjoint in redshift and therefore have vanishing correlations with one another. Thus, in practice, we use a coefficient $\lambda$ per superphenotype $T$. The sample variance estimate we use is the same as \citetalias{Myles:2021:DESY3}, which in turn follows the prescriptions of \citetalias{Sanchez:2020:NoiseSOM}. The latter work estimated the sample variance assuming a circular patch of sky with the same area as the redshift sample, and showed this method is accurate at the 10--20\% level, with differences arising primarily due to assumptions of the galaxy bias modelling. As mentioned above, \citetalias{Sanchez:2020:NoiseSOM} have validated the current method and choices using a set of simulations.

We direct readers to Appendix D of \citetalias{Myles:2021:DESY3} for more in-depth details on the practical implementation of the 3sDir algorithm.

\subsection{Zeropoint calibration uncertainty} \label{sec:sec:ZPuncert}

In addition to sample variance and shot noise, measurement systematics in the photometry of the deep-field galaxies also contributes an uncertainty to the $n(z)$ estimate. In particular, the uncertainty on the zeropoint calibration of the deep-field exposures can be relevant. For example, in the case where the calibration is catastrophically wrong, the color-redshift relationship learned from the deep fields (and subsequently applied to the wide fields) is incorrect. This is particularly important in our analysis setup (as is the case with \citetalias{Myles:2021:DESY3} and \citetalias{Sanchez:2023:highzY3}), as most of our redshift estimates are in the one (out of the four) DES deep field that overlaps with the \Cosmos field whereas a majority of the galaxy color data is in the rest of the three DES deep fields. Thus, we learn the color-redshift relation from the \Cosmos deep field and extrapolate it to the rest of the deep fields. For this extrapolation to be accurate, the relative zeropoint offsets between the different fields must be modeled accurately.

We account for the zeropoint calibration uncertainty through a simple Monte Carlo sampling. The calibration uncertainties are provided in Table 5 of \citet{Hartley:2022:Y3Deepfields}, and are estimated directly from the data by matching the DES Y3 deep-field objects to their DES Y3 wide-field counterparts. The Monte Carlo sampling follows \citetalias{Sanchez:2023:highzY3}, which is different from the procedure of DES Y3 \citepalias{Myles:2021:DESY3}; the latter work used a set of realistic simulations (derived from $N$-body simulations and designed specifically to mimic DES data) to estimate the effect of the zeropoints on the final redshift distributions. We do not have such a $N$-body derived simulation that mimics the \decade data and therefore opt for the Monte Carlo method which does not require such data products but can still incorporate zeropoint uncertainties into the final $n(z)$ estimates.

In practice, we implement the sampling over the zeropoint uncertainty by repeating the 3sDir step from Section~\ref{sec:sec:3sdir} a total of $N = 100$ times. With each repetition, we perturb the photometry of the deep-field galaxies by a small amount. The galaxies are then reclassified into the deep SOM,\footnote{The SOM training is always fixed --- that is, the SOM weights are computed only once, using the fiducial deep-field sample --- and only the classification is redone with each of the 100 variants of the perturbed deep-field photometry.} and the entire $n(z)$ procedure is run again. For each repeat, we randomly apply a zeropoint offset assuming a Gaussian with a standard deviation given by Table 5 of \citet{Hartley:2022:Y3Deepfields}. We draw one offset per band (8 in total) and per deep field (4 in total), which results in 32 numbers in total. The contribution of this uncertainty to the final $n(z)$ is discussed in more detail in Section \ref{sec:Results_Nz} and Table \ref{tab:z_calibration}.

\subsection{Redshift calibration uncertainty} \label{sec:sec:Zuncert}

\begin{figure}
    \centering
    \includegraphics[width=1\columnwidth]{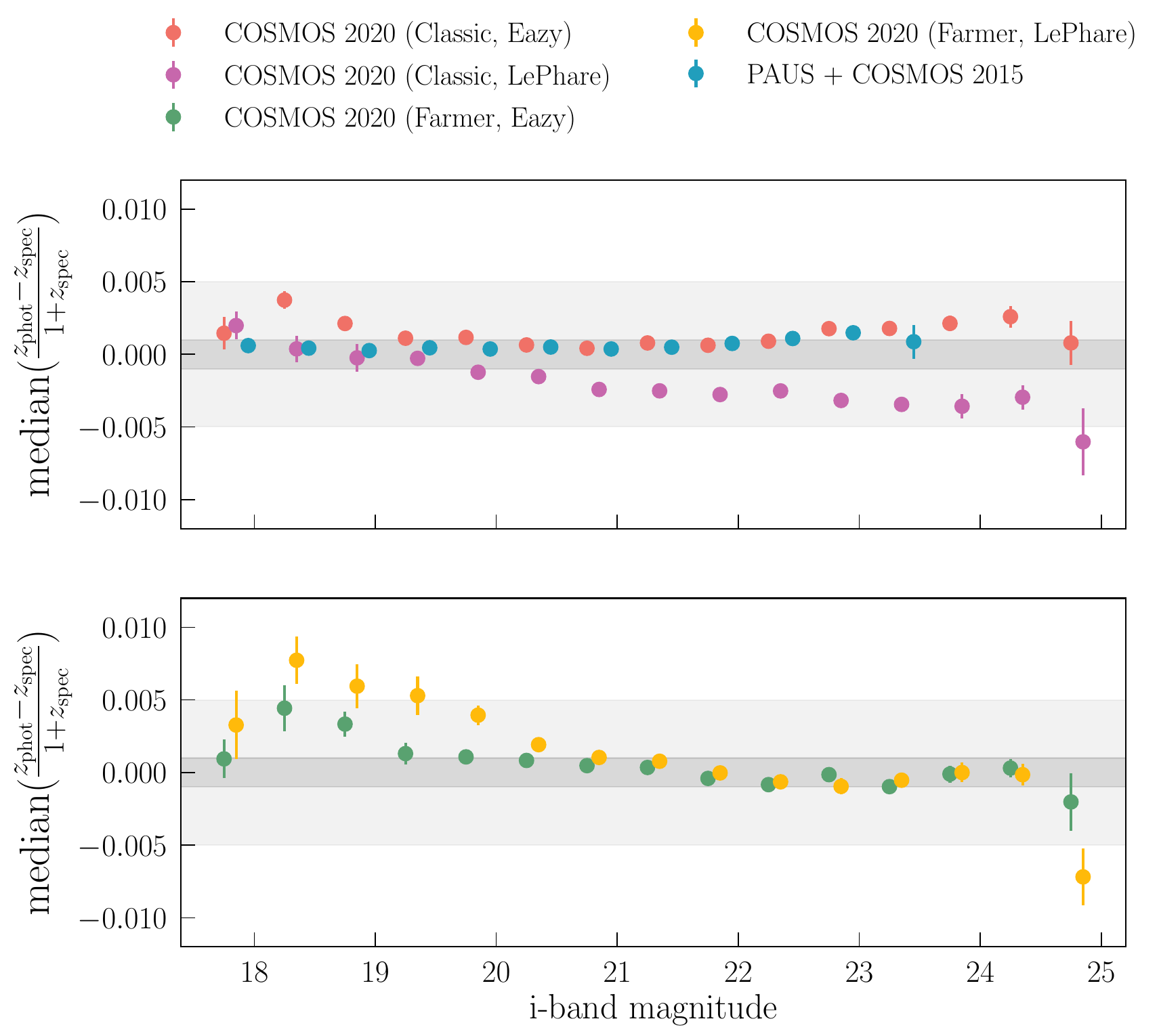}
    \caption{The median redshift bias as a function of magnitude for the two photometric redshift samples used in this work. The baseline sample is a curated list of high-quality spectroscopic redshifts, consisting of \textsc{Zcosmos} and \CTRT. This follows the choices in \citetalias{Sanchez:2023:highzY3}. The bias of \PAUS plus \Cosmos is within $0.001(1 + z)$, while that of \Cosmos is slightly higher for the brightest/faintest objects. The errors are estimated through a simple bootstrap and therefore only account for shot noise and not sample variance.}
    \label{fig:Zsample_bias}
\end{figure}

Finally, we incorporate an uncertainty on the redshift estimates found in the redshift sample (Section~\ref{sec:sec:Zsample}). Spectroscopic redshift estimates are precise enough to be treated as true redshifts, while estimates from multi-band photometric measurements --- such as those from \Cosmos or \PAUS plus \Cosmos\,--- can be biased in some systematic way. \citetalias{Sanchez:2023:highzY3} (see their Figure B1) show an estimate of this bias, as a function of $i$-band magnitude, for a slightly older version of the \Cosmos and \PAUS plus \Cosmos samples. The estimate for this work is shown in Figure~\ref{fig:Zsample_bias}. Similar to \citetalias{Sanchez:2023:highzY3}, we find that the data from \PAUS and \Cosmos is accurate to below $\Delta z < 0.001(1 + z)$ while the data from \Cosmos alone (where the available photometric bands is 30 fewer than \PAUS and \Cosmos combined) can be more biased. The latter statement depends on the exact sample being considered. As discussed in Section~\ref{sec:sec:Zsample}, the \Cosmos 2020 dataset provides catalogs with two different image processing methods (\textsc{Classic}, and \textsc{Farmer}) and two photo-z estimation methods (\textsc{LePhare}, \textsc{Eazy}). The two \textsc{Farmer} catalogs perform better (\ie have a lower bias in redshift) for fainter objects, whereas the \textsc{Classic} catalogs are better for brighter objects. In Figure~\ref{fig:Zsample_bias}, the bias is estimated using the subset of deep-field galaxies that have both a photometric redshift estimate from the sample whose bias is being estimated and a spectroscopic redshift estimate from either \CTRT or \textsc{Zcosmos}.

Similar to the zeropoint uncertainty marginalization method detailed above in Section~\ref{sec:sec:ZPuncert}, we perform 100 Monte-Carlo samplings over the possible redshift uncertainties. We perturb the galaxy redshifts by a Gaussian which has a magnitude-dependent mean given by the bias in Figure~\ref{fig:Zsample_bias} and then a magnitude-dependent standard deviation also given by the bias. We then redo our $n(z)$ estimation using the perturbed redshift sample. This is a conservative procedure as we are setting the bias as our $1\sigma$ uncertainty. We also account for variation between the different \Cosmos catalogs by randomly choosing one of the four catalogs during each Monte-Carlo iteration during our $n(z)$ estimation. Thus, our redshift calibration uncertainty accounts both for the biases within a catalog as shown in Figure~\ref{fig:Zsample_bias}, but also for the variation in photometric redshift estimates across catalogs.

Note that the tomographic binning, which also requires the redshift sample, is not re-run each time. This is primarily because redshift biases in the deep-field galaxy sample will only cause the binning to be less optimal and will not cause it to be biased. Fixing the binning allows us to isolate the impact of the redshift bias on the specific quantity of interest: the $n(z)$ of the different bins.

\subsection{Clustering redshifts (WZ)}\label{sec:sec:Wz}

It is useful to have another, independent estimate of the redshift distribution to serve as a comparison metric for the SOMPZ estimates. Following DES Y3, we choose clustering redshifts as this alternative estimate, and following DES nomenclature, we will denote this method by ``WZ'' for simplicity. This technique produces a $n(z)$ distribution by cross correlating an ``unknown'' sample with no redshifts --- which, in this work, is the \decade source-galaxy sample --- against a ``reference'' sample with known redshifts, such as the spectroscopic dataset mentioned in Section~\ref{sec:sec:reference}.

The (cross-)correlation measurement is done using the estimator of \citet{Davis:1983:2pt},
\begin{equation}\label{eqn:WZ:meas}
    w_{ur}(\theta) = \frac{N_{Rr}}{N_{Dr}} \frac{D_u D_r (\theta)}{D_u R_r(\theta)} - 1,
\end{equation}
where $N_{Dr}$ and $N_{Rr}$ are the total number of galaxies in the reference sample and in the randoms catalog, respectively, $D_u D_r$ corresponds to the weighted counts of pairs between the unknown and reference sample, and $D_u R_r$ is the weighted counts of pairs between the unknown and randoms sample. The randoms sample corresponds to the reference catalog only. The estimator of \citet{Davis:1983:2pt} is advantageous as it does not require a randoms catalog for the unknown sample; producing such a catalog for a weak lensing sample requires significant effort, accounting for survey systematics maps, observing conditions etc. It is for this reason \citetalias{Gatti:2022:WzY3} choose this estimator as well. 

We measure Equation \eqref{eqn:WZ:meas} at 20 angular scales corresponding to the physical distance $1.5 \mpc < d < 5\mpc$, which is the same choice as \citetalias{Gatti:2022:WzY3} and was chosen to avoid correlations with the galaxy clustering cosmology analysis of DES Y3.\footnote{We also note that given our choice to use the WZ estimates only as a cross-check and not to construct our fiducial redshift distribution (see Section~\ref{sec:sec:SOMPZWZ}), we can ignore any potential correlations between the WZ measurements on these scales and the fiducial, masked cosmic shear data vector used in \citetalias{paper3} and \citetalias{paper4}.} The physical scale $5\mpc$ corresponds to an angular scale range $10\arcmin < \theta < 35\arcmin$ across the redshift range of the measurements, $0.1 < z < 2$. These measurements are done using \textsc{TreeCorr} \citep{Jarvis2004TreeCorr}.

The theoretical prediction for the WZ measurement is given by,
\begin{equation}\label{eqn:WZ:theory}
    w_{ur}(\theta) = \!\!\!\int\!\!\! dz [n_u(z) n_r(z) b_u(z) b_r(z) w_{\rm DM}(z, \theta) + M(z, \theta)],
\end{equation}
where the individual terms are discussed shortly below. For the rest of this work, we will reduce both measurement and theorertical predictions of WZ to a scale-averaged quantity, defined as,
\begin{equation}\label{eqn:WZ:scale_avg}
    \Tilde{w}_{ur} = \frac{\int_{\theta_{\rm min}}^{\theta_{\rm max}} \theta^{-1} w_{ur}(\theta) d\theta}{\int_{\theta_{\rm min}}^{\theta_{\rm max}} \theta^{-1}d\theta}.
\end{equation}
Where $\theta^{-1}$ is simply a weighting function that gives the integral a logarithmic weighting, $d\theta/\theta = d\ln \theta$. This is the chosen weight function in \citetalias{Gatti:2022:WzY3}.

If the reference sample is limited to a narrow redshift range around a central redshift $z_i$, we can simply replace the term $n_r(z) \approx \delta(z - z_i)$ and then Equation \eqref{eqn:WZ:theory}, after the scale-averaging procedure, reduces to 
\begin{equation}\label{eqn:WZ:theoryavg}
    \Tilde{w}_{ur}(z_i) = n_u(z_i) b_u(z_i) b_r(z_i) \Tilde{w}_{\rm DM}(z_i) + \tilde{M}(z_i).
\end{equation}
where we use $\tilde{M}(z)$ to denote the quantity $M(z, \theta)$ after the integration over scales, $\theta$, as shown in Equation \eqref{eqn:WZ:scale_avg}.

The goal of the WZ method is making the measurement in Equation \eqref{eqn:WZ:scale_avg} and then using the prediction of Equation \eqref{eqn:WZ:theoryavg} to estimate the redshift distribution $n_u(z_i)$. If we vary the central $z_i$ of the reference sample slice, we can map out the distribution $n_u(z)$. There are a number of different terms in Equation \eqref{eqn:WZ:theoryavg}, apart from $n_u(z_i)$, that we are interested in. These can be estimated as follows:

\textbf{The reference sample's galaxy bias, $b_r(z_i)$}: Given we assume a linear galaxy biasing model for this work\footnote{Even though we use non-linear scales when evaluating Equation \eqref{eqn:WZ:scale_avg}, for an adequately narrow scale range, the bias can be treated as a constant, linear term.} we can estimate the bias for our sample as,
\begin{align}\label{eqn:WZ:br}
    \Tilde{w}_{rr}(z_i) =\,\,& \int dz [b_r(z)n_{r, i}(z)]^2\Tilde{w}_{\rm DM}(z)\nonumber\\
    =\,\, & b_r(z_i)^2\Tilde{w}_{\rm DM}(z_i)\int_{z_i - \Delta z/2}^{z_i + \Delta z/2} dz\, n^2_{r, i}(z),
\end{align}
where in the second line, we have taken the bias and dark matter (DM) correlation function to be a constant over the narrow redshift range, and $n_{r, i}$ is the redshift distribution in the $i^{\rm th}$ bin of the reference sample. Equation \eqref{eqn:WZ:br} can be rearranged to obtain $b_r(z_i)$ as a function of the measurement $\Tilde{w}_{rr}$.

\textbf{The unknown sample's galaxy bias, $b_u(z_i)$}: The approach used to estimate $b_r(z_i)$ relies on knowing which galaxies fall into the $i^{\rm th}$ redshift bin, and therefore cannot be adopted to estimate $b_u(z_i)$ since we do not know the redshifts for individual source galaxies. Instead, we follow \citetalias{Gatti:2022:WzY3} in marginalizing over the bias of the ``unknown'', source-galaxy sample using a set of systematic functions that we describe further below.

\textbf{Dark matter 2-point correlation function, $\bold{w_{\rm DM}}$:} This is the cosmological (non-linear) matter correlation function, after assuming a certain cosmology. We model the non-linear matter power-spectrum, $P_{\rm NL}(k)$, using the model of \citet{Takahashi2012}. In this work, we fix the cosmology to standard values of $\Omega_{\rm m} = 0.3$, $h = 0.7$, $\sigma_8 = 0.8$, and $n_s = 0.96$. \citetalias{Gatti:2022:WzY3} found the results are insensitive to changes in the cosmology parameters. In our work, we use WZ only as a cross-check rather than to generate our final $n(z)$ so we are even less sensitive to changes in cosmology. There are some uncertainties in modelling $P_{\rm NL}(k)$ on non-linear scales --- for example, variations across modelling approaches \citep{Takahashi2012, Mead2021b} and also corrections due to astrophysical processes \citep{Chisari2019CCL, Schneider2019Baryonification, Arico:2021:Bacco, Anbajagane:2024:Baryonification} --- but these uncertainties will be implicitly marginalized over when using the systematic functions discussed below in Section~\ref{sec:sec:SOMPZWZ}.

\textbf{Magnification signal, $M$}: The matter distributions at two different redshifts are nominally uncorrelated given they are spatially distant and therefore arise from different realizations of the initial conditions. However, observations of these distributions will be correlated due to the impact of magnification, where the foreground matter distribution magnifies the background distribution by changing the observed surface area. When observing galaxy distributions, magnification will also alter the observed brightness of the galaxies and therefore the number of detected/selected galaxies. The latter effect depends on the exact galaxy sample definition, and is captured by a magnification coefficient, $\alpha(z)$. The impact of magnification on the final redshift estimation is discussed in detail in \citetalias{Gatti:2022:WzY3}. In general, this magnification term contributes to the $w_{ur}$ measurement as shown in Equation \eqref{eqn:WZ:theory}, and is modelled as,
\begin{align}
    \tilde{M}(z_i) = &\,\, b_r(z_i)\alpha_u(z_i)\sum_{j > i}\tilde{D}_{ij}n_u(z_j) \nonumber\\
    & + b_u(z_i)\alpha_r(z_i)\sum_{j > i}\tilde{D}_{ij}n_u(z_i),
\end{align}
with the matrix $\tilde{D}_{ij}$ given as,
\begin{equation}
    \tilde{D}_{ij} = \frac{3H_0^2\Omega_m}{c^2}\tilde{w}_{\rm DM}(z_i)\frac{\chi(z_i)}{a(z_i)}\frac{\chi(z_j) - \chi(z_i)}{\chi(z_j)}\Delta \chi(z_j),
\end{equation}
where all expressions are defined assuming the redshift distribution of the reference sample is sufficiently narrow. See Equations (8--11) in \citetalias{Gatti:2022:WzY3} for the generalized expressions.

In practice, we follow \citetalias{Gatti:2022:WzY3} in simplifying the functions $\alpha_u(z_i) \rightarrow \alpha_u$ and $b_u(z_i) \rightarrow b_u$ so that they are redshift-independent constants. Furthermore, we have no estimate of $\alpha_r(z_i)$ for the reference sample. This was also the case in \citetalias{Gatti:2022:WzY3}. They assign to the reference sample estimates of $\alpha_r(z_i)$ made on a different red galaxy sample in DES Y3 \citep{ElvinPoole:2023:MagY3}. Their final results are insensitive to large changes in $\alpha_r(z_i)$ and so their choice in assuming the same magnification for two different galaxy samples is reasonable. Similarly, we find our results are insensitive to changes in these magnification parameters.

All theoretical modelling mentioned above is done using the \textsc{Core Cosmology Library} \citep[\textsc{CCL},][]{Chisari2019CCL}.

\subsection{Cross-checking SOMPZ and WZ}\label{sec:sec:SOMPZWZ}

The redshift estimates in DES Y3 used a combination of the SOMPZ and WZ methods \citepalias{Myles:2021:DESY3, Gatti:2022:WzY3}, and these were the fiducial distributions used in the cosmology analysis. In our work, we use the WZ only as a cross-check on the fiducial SOMPZ estimates. However, we also produce a combined $n(z)$, detailed in Appendix~\ref{appx:SOMPZ_WZ}, which is used as one of our variant runs in the cosmology analysis. There are multiple different ways to perform the combination of the two techniques, which vary in what parts of the WZ information is used in the combined estimate. Here, we opt for using the full measurement, following the ``full shape'' method described in \citetalias{Gatti:2022:WzY3}.

In this method, we take an $n(z)$ given by the SOMPZ method and forward model the expected $w_{ur}(z)$ measurement. This requires defining the conditional likelihood,
\begin{align}\label{eqn:SOMPZWZ:likelihood}
    & \mathcal{L}(\boldsymbol{w}_{ur} | \boldsymbol{n}_{\rm sompz}, \boldsymbol{b}_{\rm r}, \bm{\alpha}_{\rm r}, \boldsymbol{s}, \boldsymbol{p}) = \nonumber\\
    & \int d\boldsymbol{s}\, d\boldsymbol{p}\, (\boldsymbol{w}_{ur} - \boldsymbol{w}^{\rm th}_{ur})^T \mathcal{C}^{-1}(\boldsymbol{w}_{ur} - \boldsymbol{w}^{\rm th}_{ur}) \,P(\boldsymbol{s})\, P(\boldsymbol{p}),
\end{align}
where $\boldsymbol{w}_{ur}$ is the vector of WZ measurements at redshifts $\boldsymbol{z} = {z_0, z_1, ...z_N}$, $\boldsymbol{n}_{\rm SOMPZ}$ is the SOMPZ estimate of the redshift distribution, $\boldsymbol{w}^{\rm th}_{ur}$ is the theoretical prediction, and $\mathcal{C}$ is the covariance matrix of the WZ estimates. The vectors $\boldsymbol{b}_{\rm r}, \bm{\alpha}_{\rm r}$ are the bias and magnification coefficients of the reference sample, as a function of redshift. We estimate the former directly from the data and use the same coefficients as \citetalias{Gatti:2022:WzY3} for the latter. The vector $\boldsymbol{p}$ is a set of systematic parameters we marginalize over, and corresponds to a basis of systematic functions that we describe further below. The vector $\boldsymbol{s} = \{b_u, \alpha_u\}$ are the bias and magnification coefficient of the unknown sample, which we also marginalize over. We estimate $\mathcal{C}$ from the data using a jackknife procedure with 600 patches as defined by \textsc{TreeCorr} \citep{Jarvis2004TreeCorr}. We are able to use a high number of patches as the scales we study are much smaller than the survey footprint.\footnote{Our use of a high number of patches allows us to robustly compute the inverse of the matrix without inferring biases; the effects discussed in \citet{Hartlap2007, Dodelson:2013:Cov} are negligible in our case. As a consequence we can still use a $\chi^2$ test metric for the discussions at the end of Section \ref{sec:sec:SOMPZWZ}.}

Equation \eqref{eqn:SOMPZWZ:likelihood} can assign a likelihood to any $n(z)$ estimate. In this work, we use this likelihood for two distinct reasons: (1) to find the best fit of the nuisance parameters, $\boldsymbol{p}$ and $\boldsymbol{s}$, in order to compare the consistency of a given SOMPZ $n(z)$ realization with the WZ measurements; and (2) subselect the $\mathcal{O}(10^7)$ SOMPZ-based $n(z)$ realizations created during the uncertainty quantification steps detailed in Sections~\ref{sec:sec:3sdir}, \ref{sec:sec:ZPuncert}, and \ref{sec:sec:Zuncert}. This subselection removes any realizations that have strong disagreements with the WZ measurements. We only use the former model in our main analysis below, and defer the latter approach to Appendix~\ref{appx:SOMPZ_WZ}.

Both approaches require a model for $w_{ur}^{\rm th}$, which we now describe. This follows from the model described in Equation \eqref{eqn:WZ:theoryavg} with one key addition in the form of systematic functions. These functions account for known and potentially hidden systematic effects in the measurement and subsequent modeling. They are marginalized over to ensure the inference of redshift distributions is not biased due to these systematics. We follow \citetalias{Gatti:2022:WzY3} in our implementation of these functions. The theory model is modified as,
\begin{align}\label{eqn:sysfunc}
    \Tilde{w}_{\rm DM}(z) & \rightarrow \Tilde{w}_{\rm DM}(z) \times {\rm Sys}\nonumber\\
    & \rightarrow \Tilde{w}_{\rm DM}(z) \times \exp\bigg[\sum_{i = 0}^M \frac{\sqrt{2i + 1}}{0.85}s_iP_i(x)\bigg],
\end{align}
where $P_i$ is the Legendre polynomial of order $i$ and $x$ is a rescaled redshift,
\begin{equation}\label{eqn:sysfunc:x}
    x = 0.85 \times \bigg(-1 + 2\frac{z - z_{\rm min}}{z_{\rm max} - z_{\rm min}}\bigg).
\end{equation}
See Appendix A of \citetalias{Gatti:2022:WzY3} for the choices leading to this specific form for both the systematic functions in Equation \eqref{eqn:sysfunc} and the function argument, $x$, in Equation \eqref{eqn:sysfunc:x}. Notably, these systematic functions do not affect our model for the magnification, $\Tilde{M}(z_i)$, in Equation~\eqref{eqn:WZ:theoryavg}. The amplitudes of the two magnification terms are separate nuisance parameters that we marginalize over. The systematic functions primarily marginalize over the uncertainty in $b_u(z)$.

In practice, we must marginalize over all parameters of the systematics functions. Naively, this would require running a chain per $n(z)$ realization which is unfeasible for this analysis as we have $\mathcal{O}(10^7)$ realizations. Instead, we can analytically marginalize over the systematic functions, parameterized by $\boldsymbol{s}$, and also over the source galaxy bias and magnification, defined as the vector $\boldsymbol{p}$ in Equation \eqref{eqn:SOMPZWZ:likelihood}. This is done following the approach in Appendix A of \citetalias{Gatti:2022:WzY3}, which involves linearizing the likelihood function in $\boldsymbol{s}$ and then analytically evaluating the multidimensional Gaussian integral.

Let us define the vector $\boldsymbol{q} = \{\boldsymbol{s}, \boldsymbol{p}\}$ which contains the coefficients of the systematic functions, and also the source galaxy bias and magnification, $\boldsymbol{p} = \{b_u, \alpha_u\}$. The model prediction, $w_{ur}^{\rm th}$, is already linear in the parameters $\boldsymbol{p}$. Thus, we only need to modify the systematic functions, which we do through a simply Taylor expansion to linear order,
\begin{align}\label{eqn:wz:linearize}
    {\rm Sys} (z_i, \boldsymbol{s}) &  \approx  {\rm Sys} (z_i, \boldsymbol{s}_0) \times \nonumber\\
    & \left[1 + \sum_{k=0}^{M} \frac{\sqrt{2k+1}}{0.85} P_k(x) s_{k,0} (s_k - s_{k,0})\right].
\end{align}
The residuals between the data and the model for the WZ measurement can now be written as,
\begin{equation}\label{eqn:wz:linear_res}
    \boldsymbol{w}_{ur} - \boldsymbol{w}^{\rm th}_{ur} = \boldsymbol{c}(\boldsymbol{q}_0) + A\boldsymbol{q},
\end{equation}
where $\boldsymbol{c}(\boldsymbol{q}_0) = \boldsymbol{w}_{ur} - \boldsymbol{w}^{\rm th}_{ur}(\boldsymbol{q}_0) + A\boldsymbol{q}_0$ is a constant vector defined using $\boldsymbol{q}_0$, the maximum-likelihood values for the parameters $\boldsymbol{q}$, and $A$ is a derivative matrix, $A_{ij} = dw^{\rm th}_{{\rm ur}, i}/dq_j$. With the form of the residuals in Equation \eqref{eqn:wz:linear_res}, we can rewrite the likelihood of Equation \eqref{eqn:SOMPZWZ:likelihood} as,
\begin{align}\label{eqn:wz:new_likelihood}
\mathcal{L} \propto  \int & \,\, \mathrm{d}\boldsymbol{q} \, \exp\left[ -\frac{1}{2} (\mathbf{c} - \mathbf{Aq})^T \Sigma_{w_{ur}}^{-1} (\mathbf{c} - \mathbf{Aq}) \right] \nonumber\\
& \times 
\exp\left[ -\frac{1}{2} (\mathbf{q} - \mu_q)^T \Sigma_q^{-1} (\mathbf{q} - \mu_q) \right].
\end{align}
We use the proportionality sign (and drop all constant factors) as we are insensitive to the normalization of the likelihood when cross-checking/combing SOMPZ with WZ. The (Gaussian) integral of Equation \eqref{eqn:wz:new_likelihood} can be solved through simple matrix evaluation,
\begin{align}
    \mathcal{L} & \propto \exp\bigg[\frac{1}{2} \boldsymbol{d}^T B^{-1} \boldsymbol{d}^T\bigg]\label{eqn:wz:final_likelihood}\\
    B & = A^T \Sigma_{w_{ur}}^{-1}A+ \Sigma_{\rm q}^{-1}\label{eqn:wz:Bmat}\\
    \boldsymbol{d} & = A^T\Sigma_{w_{ur}}^{-1} \boldsymbol{c} + \Sigma_{\rm q}^{-1} \boldsymbol{\mu}_{\rm q}.\label{eqn:wz:dvec}
\end{align}
The parameter choice, $\boldsymbol{q}_0$, that maximizes the likelihood is obtained trivially by setting $d\mathcal{L}/d\boldsymbol{q} = 0$ and solving for $\boldsymbol{q}$. The resulting expression is $q_{\rm 0} = B^{-1}\boldsymbol{d}$. The evaluation of Equation \eqref{eqn:wz:final_likelihood} is done iteratively through the following steps:
\begin{enumerate}[label=(\roman*)]
    \item Initialize $\boldsymbol{q}_0 = \boldsymbol{\mu}_q$, which is the mean value of the parameters, $\boldsymbol{q}$. We use $\boldsymbol{s} = 0$, $b_u = 1$ and $\alpha_u = 1.5$. In future iterations, $\boldsymbol{q}_0$ is set by solving for the maximum-likelihood point as mentioned above.\vspace{5pt}
    
    \item Compute the matrix $A$ at the chosen $\boldsymbol{q}_0$. Obtain $\boldsymbol{c}$ by rearranging Equation \eqref{eqn:wz:linear_res}.\vspace{5pt}
    
    \item Evaluate the likelihood in Equation \eqref{eqn:wz:final_likelihood}.\vspace{5pt}
    
    \item Find the maximum-likelihood estimate $\boldsymbol{q}_0 = B^{-1} \boldsymbol{d}$.\vspace{5pt}
    
     \item Repeat steps (i) to (iii) using the new estimate of $\boldsymbol{q}_0$ and recalculate the maximum-likelihood estimate in (iv).
\end{enumerate}
The above steps are repeated until convergence or until we perform $N_{\rm iter} = 30$ iterations. Convergence is achieved if the log-likelihood changes by $|\Delta \ln\mathcal{L}| < 0.05$, which is a $< 5\%$ change in $\mathcal{L}$. Less than $0.0001\%$  of the samples require more than 30 iterations to converge.

We can now check the consistency between the SOMPZ and WZ methods using the following procedure: (i) for each $n(z)$ sample from SOMPZ, obtain a best fit of the eight systematic parameters by solving the linear equations described above, (ii) compute the $\chi^2$ between the WZ measurement and this best fit prediction, and finally, (iii) check that the $\chi^2$ for most of the SOMPZ-derived $n(z)$ samples is acceptable. Our WZ measurement has $40$ datapoints, but we only use the 30 points between $0 < z < 1.6$ to compare against the SOMPZ $n(z)$ as the latter distributions have no support above this redshift. There are eight free parameters in our forward model for transforming the SOMPZ $n(z)$ to a WZ model, so $N_{\rm dof} = 30 - 8 = 22$\footnote{In practice, this is a conservative choice. If parameters are degenerate with each other, the effective number of parameters will necessarily be lower than the $N_{\rm params} = 8$ used here. Our conservative choices results in a stricter test as it reduces the total degrees of freedom and therefore decreases the resulting $\chi^2$ threshold.} and then $\chi^2 < 44.6$ corresponds to the $n(z)$ sample being within $3\sigma$ of the WZ measurement. This procedure is performed on all $n(z)$ realizations. We have also checked the statistical power of this technique by artificially modifying our SOMPZ estimates --- by shifting them with $\Delta \langle z \rangle = \pm 0.1$ --- and found that in this case less than $5\%$ of the modified samples passed the $\chi^2$ test described above. This implies our test can indicate inconsistencies between SOMPZ and WZ at the $\Delta \langle z \rangle \approx 0.1$ level.\footnote{While this is a generally large shift relative to the uncertainties on our SOMPZ estimates in Table \ref{tab:z_calibration}, we still find it a valuable check since the impact of spatially varying systematics will be different on the two estimates and so the cross-check is a beneficial check in avoiding catastrophic errors due to such variations.}

When computing the $\chi^2$ above for a given $n(z)$ sample, we use a covariance matrix that accounts for two uncertainty contributions. First is the uncertainty in the WZ measurement, denoted by $\Sigma_{w_{ur}}$. The other is the uncertainty in the SOMPZ $n(z)$, denoted here as $\Sigma_{n(z)}$ and computed across the $10^7$ samples obtained from the procedures discussed above. Then, this covariance matrix can be linearly transformed from the $n(z)$ basis into the WZ basis using the function $v(z) = [b_u b_r \Tilde{w}_{\rm DM} \times \text{Sys}](z)$, which is simply the first term in Equation \eqref{eqn:WZ:theoryavg} with the inclusion of the systematic functions as shown in Equation \eqref{eqn:sysfunc}. Our covariance matrix for the $\chi^2$ calculation is then,
\begin{equation}\label{eqn:cov_matrix}
    \Sigma_{\rm tot} = \Sigma_{w_{ur}} + \bold{v}^T \Sigma_{n(z)} \bold{v},
\end{equation}
which now accounts for the noise in the WZ measurement and in the exact SOMPZ $n(z)$ sample being considered. The latter is particularly relevant as the SOMPZ-based uncertainties on the $n(z)$, in a given redshift interval, are often larger than the uncertainties on the corresponding WZ measurement uncertainty in that interval; particularly at lower redshifts. See Figure~\ref{fig:n_of_z} below.\footnote{Note that in practice, we can compare WZ measurement uncertainty with SOMPZ uncertainty only after forward modelling the SOMPZ $n(z)$ to a WZ model; see Equation \eqref{eqn:cov_matrix}. We have done so and confirmed the SOMPZ uncertainty is indeed larger than the WZ measurements even in this case. However, for simplicity, we still use Figure~\ref{fig:n_of_z} as a rough demonstration of the relative uncertainties.}

We reiterate that our fiducial $n(z)$ estimate derives from the SOMPZ-only method, with an additional cross-check with the WZ measurements. In Appendix~\ref{appx:SOMPZ_WZ}, we generate an alternative set of $n(z)$ samples that combine the SOMPZ and WZ\footnote{The DES Y3 analysis also included additional information from the ``shear ratio'' measurements \citep{Sanchez2022}. These isolate the geometric information of the lensing observable and have been shown to improve constraints on the redshift calibrations. However, we do not use this in our work as we do not have a galaxy clustering ``lens'' samples for the \decade data at this time.} estimates using a sub-sampling step. Our cosmology analysis in \citetalias{paper4} includes an analysis variant using this alternative $n(z)$ and shows our results are consistent across the two $n(z)$ choices.

\section{\textsc{Balrog}: Estimating the Survey transfer function}\label{sec:Balrog}

A necessary input to the SOMPZ method is the transfer function, which can be represented by the joint probability $p(c, \chat)$, can accurately convert the color-redshift relation learned from the deep-field data into one for the wide-field data. In DES, this transfer function is estimated using a source injection pipeline, referred to as \Balrog \citep{Suchyta:2016, Everett:2022, Anbajagane:2025:Y6Balrog}. Other surveys have also implemented distinct source injection pipelines, such as the \textsc{Obi-wan} pipeline \citep{Kong:2024:ObiWan} for the Dark Energy Spectroscopic Instrument (DESI), the \textsc{SynPipe} software used in the Hyper Suprime-Cam (HSC) data \citep{Miyazaki:2018:HSC}, or the \textsc{CosmoDC} datasets \citep{Sanchez:2020:CosmoDC2} used in the Dark Energy Science Collaboration \citep[DESC,][]{DESC:2018:SRD}. In this work, we follow DES in using source injection to characterize the transfer function, and also follow the same terminology in referring to our source injection pipeline as \Balrog. Our pipeline is constructed by modifying the \decade image simulation pipelines presented in \citetalias{paper1} (see their Section 5.1). We detail below the modifications to that pipeline needed for the source injection requirements of this work. For the sake of completeness, we also briefly describe the common features between the two pipeline.

As discussed above, the imaging data in the \decade survey exhibit a significant variation in data quality (depth, seeing etc.) relative to other surveys such as DES. The impact of varying image quality on the mean $n(z)$ (averaged over the full survey) can be accounted for through source injection. As \Balrog utilizes the actual image data from the survey, its estimate of the distribution $p(c, \chat)$ already incorporates this varying imaging data quality across the full survey. Thus, \Balrog provides a data-driven approach to account for such variations when estimating this survey-averaged $n(z)$.

\subsection{Processing pipeline}\label{sec:sec:Pipeline}

As a brief overview, the pipeline starts from the CCD images in the \decade dataset, and injects synthetic galaxies into them. The entire DESDM pipeline is run on these images to reproduce coadd images, \textsc{SourceExtractor} detection catalogs, \textsc{Metacalibration} shear estimates, etc. We now describe each piece in more detail. To aid the reader, any steps that are modified --- when compared to the image simulation pipeline of \citetalias{paper1} (see their Section 5.1) --- are denoted with a $^\star$:

\textbf{Coadd tile selection$^\star$:} We randomly select 1400 coadd tiles from across the footprint. A portion of this data is removed after accounting for the foreground mask introduced in \citetalias{paper1}. The list of tiles used here is a different random subset compared to that used in the image simulations pipeline of \citetalias{paper1}.

\textbf{Galaxy sample$^\star$:} The image simulation pipeline uses deep-field galaxies from the \Cosmos field. For \Balrog we use all available DES deep fields. This includes the other three ``supernovae (SN)'' fields. See \citet{Hartley:2022:Y3Deepfields} for more details on the catalog. We continue to only select galaxies brighter than $i < 25.5$, given that limit is two magnitudes fainter than the average survey depth of $i \approx 23.5$ (see Figure 3 of \citetalias{paper1}). The \textsc{ngmix} \citep{Sheldon:2015:ngmix} and \textsc{Galsim} \citep{Rowe:2015:galsim} software packages are used to ingest the galaxy morphology provided in the catalogs from \citet{Hartley:2022:Y3Deepfields} and then render an image of the galaxy. The exact cuts on the sample are as follows:
\begin{align}\label{eqn:DFSampleCuts}
     & \texttt{flags} == 0 \nonumber\\
     \texttt{AND} \,\,\,\,& \texttt{mask\_flags} == 0 \nonumber\\
     \texttt{AND} \,\,\,\,& \texttt{flags\_NIR} == 0 \nonumber\\
     \texttt{AND} \,\,\,\,& \texttt{SNR\_{griz}} > -3 \nonumber\\
     \texttt{AND} \,\,\,\,& \texttt{bdf\_mag\_i} < 25.5 \nonumber\\
     \texttt{AND} \,\,\,\,& \texttt{bdf\_T} < 100\,\,\texttt{arcsec}^2\nonumber\\
     \texttt{AND} \,\,\,\,& \texttt{KNN\_CLASS} == 1,
\end{align}
where all quality cuts, other than the final one, follow directly from \citet[][see their Section 3.4 for the origin of these cuts]{Everett:2022}. In brief, the first four cuts remove objects with artifacts/failures during fitting or image processing, the cut on \texttt{bdf\_mag\_i} removes objects that are too faint to be detected in the survey more than 1\% of the time, and the cut on $\texttt{bdf\_T}$ limits blending amongst \Balrog-injected sources (as noted below, we place injections on a hexagonal grid with 10\arcsec spacing). The final cut removes stars from our injection catalog, using the k-Nearest Neighbor, photometric classifier from \citet{Hartley:2022:Y3Deepfields}; their Figure 15 shows the star classification has high completeness down to $i \lesssim 24$.

\textbf{Galaxy counts, shapes \& positions$^\star$:} Following \citet{Everett:2022} and \citet{Anbajagane:2025:Y6Balrog}, the galaxies are injected in a hexagonal grid whose axes are aligned with the coadd image's coordinates. The separation between neighboring injections is always $20 \arcsec$ and is chosen since the objects being injected have a maximum half-light radius of $10 \arcsec$. The separation scale prevents the injections from being blended with one another and subsequently increases the number of usable injections after postprocessing. The number of objects per tile is therefore set by the size of the tile and by this spacing. We find $\mathcal{O}(10^4)$ injections per tile. All galaxy shapes are randomly rotated prior to injection. Notably, the galaxies are not additionally sheared like they are in the image simulations.

\textbf{Simulated star catalog$^\star$:} Unlike the image simulations, there are no simulated stars injected into the pipeline. This is because in \Balrog we inject galaxies into the real images, which already contain the real distribution of stars. Thus, there is no need to add simulated stars as well.

\textbf{Weighting scheme$^\star$:} In the image simulations, we inject sources at up to two magnitudes fainter than our detection limit, as doing so increases the realism of the noise in the simulations. For \Balrog, we already have realistic noise given we start with the actual images, and thus we have freedom to only inject objects that are most relevant to our science goals. We adopt a tiered injection system, where all injections in a given tile are split into the following three types:
\begin{itemize}
    \item[1.] \textit{All sources:} One half of all injections in a given tile are of objects that are in our fiducial injection catalog (described above), with no subsampling.\vspace{5pt}

    \item[2.] \textit{Bright sample:} One-fourth of the injections are a weighted subsampling of the objects. The weights are a sigmoid function that downweight all galaxies with magnitudes fainter than $m_i > 23.5$. Such objects are not frequently detected in the actual data catalog. As shown in Figure~\ref{fig:Zsample}, our peak sensitivity (after using the lensing weights) is to galaxies of $m_i \approx 21.5$.\vspace{5pt}

    \item[3.] \textit{Bright high-quality redshifts sample:} The final fourth of the injections are similar to the above, but now with the additional requirement that the objects must have a redshift estimate from one of the samples described in Section~\ref{sec:sec:Zsample}.
\end{itemize}
\noindent We stress that the weighting scheme only serves to increase the optimality of the \Balrog-based measurements in the SOMPZ pipeline. The scheme used in this work closely follows that used in the DES Y6 \Balrog runs \citep{Anbajagane:2025:Y6Balrog}.

\textbf{Extinction:} All objects are reddened during injection, with the interstellar extinction coefficients, $E(B - V)$, taken from the map of \citet{Schlegel:1998:Dust}. The reddening coefficients, $A_b$, for band $b$, are 2.140, 1.569, 1.196 for the $riz$ bands, respectively \citep[][see their Section 4.2]{DES2018}. We obtain the reddened magnitudes, $m_{b} = m_{b,{\rm 0}} + A_b \times E(B-V)$, by computing $A_b \times E(B-V)$ at the sky location of each object we inject.

\textbf{Simulating individual CCD images$^\star$:} Once we have defined the shapes and location of galaxies on the coadd tile, we simulate the individual CCD images that contribute to the coadd image. The objects are rendered as \textsc{Galsim} models and then convolved with the PSF model of that CCD image, where the latter model is evaluated at the location of the injection in the CCD. The convolved galaxy models are then injected into the image, with all the actual sources, background, and noise still present. Thus, we do not separately add the background image and a noise image as was done in the image simulations.

\textbf{Generating coadd image:} The individual CCD images are coadded using the \textsc{Swarp} \citep{Bertin:2010:Swarp} algorithm to form coadd images for each of the three bands ($r, i, z$). The images from the three bands are then coadded again, using \textsc{Swarp}, to form the detection coadd.

\textbf{\textsc{SourceExtractor} object catalogs and \textsc{MEDS} files:} The $r+i+z$ detection coadd is analyzed using \textsc{SourceExtractor} \citep{Bertin:1996:SrcExt} to obtain the object catalog. The algorithm is run three times in ``dual mode'' to get the object properties (photometry, size etc.) in each of the three bands. This catalog, alongside the simulated images, is used to create a  \textsc{MEDS} file \citep{Jarvis2016} for the coadd tile.

\textbf{\textsc{Metacalibration}}: We then continue to process the simulated \textsc{MEDS} file through the same shape measurement pipeline as that used on the data (see Section  3.1 of \citetalias{paper1}).

\textbf{Matching injections to detections$^\star$:} Finally, the detected objects must be matched back to the injections they originated from. Here, ``detected object'' refers to the objects found by \textsc{SourceExtractor}. This match is done via a simple position match on the sky, using the \textsc{Balltree} implementation in the \textsc{scikit-learn} package \citep{Pedregosa2012Sklearn}. In practice, we consider every object in the injection catalog, and find the nearest detection to it on the sky. We consider it a match if the objects are within $0.5 \arcsec$ of each other. This follows the choices of \citet{Everett:2022} and \citet{Anbajagane:2025:Y6Balrog}. We additionally also search for the nearest real (\ie not synthetic) source in the original, unaltered image. If this source is within $1.5 \arcsec$ of the injection, then we consider the injection a blended object and do not use it in our analyses. Once again, this follows the approach of DES Y6 \citep{Anbajagane:2025:Y6Balrog}. Thus, in our case we remove all blends regardless of the flux ratios between the injected source and real source.

\subsection{Validation}\label{sec:sec:BalrogValidation}

There are two classes of validation relevant to synthetic source injection. The first is to verify that the image processing pipeline used by \Balrog reproduces the same results as the pipeline run on the real data. The second is to verify that the distribution of object properties in the synthetic \Balrog catalog matches those of the data catalog. While the latter validation is done for the image simulations in \citetalias{paper1} (see their Figure 15), those simulations use a different injection catalog; they only inject objects from the \Cosmos field, as that dataset is uniquely advantageous for performing simulations focusing on galaxy shapes. Since we inject objects from a wider set of deep fields, we redo that test again for the entire \Balrog dataset.

\begin{figure}
    \centering
    \includegraphics[width = 1.08\columnwidth]{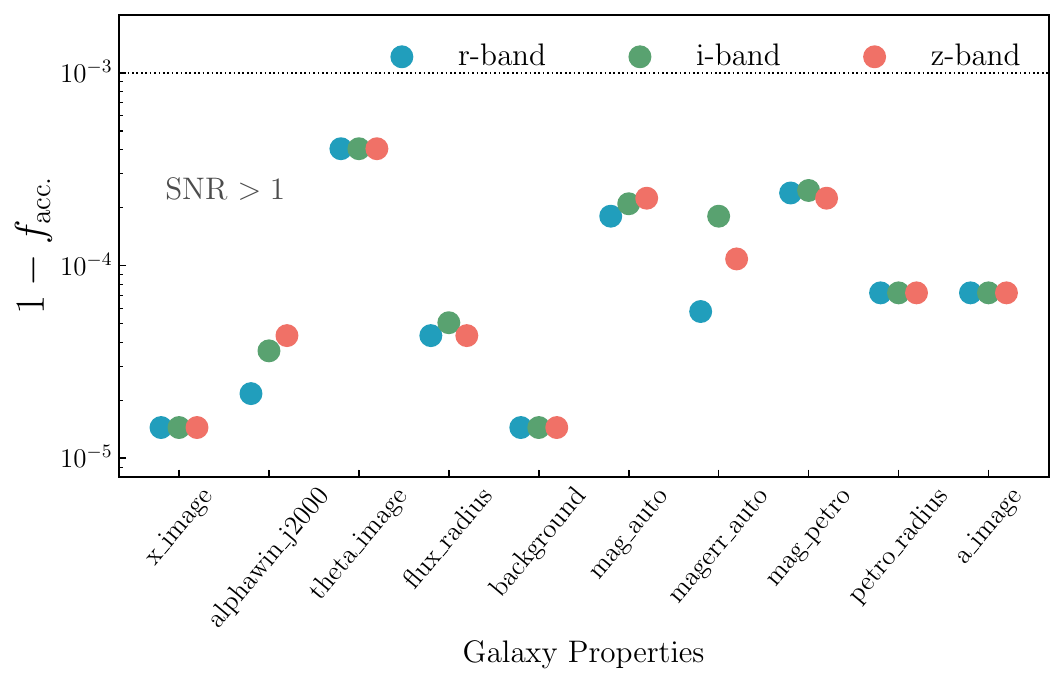}
    \caption{The accuracy of measurements from the \Balrog image processing pipeline compared those from the fiducial pipeline, for three coadd tiles (0804-0624, 0834-0833, 1455+1126). The y-axis denotes the fraction of objects, per band, exceeding an accuracy threshold; see text for details on the thresholds. The x-axis denotes the galaxy quantities from \textsc{SourceExtractor}, and from left to right are: the position in image coordinates, right ascension in sky coordinates, the position angle in image coordinates, the flux radius, the background, the magnitude, magnitude error, petrosian magnitude, petrosian radius, and length of the major axis. In all cases, at least $99.9\%$ of all objects are accurate, while in most cases the accuracy is adequate for over $99.99\%$ of the sample.}
    \label{fig:DESDMvsBalrog}
\end{figure}

\textbf{First, we validate the pipeline}, which is done by running the entire \Balrog pipeline but with no synthetic injections. In this case, we are simply rerunning the image processing pipeline on the raw images, and should recover the same outputs as the \decade catalogs. This check is shown in Figure~\ref{fig:DESDMvsBalrog}. We take the objects from the \Balrog pipeline processing and match them to those from the DESDM processing using their positions on the sky. All objects in the \Balrog catalog have an exact match in the data catalog. We then compare the properties of the different pairs. We define adequate accuracy based on the nature of the quantity: for any quantity in units of number of pixels (image position, radius, major axis length) the measurement are consistent if they are within $0.1$ pixels. For the position angle (\ie orientation), the limit is $1\arcmin$. For all magnitudes and magnitude errors, we take the limit as 1 milli-mag. In general we find that $\approx\! 99.99\%$ of all objects pass our accuracy threshold, for all three bands. This level of agreement is adequate for using \Balrog to estimate the transfer function.

\begin{figure}
    \centering
    \includegraphics[width = \columnwidth]{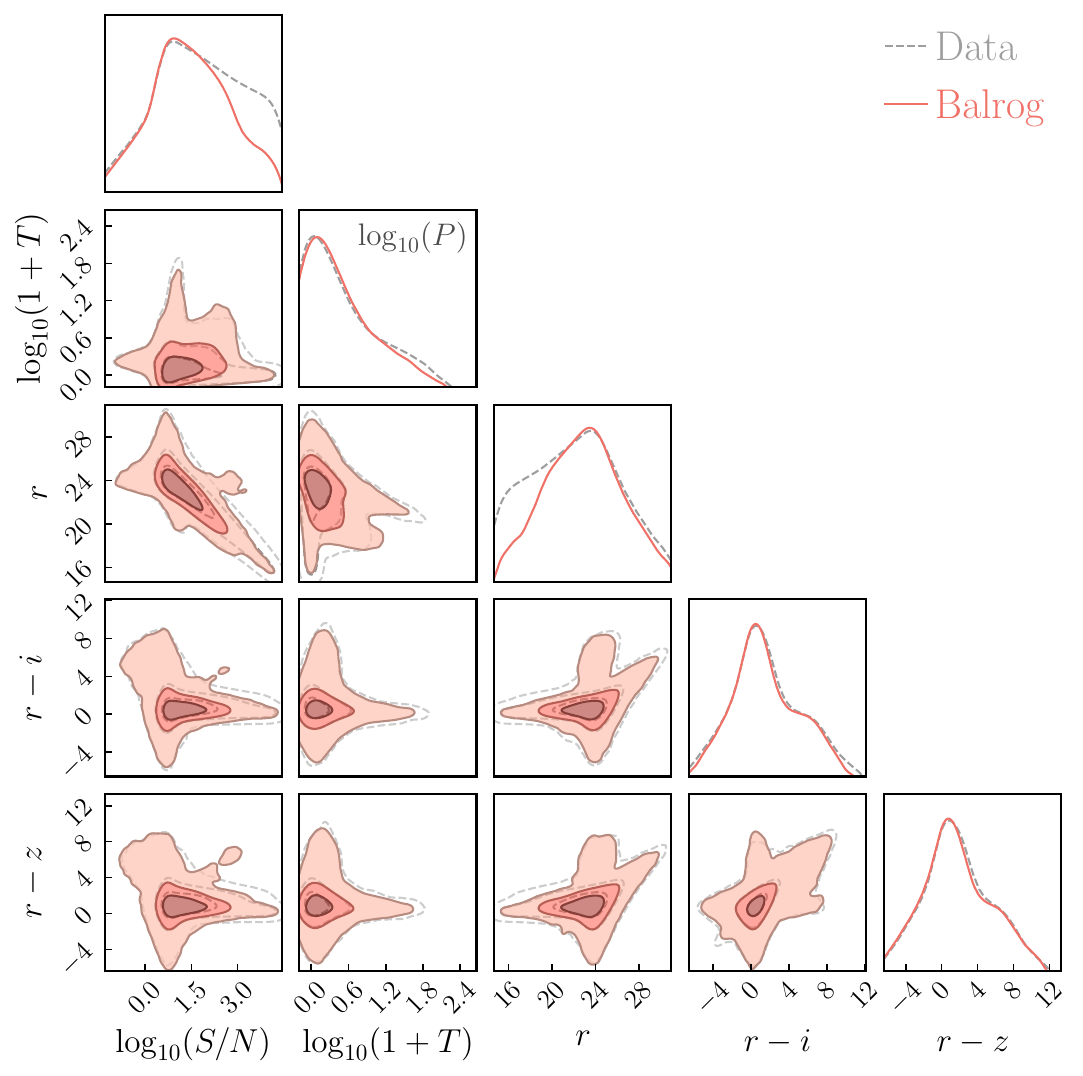}
    \caption{The distribution of \textsc{Metacalibration} properties for objects in the \Balrog (red) and the data (gray) catalogs. In both cases, we do not place the full weak lensing selection cut on the sample, and instead only require SNR $> 0$, $r > 0$, and $T > -1$. The \Balrog sample closely matches the data. There is a distinct lack of bright, high SNR objects in the former. These are due to stars which are present in the data but are not in \Balrog given we only inject galaxies in the latter. We have checked the number density distributions of \Balrog and data, and confirmed there is no visible stellar locus in the \Balrog sample. Note that the 1D marginal distributions are shown on a logarithmic scale.}
    \label{fig:BalrogVsData}
\end{figure}

\textbf{Second, we validate the dataset}, by comparing the distribution of \textsc{Metacalibration} properties of \Balrog objects with those of the \decade data. Figure~\ref{fig:BalrogVsData} shows the result of this comparison, with no selection cuts on either sample. The property distributions are consistent with one major, but expected, difference: \Balrog has a clear lack of bright, high signal-to-noise objects. This is expected as the \Balrog injections only contain galaxies and no stars, and the latter (which is in the data as we have not performed any selection cuts) would cause this mismatch. We intentionally present the raw catalog, without selections, to highlight this. We also note that the distribution of $T$ agrees much better in Figure~\ref{fig:BalrogVsData} than it does in the similar comparison between image simulations and data; see Figure 15 in \citetalias{paper1}. This mismatch is also expected and arises from a lack of bright and big objects in the \Cosmos field, which is the sole sample used for injections in the image simulations. The \Balrog sample does not face this issue given it is also supplemented by an additional three deep fields. We have also confirmed that the difference in the distribution of bright objects ($r \lesssim 20$) in \Balrog and data is alleviated if the stellar locus is removed from the latter (which we perform using a rough cut in size and magnitude).

\begin{figure}
    \centering
    \includegraphics[width = \columnwidth]{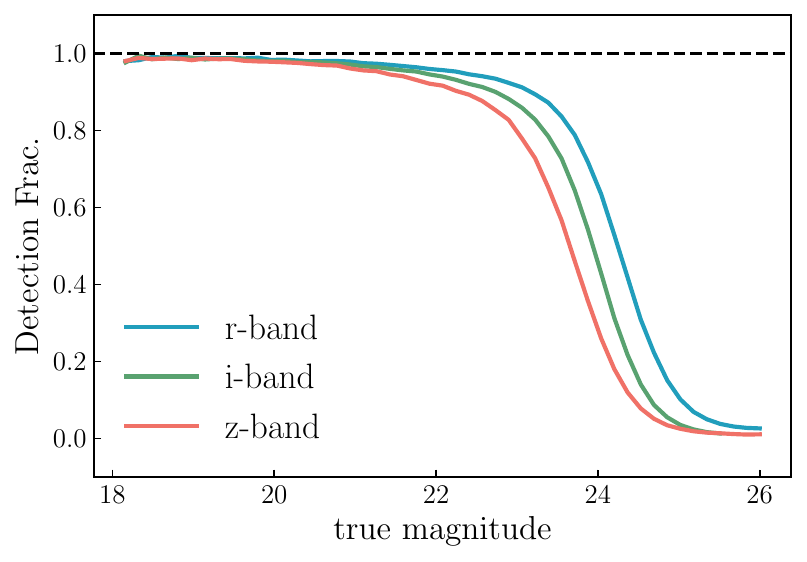}
    \caption{The detection rate of a galaxy as a function of its magnitude in a given band. The rate asymptotes to 1 for bright objects and then drops for faint objects. There are no selections applied beyond the simple detection threshold performed within \textsc{SourceExtractor}.}
    \label{fig:BalrogCompleteness}
\end{figure}

Finally, we also check the detection rate of a galaxy as computed by \Balrog. Figure~\ref{fig:BalrogCompleteness} shows this rate as a function of the true magnitude (obtained from \textsc{Fitvd}) in different bands. As expected, the rate asymptotes to 1 for very bright objects and drops for fainter objects. Note that we do not place any selection cuts or signal-to-noise cuts when measuring this rate from \Balrog. The completeness measured from this Figure will be deeper than the numbers quoted in our introduction above as the latter are defined at a signal-to-noise of 10. The detection probability asymptotes to $\approx 99\%$ (rather than 100\%) at the bright end. This is expected as there is still some non-zero probability of even bright objects not being detected due to complexities in the image, such as artifacts, saturation from nearby bright objects, masking, etc.

\section{Redshift distributions and uncertainty estimates}\label{sec:Results_Nz}

We now describe the $n(z)$ estimates from the SOMPZ method in Section~\ref{sec:Results_Nz:SOMPZ}, the estimates from the WZ measurements in Section~\ref{sec:Results_Nz:WZ}, and the cross-check between the two methods in Section~\ref{sec:Results_Nz:Crosscheck}. The combination of SOMPZ and WZ is detailed in Appendix~\ref{appx:SOMPZ_WZ}.

\subsection{SOMPZ} \label{sec:Results_Nz:SOMPZ}

\begin{figure*}
    \centering
    \includegraphics[width = 2\columnwidth]{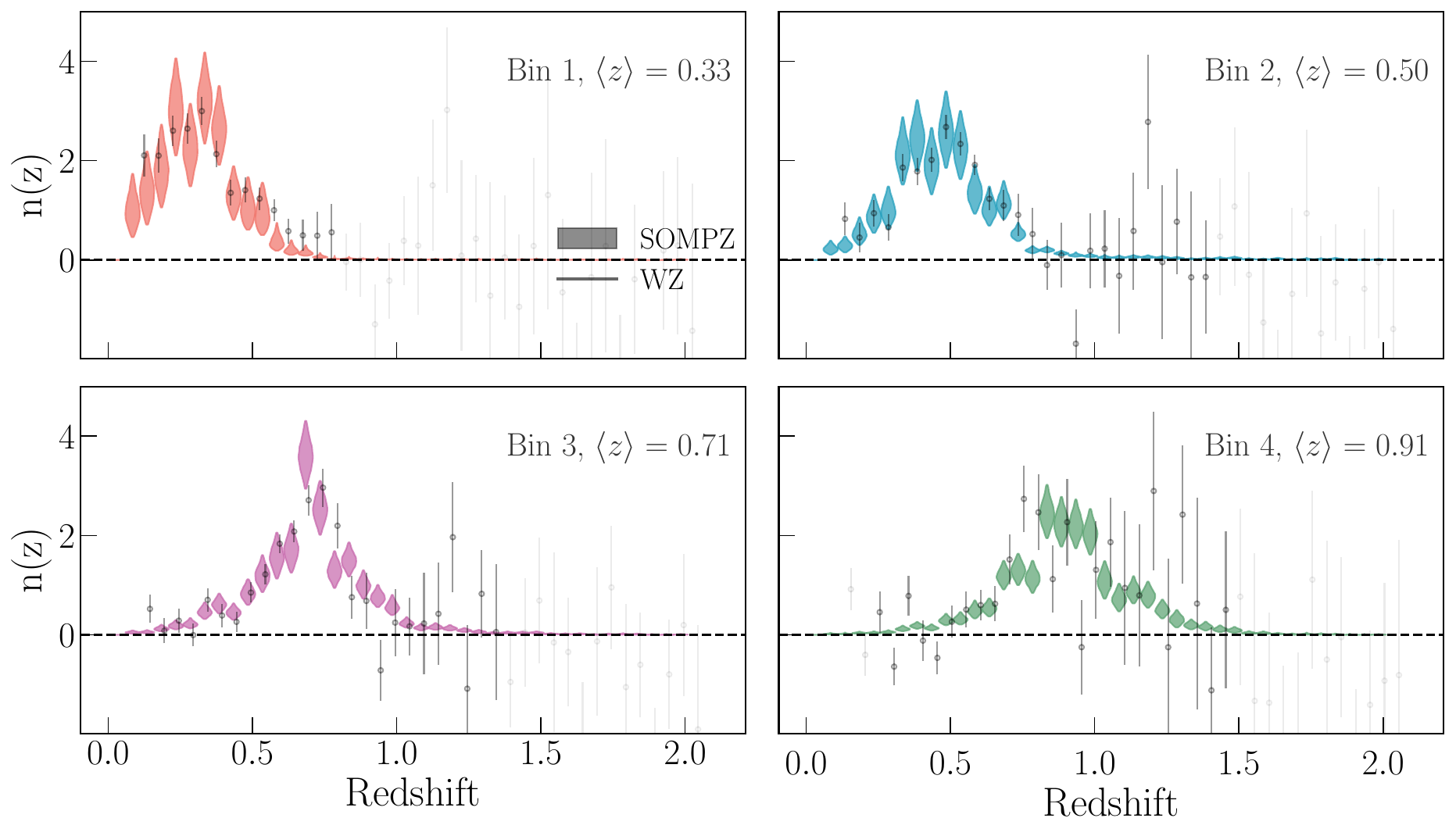}
    \caption{The fiducial redshift distributions from the SOMPZ method (including all uncertainty estimates detailed in Sections~\ref{sec:sec:3sdir}, \ref{sec:sec:ZPuncert}, and \ref{sec:sec:Zuncert}) compared with an estimate from the clustering redshift measurements. See Section~\ref{sec:Results_Nz:WZ} for details on this estimate and how it is presented here. The violins show the $95\%$ range of the $n(z)$ values in a given bin. The clustering redshift estimates are only available for a subset of the range probed by the SOMPZ estimate. In each panel, the WZ points outside the $95\%$ probability range of the SOMPZ distribution are shown as translucent points to improve visibility. We perform a ramping step for the SOMPZ estimate that sets $n(z \rightarrow 0) \rightarrow 0$; see Section~\ref{sec:Results_Nz:SOMPZ}. A detailed comparison of the SOMPZ- and WZ-based results is discussed in Section~\ref{sec:Results_Nz:Crosscheck}.}
    \label{fig:n_of_z}
\end{figure*}

We derive our fiducial $n(z)$ samples following the SOMPZ-based methodology detailed in Sections~\ref{sec:sec:SOMPZ}, \ref{sec:sec:ZPuncert}, and \ref{sec:sec:Zuncert}. These are presented in Figure~\ref{fig:n_of_z}, alongside WZ measurements that are described below in Section~\ref{sec:Results_Nz:WZ}. The violin plot shows the distribution of $n(z)$ values in a given redshift interval, estimated using the $10^7$ realizations. The mean redshifts of our $n(z)$ are $z_{\rm mean} \in \{0.33, 0.50, 0.71, 0.91\}$, which is similar to the DES Y3 results of \citetalias{Myles:2021:DESY3}, $z_{\rm mean} \in \{0.33, 0.52, 0.75, 0.94\}$. The \decade data is slightly shallower for the last two tomographic bins.

The distributions in Figure~\ref{fig:n_of_z} have already been processed with the ``ramping'' step discussed in Section~\ref{sec:sec:SOMPZ}, which is why the lowest-redshift interval shows $n(z) \approx 0$. The $n(z)$ in the fourth tomographic bin drops to 0 for $z \gtrsim 1.5$, which is similar to the result of \citetalias{Myles:2021:DESY3} (see their Figure 11). The $n(z)$ in that fourth bin shows a fluctuation around $z \approx 1.1$, which is also found in their work. This is because some stochastic features in the $n(z)$ originate primarily from fluctuations in number counts --- arising from sample variance and shot noise --- of the deep-field sample and the redshift sample (see Section~\ref{sec:sec:deep} and \ref{sec:sec:Zsample} for sample definitions). The fluctuations are shared between our analysis and that of \citetalias{Myles:2021:DESY3} since both use the same deep-field data and very similar redshift data (see Section~\ref{sec:sec:Zsample}). Our uncertainty quantification formalism (Section \ref{sec:sec:3sdir}) accounts for such fluctuations in the deep-field/redshift sample and propagates these effects into the uncertainties of the presented $n(z)$ estimates.

Table~\ref{tab:z_calibration} lists the uncertainty on the mean redshift, when using different combinations of the uncertainty quantification methods from Sections~\ref{sec:sec:3sdir}, \ref{sec:sec:ZPuncert}, and \ref{sec:sec:Zsample}. Similar to \citetalias{Myles:2021:DESY3} (see their Table 2), the photometric zeropoint uncertainty (ZP) is the dominant contribution for the lowest redshift bins. We also find our uncertainty from just shot noise is similar to theirs.

\begin{table*}
    \centering
    \begin{tabular}{ccccccc}
        \hline
        & & Bin 1 & Bin 2 & Bin 3 & Bin 4  \\[5pt]
        & $z_{\rm bin}$ range & 0 -- 0.381 & 0.381 -- 0.619 & 0.619 -- 0.8030 & 0.8030 -- 2.0\\[2pt]
        \hline
        \hline
        \vspace{10pt}
        \multirow{2}{*}{$\langle z \rangle$}& SOMPZ only & 0.3332 & 0.5013 & 0.7064 & 0.9056\\
        & SOMPZ + WZ & 0.3304 & 0.5079 & 0.7177 & 0.9106\\
        \hline
        \hline
        \multirow{4}{*}{Uncertainty, $\sigma(\langle z \rangle)$}& 3sDir & 0.0049 & 0.0055 & 0.0050 & 0.0063\\
        & 3sDir + ZP & 0.0161 & 0.0132 & 0.0097 & 0.0097\\
        & 3sDir + ZB & 0.0055 & 0.0068 & 0.0058 & 0.0089\\
        & \textbf{3sDir + ZP + ZB} & \textbf{0.0163} & \textbf{0.0139} & \textbf{0.0101} & \textbf{0.0117}\\
        & 3sDir + ZP + ZB + WZ & 0.0132 & 0.0146 & 0.0111 & 0.0098\\
        \hline
    \end{tabular}
    \caption{A summary of the mean redshift, and associated uncertainties, of the $n(z)$. The first row shows the minimum/maximum $\langle z | \chat \rangle$ (mean redshift given wide cell) of all wide cells assigned to this tomographic bin. The leftmost (rightmost) edge of the first (fourth) tomographic bin is assigned as $z = 0$ ($z = 2$). The next two rows are the mean redshifts for the $n(z)$ estimates using only SOMPZ and using SOMPZ plus WZ. The remaining rows show the uncertainty on the mean redshifts, for different setups involving the sample variance and shot noise from 3sDir, the photometric zeropoint (ZP) uncertainty, redshift bias uncertainty (ZB), and inclusion of clustering redshift information (see Appendix~\ref{appx:SOMPZ_WZ}). The bolded numbers are the fiducial prior used in the cosmic shear analysis \citepalias{paper3}.}
    \label{tab:z_calibration}
\end{table*}

\subsection{Clustering redshifts} \label{sec:Results_Nz:WZ}

We now focus on the WZ measurements. The $w_{ur}$ estimates cannot be trivially converted to $n(z)$ estimates; hence the multi-faceted forward modelling approach detailed in Sections~\ref{sec:sec:Wz} and \ref{sec:sec:SOMPZWZ}. We instead show a naive estimate of the $n(z)$ obtained by inverting Equation \eqref{eqn:WZ:theoryavg}. We will assume magnification is negligible, which is reasonable for the bulk of the distribution but not the tails \citepalias{Gatti:2022:WzY3}. The amplitude of this estimate is then uncertain by factors of order $\mathcal{O}(1-10)$ given the remaining uncertainties in the bias of the unknown (source) sample. For visual purposes, we rescale our naive WZ-based $n(z)$ estimate by a single redshift-independent factor such that the estimate has a similar range of values as the SOMPZ estimate. This enables an easier (qualitative) comparison of the different shapes of the SOMPZ-based and WZ-based $n(z)$ estimates. We intentionally use the raw WZ measurement to showcase the level of this qualitative agreement between the SOMPZ and WZ methods prior to any systematics marginalization detailed in the forward model of Section~\ref{sec:sec:SOMPZWZ}.

Figure~\ref{fig:n_of_z} presents these WZ-based $n(z)$ estimates and shows they generally have the same shape as the SOMPZ-based estimates. Note that we still cannot make any quantitative, statistical statements about agreement between the two estimates given that converting the WZ measurement to an $n(z)$ estimate requires marginalizing over a number of systematics parameters. A formal, statistical comparison of the two is found below in Section~\ref{sec:Results_Nz:Crosscheck}. Here, we only discuss qualitative aspects of the measurement.

The results of Figure~\ref{fig:n_of_z} show the available information from WZ is negligible above $z \gtrsim 1.5$. This higher redshift regime is not well-constrained due to the low number of reference galaxies in that range. The \Boss sample nominally contains many galaxies above that redshift but the sample was not constructed uniformly across the entire sky footprint of the \Boss survey and is instead localized to specific areas of the sky (such as the northern strip in the DES footprint) that have minimal or no overlap with our dataset. See Section~\ref{sec:sec:reference} for more details on the statistical power of our reference sample compared to \citetalias{Gatti:2022:WzY3}. Still, we find the WZ measurements provide sufficient information across most of the redshift range spanned by our source-galaxy sample.

\subsection{Combination of SOMPZ and WZ} \label{sec:Results_Nz:Crosscheck}

Finally, we perform the procedure detailed in Section~\ref{sec:sec:SOMPZWZ} to estimate the consistency of the SOMPZ estimates with the WZ measurement. Briefly, this involves (i) estimating the best-fit WZ prediction for a specific SOMPZ-based $n(z)$ realization after marginalizing over all the systematic functions and magnification terms, and (ii) calculating the $\chi^2$ between this best-fit model and the WZ measurement. We use thirty WZ datapoints and the model has eight free parameters, resulting in $N_{\rm dof} = 22$. Therefore, values of $\chi^2 < 44.6$ indicate consistency between the two estimates at better than the $3\sigma$ level.

\begin{figure}
    \centering
    \includegraphics[width = \columnwidth]{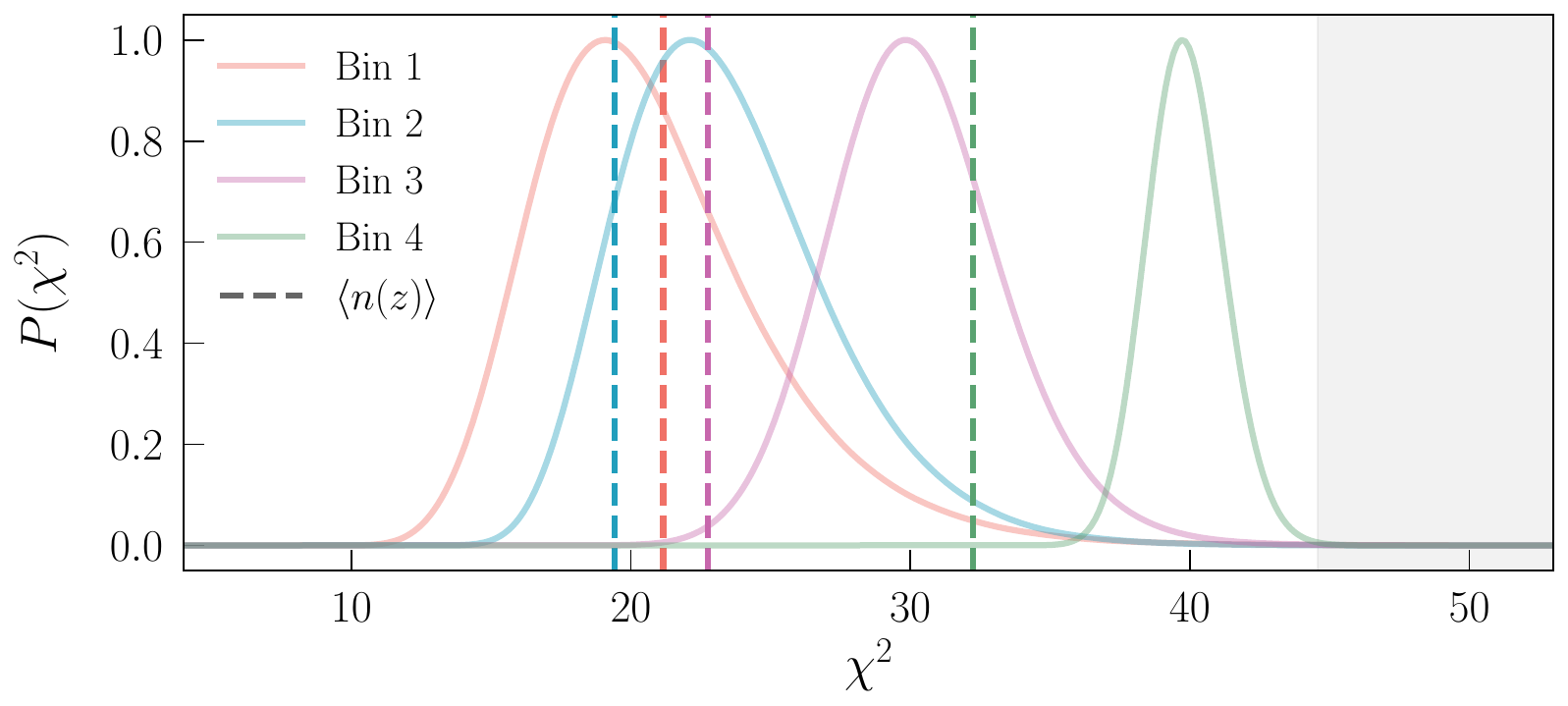}
    \caption{The distribution of $\chi^2$ values from comparing the SOMPZ-based $n(z)$, for each tomographic bin, against the WZ measurement in that bin. The $\chi^2$ is obtained after marginalizing over all systematic functions, using the techniques in Section~\ref{sec:sec:SOMPZWZ}. The mean $n(z)$, shown as dashed lines, follows $\chi^2 < 44.6$. The threshold is the $3\sigma$ limit of a $\chi^2$ distribution with $N_{\rm dof} = 22$. We compute the same for the individual SOMPZ samples, and show all values are also below the $3\sigma$ threshold. In summary, the SOMPZ-based $n(z)$ is consistent with the WZ measurement.}
    \label{fig:nz_chi2}
\end{figure}

Figure~\ref{fig:nz_chi2} shows the distribution of $\chi^2$ values for all $10^7$ SOMPZ-based $n(z)$ samples. The gray region demarcates the range of $\chi^2$ values corresponding to inconsistencies at worse than $3\sigma$. All samples are below the $3\sigma$ threshold. As a reminder, our test discussed in Section~\ref{sec:sec:SOMPZWZ} found less than $5\%$ of the samples pass the $\chi^2$ threshold if we intentionally offset the SOMPZ-based $n(z)$ by $\Delta z = 0.1$. For our main test, we repeat the $\chi^2$ estimate with the mean $n(z)$ --- obtained by averaging over all SOMPZ-based samples --- rather than the individual samples. This is a more pertinent test as the mean $n(z)$, and not the individual samples, are the actual input to our cosmology analysis. The resulting $\chi^2$ is shown in the vertical dashed lines of Figure~\ref{fig:nz_chi2}. The first three bins are consistent within $0.4\sigma$ and fourth bin is consistent at $1.4\sigma$. We have also confirmed that the $\chi^2$ estimates are within $2\sigma$ even if we ignore the $\Sigma_{n(z)}$ term in the covariance (Equation~\ref{eqn:cov_matrix}). This indicates that our final result --- namely, that the average SOMPZ $n(z)$ realization and the WZ measurements are consistent --- is not sensitive to our assumptions about the $n(z)$ covariance. We also note that in Figure \ref{fig:nz_chi2}, the individual $n(z)$ samples can exhibit larger $\chi^2$ compared to the mean $n(z)$ as the fluctuations in the former can be non-Gaussian in nature (see Figure \ref{fig:n_of_z}) whereas the formalism of Section \ref{sec:sec:SOMPZWZ} is limited to a Gaussian covariance matrix. This limitation underestimates the total variance, and leads to a stricter test, which all samples still pass.

Given the agreement mentioned above, we conclude that our SOMPZ estimates are consistent with the WZ measurement to within $\Delta z \approx 0.1$. This $\Delta z$ is larger than the uncertainty on the mean redshift (see Table~\ref{tab:z_calibration}) but is nonetheless vital in asserting that our redshift estimates do not suffer from any catastrophic errors.

\textbf{In summary}, Table~\ref{tab:z_calibration} shows the mean redshift and its associated uncertainty for the different $n(z)$ estimates we use in this work. We also show the estimates from the combination (not cross-check, as is done in this section) of the SOMPZ and WZ information; the procedure for obtaining these combined samples is detailed in Appendix~\ref{appx:SOMPZ_WZ}. This alternative estimate has a mean redshift that is within the quoted $1\sigma$ uncertainties of the fiducial $n(z)$ samples. The uncertainties are also shown for a variety of different cases, and their estimates are all generally similar. The dominant uncertainty contributions are from shot noise, sample variance, and the zeropoint uncertainties. The redshift sample bias is much lower than in \citetalias{Myles:2021:DESY3} given our updated dataset exhibits a lower bias (see Figure~\ref{fig:Zsample_bias}). The bolded numbers in Table~\ref{tab:z_calibration} are used in our cosmic shear analysis \citepalias{paper4}. 

\textbf{Area-dependent $\boldsymbol{n(z)}$:} We now briefly discuss the sensitivity of our results to survey inhomogeneity. As mentioned before, the \decade image data spans a wider range of image quality compared to surveys like DES (\eg see Figure 1 in \citetalias{paper3}), and exhibits a more spatially inhomogeneous distribution for the observing conditions (depth, seeing etc.). We perform a detailed study in \citetalias{paper3}, where we repeat our cosmology inference, including all necessary shear and redshift calibrations, for subsets of the \decade catalog. We define thirty subsets, by selecting regions of the survey footprint with higher/lower values for different observing conditions such as depth, seeing, extinction etc. This split also probes the sensitivity of our final results to the amplitude of the variations, as the two splits (higher/lower) vary by a factor of two in the exhibited variations of the survey property values (see Figure 5 in \citetalias{paper3}). The cosmology constraints for the different subsets (Figure 4 in \citetalias{paper3}) are statistically consistent with each other and with the fiducial cosmology result. This is a data-driven test (as detailed at the end of Section 6.2 of \citetalias{paper3}) which shows our final cosmology constraints are not incur any significant bias due to the presence of such spatial inhomogeneity. We also test an additional sixteen subsets where the catalog is split on object measurements like size, magnitude, and color. Figure 6 in \citetalias{paper3} shows the cosmology constraints from these splits are also consistent with each other, while the Figure B2 in that work shows the mean redshift changes significantly across some of the split definitions.

\section{Conclusions} \label{sec:conclusions}

In this paper, we present the redshift calibration of the DECADE galaxy shape catalog containing $107$ million galaxies covering $5,\!412 \deg^2$ in the northern Galactic cap. This dataset was constructed using public DECam data from a wide array of both large survey programs and standard community observing proposals.

We split the galaxy catalog into four tomographic bins, each containing a quarter of the total source galaxies. The redshift distribution of all galaxies in each bin is estimated using a fully Bayesian formalism (``SOMPZ'') that follows that of DES Y3 \citepalias{Myles:2021:DESY3} while using some updates described in \citetalias{Sanchez:2023:highzY3}. This method relies on using self-organizing maps (SOMs) to discretize galaxies into different phenotypes, and using a deep-field sample with low-noise photometry and redshift information to learn a color-redshift relation that can be transferred to the noisier wide-field dataset. We build a full source injection pipeline (``\Balrog'') for the \decade dataset, following that of DES \citep{Suchyta:2016, Everett:2022, Anbajagane:2025:Y6Balrog} and this dataset helps map the low-noise deep-field measurements to noisy wide-field measurements in the \decade data. Our SOMPZ methodology also incorporates uncertainties in the deep-field sample due to sample variance, shot noise, photometric zeropoint uncertainties, and redshift biases.

We also use the spatial cross-correlations of the source galaxies with spectroscopic data to make an independent estimate of the redshift distributions (``WZ'', or clustering redshifts). We account for unknown systematics in the WZ measurements through a series of systematic functions, once again following DES Y3 \citepalias{Gatti:2022:WzY3}, and show that our SOMPZ and WZ estimates are statistically consistent with each other.

Our fiducial redshift distributions (Figure~\ref{fig:n_of_z}) are fairly similar to DES Y3, with a tendency for our bin 3 and 4 to be slightly shallower than DES Y3. Our redshift uncertainties are $\sigma_{\langle z\rangle} \approx 0.01$ (depending on the bin), which is comparable to other state-of-the-art weak lensing catalogs. Finally, we note that this work is part of a series of papers. The others detail the shear catalog and all associated systematic checks \citepalias{paper1}, the survey inhomogeneity tests mentioned above \citepalias{paper3}, and the cosmology results for the \decade catalog \citepalias{paper4}. Results from combining this \decade dataset with DES Y3, and with another $\approx 3,\!300\deg^2$ of \decade data in the southern Galactic cap, can be found in \href{\#cite.paper5}{Anbajagane \& Chang et al. (\citeyear{paper5})}.

\section*{Acknowledgements}

DA is supported by the National Science Foundation (NSF) Graduate Research Fellowship under Grant No.\ DGE 1746045. 
CC is supported by the Henry Luce Foundation and Department of Energy (DOE) grant DE-SC0021949. 
The DECADE project is supported by NSF AST-2108168 and AST-2108169.
The DELVE Survey gratefully acknowledges support from Fermilab LDRD (L2019.011), the NASA {\it Fermi} Guest Investigator Program Cycle 9 (No.\ 91201), and the NSF (AST-2108168, AST-2108169, AST-2307126,  AST-2407526, AST-2407527, AST-2407528). This work was completed in part with resources provided by the University of Chicago’s Research Computing Center. The project that gave rise to these results received the support of a fellowship from "la Caixa" Foundation (ID 100010434). The fellowship code is LCF/BQ/PI23/11970028. C.E.M.-V. is supported by the international Gemini Observatory, a program of NSF NOIRLab, which is managed by the Association of Universities for Research in Astronomy (AURA) under a cooperative agreement with the U.S. National Science Foundation, on behalf of the Gemini partnership of Argentina, Brazil, Canada, Chile, the Republic of Korea, and the United States of America.

Funding for the DES Projects has been provided by the U.S. Department of Energy, the U.S. National Science Foundation, the Ministry of Science and Education of Spain, 
the Science and Technology Facilities Council of the United Kingdom, the Higher Education Funding Council for England, the National Center for Supercomputing 
Applications at the University of Illinois at Urbana-Champaign, the Kavli Institute of Cosmological Physics at the University of Chicago, 
the Center for Cosmology and Astro-Particle Physics at the Ohio State University,
the Mitchell Institute for Fundamental Physics and Astronomy at Texas A\&M University, Financiadora de Estudos e Projetos, 
Funda{\c c}{\~a}o Carlos Chagas Filho de Amparo {\`a} Pesquisa do Estado do Rio de Janeiro, Conselho Nacional de Desenvolvimento Cient{\'i}fico e Tecnol{\'o}gico and 
the Minist{\'e}rio da Ci{\^e}ncia, Tecnologia e Inova{\c c}{\~a}o, the Deutsche Forschungsgemeinschaft and the Collaborating Institutions in the Dark Energy Survey. 

The Collaborating Institutions are Argonne National Laboratory, the University of California at Santa Cruz, the University of Cambridge, Centro de Investigaciones Energ{\'e}ticas, 
Medioambientales y Tecnol{\'o}gicas-Madrid, the University of Chicago, University College London, the DES-Brazil Consortium, the University of Edinburgh, 
the Eidgen{\"o}ssische Technische Hochschule (ETH) Z{\"u}rich, 
Fermi National Accelerator Laboratory, the University of Illinois at Urbana-Champaign, the Institut de Ci{\`e}ncies de l'Espai (IEEC/CSIC), 
the Institut de F{\'i}sica d'Altes Energies, Lawrence Berkeley National Laboratory, the Ludwig-Maximilians Universit{\"a}t M{\"u}nchen and the associated Excellence Cluster Universe, 
the University of Michigan, NSF's NOIRLab, the University of Nottingham, The Ohio State University, the University of Pennsylvania, the University of Portsmouth, 
SLAC National Accelerator Laboratory, Stanford University, the University of Sussex, Texas A\&M University, and the OzDES Membership Consortium.

The DES data management system is supported by the National Science Foundation under Grant Numbers AST-1138766 and AST-1536171.
The DES participants from Spanish institutions are partially supported by MICINN under grants ESP2017-89838, PGC2018-094773, PGC2018-102021, SEV-2016-0588, SEV-2016-0597, and MDM-2015-0509, some of which include ERDF funds from the European Union. IFAE is partially funded by the CERCA program of the Generalitat de Catalunya.
Research leading to these results has received funding from the European Research
Council under the European Union's Seventh Framework Program (FP7/2007-2013) including ERC grant agreements 240672, 291329, and 306478.
We  acknowledge support from the Brazilian Instituto Nacional de Ci\^encia
e Tecnologia (INCT) do e-Universo (CNPq grant 465376/2014-2).

Based in part on observations at Cerro Tololo Inter-American Observatory at NSF's NOIRLab, which is managed by the Association of Universities for Research in Astronomy (AURA) under a cooperative agreement with the National Science Foundation.

This work has made use of data from the European Space Agency (ESA) mission {\it Gaia} (\url{https://www.cosmos.esa.int/gaia}), processed by the {\it Gaia} Data Processing and Analysis Consortium (DPAC, \url{https://www.cosmos.esa.int/web/gaia/dpac/consortium}).
Funding for the DPAC has been provided by national institutions, in particular the institutions participating in the {\it Gaia} Multilateral Agreement.

This paper is based on data collected at the Subaru Telescope and retrieved from the HSC data archive system, which is operated by the Subaru Telescope and Astronomy Data Center (ADC) at NAOJ. Data analysis was in part carried out with the cooperation of Center for Computational Astrophysics (CfCA), NAOJ. We are honored and grateful for the opportunity of observing the Universe from Maunakea, which has the cultural, historical and natural significance in Hawaii. 

This research uses services or data provided by the Astro Data Lab, which is part of the Community Science and Data Center (CSDC) Program of NSF NOIRLab. NOIRLab is operated by the Association of Universities for Research in Astronomy (AURA), Inc. under a cooperative agreement with the U.S. National Science Foundation.

This manuscript has been authored by Fermi Forward Discovery Group, LLC under Contract No.\ 89243024CSC000002 with the U.S. Department of Energy, Office of Science, Office of High Energy Physics.

All analysis in this work was enabled greatly by the following software: \textsc{Pandas} \citep{Mckinney2011pandas}, \textsc{NumPy} \citep{vanderWalt2011Numpy}, \textsc{SciPy} \citep{Virtanen2020Scipy}, and \textsc{Matplotlib} \citep{Hunter2007Matplotlib}. We have also used
the Astrophysics Data Service (\href{https://ui.adsabs.harvard.edu/}{ADS}) and \href{https://arxiv.org/}{\texttt{arXiv}} preprint repository extensively during this project and the writing of the paper.

\section*{Data Availability}

All catalogs and derived data products (data vectors, redshift distributions, calibrations etc.) for the cosmology analysis are now publicly available through the Noirlab Datalab portal \citep{Fitzpatrick:2014:DataLab, Nikutta:2020:DataLab} as well as through Globus and other avenues. Please visit \url{dhayaaanbajagane.github.io/data_release/decade} for a list of the available dataproducts and their corresponding data access. Our intention is to make all useful products immediately available to the community. Please reach out to DA if a data product of interest to you is not on the above list.

\bibliographystyle{mnras}
\bibliography{References}

\begin{thebibliography}{}
\makeatletter
\relax
\def\mn@urlcharsother{\let\do\@makeother \do\$\do\&\do\#\do\^\do\_\do\%\do\~}
\def\mn@doi{\begingroup\mn@urlcharsother \@ifnextchar [ {\mn@doi@} {\mn@doi@[]}}
\def\mn@doi@[#1]#2{\def\@tempa{#1}\ifx\@tempa\@empty \href {http://dx.doi.org/#2} {doi:#2}\else \href {http://dx.doi.org/#2} {#1}\fi \endgroup}
\def\mn@eprint#1#2{\mn@eprint@#1:#2::\@nil}
\def\mn@eprint@arXiv#1{\href {http://arxiv.org/abs/#1} {{\tt arXiv:#1}}}
\def\mn@eprint@dblp#1{\href {http://dblp.uni-trier.de/rec/bibtex/#1.xml} {dblp:#1}}
\def\mn@eprint@#1:#2:#3:#4\@nil{\def\@tempa {#1}\def\@tempb {#2}\def\@tempc {#3}\ifx \@tempc \@empty \let \@tempc \@tempb \let \@tempb \@tempa \fi \ifx \@tempb \@empty \def\@tempb {arXiv}\fi \@ifundefined {mn@eprint@\@tempb}{\@tempb:\@tempc}{\expandafter \expandafter \csname mn@eprint@\@tempb\endcsname \expandafter{\@tempc}}}

\bibitem[\protect\citeauthoryear{{Abbott} et~al.,}{{Abbott} et~al.}{2018}]{DES2018}
{Abbott} T.~M.~C.,  et~al., 2018, \mn@doi [\apjs] {10.3847/1538-4365/aae9f0}, \href {https://ui.adsabs.harvard.edu/abs/2018ApJS..239...18A} {239, 2, 18}

\bibitem[\protect\citeauthoryear{{Aihara} et~al.,}{{Aihara} et~al.}{2018}]{Aihara2018a}
{Aihara} H.,  et~al., 2018, \mn@doi [\pasj] {10.1093/pasj/psx081}, \href {https://ui.adsabs.harvard.edu/abs/2018PASJ...70S...8A} {70, S8}

\bibitem[\protect\citeauthoryear{{Alarcon} et~al.,}{{Alarcon} et~al.}{2021}]{Alarcon:2021:PausCosmos}
{Alarcon} A.,  et~al., 2021, \mn@doi [\mnras] {10.1093/mnras/staa3659}, \href {https://ui.adsabs.harvard.edu/abs/2021MNRAS.501.6103A} {501, 4, 6103}

\bibitem[\protect\citeauthoryear{Alfeld}{Alfeld}{1984}]{Alfeld:1984:Interp}
Alfeld P.,  1984, \mn@doi [Computer Aided Geometric Design] {https://doi.org/10.1016/0167-8396(84)90029-3}, 1, 2, 169

\bibitem[\protect\citeauthoryear{{Amon} et~al.,}{{Amon} et~al.}{2022}]{Amon2022}
{Amon} A.,  et~al., 2022, \mn@doi [\prd] {10.1103/PhysRevD.105.023514}, \href {https://ui.adsabs.harvard.edu/abs/2022PhRvD.105b3514A} {105, 2, 023514}

\bibitem[\protect\citeauthoryear{{Anbajagane} \& {Lee}}{{Anbajagane} \& {Lee}}{2025a}]{Primordial1}
{Anbajagane} D.,  {Lee} H.,  2025a, \mn@doi [arXiv e-prints] {10.48550/arXiv.2509.02693}, \href {https://ui.adsabs.harvard.edu/abs/2025arXiv250902693A} {p. arXiv:2509.02693}

\bibitem[\protect\citeauthoryear{{Anbajagane} \& {Lee}}{{Anbajagane} \& {Lee}}{2025b}]{Primordial2}
{Anbajagane} D.,  {Lee} H.,  2025b, \mn@doi [arXiv e-prints] {10.48550/arXiv.2509.02695}, \href {https://ui.adsabs.harvard.edu/abs/2025arXiv250902695A} {p. arXiv:2509.02695}

\bibitem[\protect\citeauthoryear{{Anbajagane \& Tabbutt} et~al.,}{{Anbajagane \& Tabbutt} et~al.}{2025}]{Anbajagane:2025:Y6Balrog}
{Anbajagane \& Tabbutt} et~al., 2025, \mn@doi [arXiv e-prints] {10.48550/arXiv.2501.05683}, \href {https://ui.adsabs.harvard.edu/abs/2025arXiv250105683A} {p. arXiv:2501.05683}

\bibitem[\protect\citeauthoryear{{Anbajagane}, {Pandey}  \& {Chang}}{{Anbajagane} et~al.}{2024a}]{Anbajagane:2024:Baryonification}
{Anbajagane} D.,  {Pandey} S.,   {Chang} C.,  2024a, \mn@doi [The Open Journal of Astrophysics] {10.33232/001c.126788}, \href {https://ui.adsabs.harvard.edu/abs/2024OJAp....7E.108A} {7, 108}

\bibitem[\protect\citeauthoryear{{Anbajagane}, {Chang}, {Lee}  \& {Gatti}}{{Anbajagane} et~al.}{2024b}]{Anbajagane2023Inflation}
{Anbajagane} D.,  {Chang} C.,  {Lee} H.,   {Gatti} M.,  2024b, \mn@doi [\jcap] {10.1088/1475-7516/2024/03/062}, \href {https://ui.adsabs.harvard.edu/abs/2024JCAP...03..062A} {2024, 3, 062}

\bibitem[\protect\citeauthoryear{{Anbajagane} et~al.,}{{Anbajagane} et~al.}{2025a}]{paper1}
{Anbajagane} D.,  et~al., 2025a, \mn@doi [arXiv e-prints] {10.48550/arXiv.2502.17674}, \href {https://ui.adsabs.harvard.edu/abs/2025arXiv250217674A} {p. arXiv:2502.17674}

\bibitem[\protect\citeauthoryear{{Anbajagane} et~al.,}{{Anbajagane} et~al.}{2025b}]{paper3}
{Anbajagane} D.,  et~al., 2025b, \mn@doi [arXiv e-prints] {10.48550/arXiv.2502.17676}, \href {https://ui.adsabs.harvard.edu/abs/2025arXiv250217676A} {p. arXiv:2502.17676}

\bibitem[\protect\citeauthoryear{{Anbajagane} et~al.,}{{Anbajagane} et~al.}{2025c}]{paper4}
{Anbajagane} D.,  et~al., 2025c, \mn@doi [arXiv e-prints] {10.48550/arXiv.2502.17677}, \href {https://ui.adsabs.harvard.edu/abs/2025arXiv250217677A} {p. arXiv:2502.17677}

\bibitem[\protect\citeauthoryear{{Anbajagane} et~al.,}{{Anbajagane} et~al.}{2025d}]{paper5}
{Anbajagane} D.,  et~al., 2025d, \mn@doi [arXiv e-prints] {10.48550/arXiv.2509.03582}, \href {https://ui.adsabs.harvard.edu/abs/2025arXiv250903582A} {p. arXiv:2509.03582}

\bibitem[\protect\citeauthoryear{{Aric{\`o}}, {Angulo}, {Contreras}, {Ondaro-Mallea}, {Pellejero-Iba{\~n}ez}  \& {Zennaro}}{{Aric{\`o}} et~al.}{2021}]{Arico:2021:Bacco}
{Aric{\`o}} G.,  {Angulo} R.~E.,  {Contreras} S.,  {Ondaro-Mallea} L.,  {Pellejero-Iba{\~n}ez} M.,   {Zennaro} M.,  2021, \mn@doi [\mnras] {10.1093/mnras/stab1911}, \href {https://ui.adsabs.harvard.edu/abs/2021MNRAS.506.4070A} {506, 3, 4070}

\bibitem[\protect\citeauthoryear{{Asgari} et~al.,}{{Asgari} et~al.}{2021}]{Asgari2021}
{Asgari} M.,  et~al., 2021, \mn@doi [\aap] {10.1051/0004-6361/202039070}, \href {https://ui.adsabs.harvard.edu/abs/2021A&A...645A.104A} {645, A104}

\bibitem[\protect\citeauthoryear{{Bartelmann} \& {Schneider}}{{Bartelmann} \& {Schneider}}{2001}]{Bartelmann2001}
{Bartelmann} M.,  {Schneider} P.,  2001, \mn@doi [\physrep] {10.1016/S0370-1573(00)00082-X}, \href {http://adsabs.harvard.edu/abs/2001PhR...340..291B} {340, 291}

\bibitem[\protect\citeauthoryear{{Ben{\'\i}tez}}{{Ben{\'\i}tez}}{2000}]{Benitez2000}
{Ben{\'\i}tez} N.,  2000, \mn@doi [\apj] {10.1086/308947}, \href {https://ui.adsabs.harvard.edu/abs/2000ApJ...536..571B} {536, 2, 571}

\bibitem[\protect\citeauthoryear{{Bertin}}{{Bertin}}{2010}]{Bertin:2010:Swarp}
{Bertin} E.,  2010, {SWarp: Resampling and Co-adding FITS Images Together}, Astrophysics Source Code Library, record ascl:1010.068

\bibitem[\protect\citeauthoryear{{Bertin} \& {Arnouts}}{{Bertin} \& {Arnouts}}{1996}]{Bertin:1996:SrcExt}
{Bertin} E.,  {Arnouts} S.,  1996, \mn@doi [\aaps] {10.1051/aas:1996164}, \href {https://ui.adsabs.harvard.edu/abs/1996A&AS..117..393B} {117, 393}

\bibitem[\protect\citeauthoryear{{Buchs \& Davis} et~al.,}{{Buchs \& Davis} et~al.}{2019}]{Buchs2019}
{Buchs \& Davis} et~al., 2019, \mn@doi [\mnras] {10.1093/mnras/stz2162}, \href {https://ui.adsabs.harvard.edu/abs/2019MNRAS.489..820B} {489, 1, 820}

\bibitem[\protect\citeauthoryear{{Campos} et~al.,}{{Campos} et~al.}{2024}]{Campos:2024:PZ}
{Campos} A.,  et~al., 2024, \mn@doi [arXiv e-prints] {10.48550/arXiv.2408.00922}, \href {https://ui.adsabs.harvard.edu/abs/2024arXiv240800922C} {p. arXiv:2408.00922}

\bibitem[\protect\citeauthoryear{{Carrasco Kind} \& {Brunner}}{{Carrasco Kind} \& {Brunner}}{2014}]{Carrasco:2014:SOM}
{Carrasco Kind} M.,  {Brunner} R.~J.,  2014, \mn@doi [\mnras] {10.1093/mnras/stt2456}, \href {https://ui.adsabs.harvard.edu/abs/2014MNRAS.438.3409C} {438, 4, 3409}

\bibitem[\protect\citeauthoryear{{Chisari} et~al.,}{{Chisari} et~al.}{2019}]{Chisari2019CCL}
{Chisari} N.~E.,  et~al., 2019, \mn@doi [\apjs] {10.3847/1538-4365/ab1658}, \href {https://ui.adsabs.harvard.edu/abs/2019ApJS..242....2C} {242, 1, 2}

\bibitem[\protect\citeauthoryear{{Dalal} et~al.,}{{Dalal} et~al.}{2023}]{Dalal2023}
{Dalal} R.,  et~al., 2023, \mn@doi [arXiv e-prints] {10.48550/arXiv.2304.00701}, \href {https://ui.adsabs.harvard.edu/abs/2023arXiv230400701D} {p. arXiv:2304.00701}

\bibitem[\protect\citeauthoryear{{Dalton} et~al.,}{{Dalton} et~al.}{2006}]{Dalton2006}
{Dalton} G.~B.,  et~al., 2006, in {McLean} I.~S.,  {Iye} M.,  eds,  Society of Photo-Optical Instrumentation Engineers (SPIE) Conference Series Vol. 6269, Ground-based and Airborne Instrumentation for Astronomy. p. 62690X, \mn@doi{10.1117/12.670018}

\bibitem[\protect\citeauthoryear{{Davis} \& {Peebles}}{{Davis} \& {Peebles}}{1983}]{Davis:1983:2pt}
{Davis} M.,  {Peebles} P.~J.~E.,  1983, \mn@doi [\apj] {10.1086/160884}, \href {https://ui.adsabs.harvard.edu/abs/1983ApJ...267..465D} {267, 465}

\bibitem[\protect\citeauthoryear{{Davis} et~al.,}{{Davis} et~al.}{2017}]{Davis2017}
{Davis} C.,  et~al., 2017, \mn@doi [arXiv e-prints] {10.48550/arXiv.1710.02517}, \href {https://ui.adsabs.harvard.edu/abs/2017arXiv171002517D} {p. arXiv:1710.02517}

\bibitem[\protect\citeauthoryear{{Dawson} et~al.,}{{Dawson} et~al.}{2013}]{Dawson:2013:BOSS}
{Dawson} K.~S.,  et~al., 2013, \mn@doi [\aj] {10.1088/0004-6256/145/1/10}, \href {https://ui.adsabs.harvard.edu/abs/2013AJ....145...10D} {145, 1, 10}

\bibitem[\protect\citeauthoryear{{Dawson} et~al.,}{{Dawson} et~al.}{2016}]{Dawson:2016:eBOSS}
{Dawson} K.~S.,  et~al., 2016, \mn@doi [\aj] {10.3847/0004-6256/151/2/44}, \href {https://ui.adsabs.harvard.edu/abs/2016AJ....151...44D} {151, 2, 44}

\bibitem[\protect\citeauthoryear{{Dodelson} \& {Schneider}}{{Dodelson} \& {Schneider}}{2013}]{Dodelson:2013:Cov}
{Dodelson} S.,  {Schneider} M.~D.,  2013, \mn@doi [\prd] {10.1103/PhysRevD.88.063537}, \href {https://ui.adsabs.harvard.edu/abs/2013PhRvD..88f3537D} {88, 6, 063537}

\bibitem[\protect\citeauthoryear{{Edge}, {Sutherland}, {Kuijken}, {Driver}, {McMahon}, {Eales}  \& {Emerson}}{{Edge} et~al.}{2013}]{Edge:2013:VIKING}
{Edge} A.,  {Sutherland} W.,  {Kuijken} K.,  {Driver} S.,  {McMahon} R.,  {Eales} S.,   {Emerson} J.~P.,  2013, The Messenger, \href {https://ui.adsabs.harvard.edu/abs/2013Msngr.154...32E} {154, 32}

\bibitem[\protect\citeauthoryear{{Elvin-Poole \& MacCrann} et~al.,}{{Elvin-Poole \& MacCrann} et~al.}{2023}]{ElvinPoole:2023:MagY3}
{Elvin-Poole \& MacCrann} et~al., 2023, \mn@doi [\mnras] {10.1093/mnras/stad1594}, \href {https://ui.adsabs.harvard.edu/abs/2023MNRAS.523.3649E} {523, 3, 3649}

\bibitem[\protect\citeauthoryear{{Everett} et~al.,}{{Everett} et~al.}{2022}]{Everett:2022}
{Everett} S.,  et~al., 2022, \mn@doi [\apjs] {10.3847/1538-4365/ac26c1}, \href {https://ui.adsabs.harvard.edu/abs/2022ApJS..258...15E} {258, 1, 15}

\bibitem[\protect\citeauthoryear{{Fitzpatrick} et~al.,}{{Fitzpatrick} et~al.}{2014}]{Fitzpatrick:2014:DataLab}
{Fitzpatrick} M.~J.,  et~al., 2014, in {Peck} A.~B.,  {Benn} C.~R.,   {Seaman} R.~L.,  eds,  Society of Photo-Optical Instrumentation Engineers (SPIE) Conference Series Vol. 9149, Observatory Operations: Strategies, Processes, and Systems V. p. 91491T, \mn@doi{10.1117/12.2057445}

\bibitem[\protect\citeauthoryear{{Flaugher}}{{Flaugher}}{2005}]{Flaugher2005}
{Flaugher} B.,  2005, \mn@doi [International Journal of Modern Physics A] {10.1142/S0217751X05025917}, \href {http://adsabs.harvard.edu/abs/2005IJMPA..20.3121F} {20, 3121}

\bibitem[\protect\citeauthoryear{{Gatti \& Giannini} et~al.,}{{Gatti \& Giannini} et~al.}{2022}]{Gatti:2022:WzY3}
{Gatti \& Giannini} et~al., 2022, \mn@doi [\mnras] {10.1093/mnras/stab3311}, \href {https://ui.adsabs.harvard.edu/abs/2022MNRAS.510.1223G} {510, 1, 1223}

\bibitem[\protect\citeauthoryear{{Gatti \& Vielzeuf} et~al.,}{{Gatti \& Vielzeuf} et~al.}{2018}]{Gatti:2018:WZ}
{Gatti \& Vielzeuf} et~al., 2018, \mn@doi [\mnras] {10.1093/mnras/sty466}, \href {https://ui.adsabs.harvard.edu/abs/2018MNRAS.477.1664G} {477, 2, 1664}

\bibitem[\protect\citeauthoryear{{Giannini} et~al.,}{{Giannini} et~al.}{2022}]{Giannini2022}
{Giannini} G.,  et~al., 2022, \mn@doi [arXiv e-prints] {10.48550/arXiv.2209.05853}, \href {https://ui.adsabs.harvard.edu/abs/2022arXiv220905853G} {p. arXiv:2209.05853}

\bibitem[\protect\citeauthoryear{{Goldstein}, {Philcox}, {Hill}  \& {Hui}}{{Goldstein} et~al.}{2024}]{Goldstein:2024:inflation}
{Goldstein} S.,  {Philcox} O. H.~E.,  {Hill} J.~C.,   {Hui} L.,  2024, \mn@doi [\prd] {10.1103/PhysRevD.110.083516}, \href {https://ui.adsabs.harvard.edu/abs/2024PhRvD.110h3516G} {110, 8, 083516}

\bibitem[\protect\citeauthoryear{{Hartlap}, {Simon}  \& {Schneider}}{{Hartlap} et~al.}{2007}]{Hartlap2007}
{Hartlap} J.,  {Simon} P.,   {Schneider} P.,  2007, \mn@doi [\aap] {10.1051/0004-6361:20066170}, \href {https://ui.adsabs.harvard.edu/abs/2007A&A...464..399H} {464, 1, 399}

\bibitem[\protect\citeauthoryear{{Hartley \& Choi} et~al.,}{{Hartley \& Choi} et~al.}{2022}]{Hartley:2022:Y3Deepfields}
{Hartley \& Choi} et~al., 2022, \mn@doi [\mnras] {10.1093/mnras/stab3055}, \href {https://ui.adsabs.harvard.edu/abs/2022MNRAS.509.3547H} {509, 3, 3547}

\bibitem[\protect\citeauthoryear{{Hikage} et~al.,}{{Hikage} et~al.}{2019}]{Hikage2019}
{Hikage} C.,  et~al., 2019, \mn@doi [\pasj] {10.1093/pasj/psz010}, \href {https://ui.adsabs.harvard.edu/abs/2019PASJ...71...43H} {71, 2, 43}

\bibitem[\protect\citeauthoryear{{Hildebrandt} et~al.,}{{Hildebrandt} et~al.}{2020}]{Hildebrandt2020}
{Hildebrandt} H.,  et~al., 2020, \mn@doi [\aap] {10.1051/0004-6361/201834878}, \href {https://ui.adsabs.harvard.edu/abs/2020A&A...633A..69H} {633, A69}

\bibitem[\protect\citeauthoryear{{Hoyle \& Gruen} et~al.,}{{Hoyle \& Gruen} et~al.}{2018}]{Hoyle2018}
{Hoyle \& Gruen} et~al., 2018, \mn@doi [\mnras] {10.1093/mnras/sty957}, \href {https://ui.adsabs.harvard.edu/abs/2018MNRAS.478..592H} {478, 1, 592}

\bibitem[\protect\citeauthoryear{{Hu}}{{Hu}}{1999}]{Hu:1999:WL_Tomography}
{Hu} W.,  1999, \mn@doi [\apjl] {10.1086/312210}, \href {https://ui.adsabs.harvard.edu/abs/1999ApJ...522L..21H} {522, 1, L21}

\bibitem[\protect\citeauthoryear{{Huff} \& {Mandelbaum}}{{Huff} \& {Mandelbaum}}{2017}]{Huff2017}
{Huff} E.,  {Mandelbaum} R.,  2017, \mn@doi [arXiv e-prints] {10.48550/arXiv.1702.02600}, \href {https://ui.adsabs.harvard.edu/abs/2017arXiv170202600H} {p. arXiv:1702.02600}

\bibitem[\protect\citeauthoryear{{Hunter}}{{Hunter}}{2007}]{Hunter2007Matplotlib}
{Hunter} J.~D.,  2007, \mn@doi [Computing in Science and Engineering] {10.1109/MCSE.2007.55}, \href {https://ui.adsabs.harvard.edu/abs/2007CSE.....9...90H} {9, 3, 90}

\bibitem[\protect\citeauthoryear{{Jarvis}, {Bernstein}  \& {Jain}}{{Jarvis} et~al.}{2004}]{Jarvis2004TreeCorr}
{Jarvis} M.,  {Bernstein} G.,   {Jain} B.,  2004, \mn@doi [\mnras] {10.1111/j.1365-2966.2004.07926.x}, \href {https://ui.adsabs.harvard.edu/abs/2004MNRAS.352..338J} {352, 1, 338}

\bibitem[\protect\citeauthoryear{{Jarvis} et~al.,}{{Jarvis} et~al.}{2013}]{Jarvis2013}
{Jarvis} M.~J.,  et~al., 2013, \mn@doi [\mnras] {10.1093/mnras/sts118}, \href {https://ui.adsabs.harvard.edu/abs/2013MNRAS.428.1281J} {428, 2, 1281}

\bibitem[\protect\citeauthoryear{{Jarvis} et~al.,}{{Jarvis} et~al.}{2016}]{Jarvis2016}
{Jarvis} M.,  et~al., 2016, \mn@doi [\mnras] {10.1093/mnras/stw990}, \href {https://ui.adsabs.harvard.edu/abs/2016MNRAS.460.2245J} {460, 2, 2245}

\bibitem[\protect\citeauthoryear{{Kitching} et~al.,}{{Kitching} et~al.}{2012}]{Kitching2012}
{Kitching} T.~D.,  et~al., 2012, \mn@doi [\mnras] {10.1111/j.1365-2966.2012.21095.x}, \href {https://ui.adsabs.harvard.edu/abs/2012MNRAS.423.3163K} {423, 4, 3163}

\bibitem[\protect\citeauthoryear{{Kitching} et~al.,}{{Kitching} et~al.}{2013}]{Kitching2013}
{Kitching} T.~D.,  et~al., 2013, \mn@doi [\apjs] {10.1088/0067-0049/205/2/12}, \href {https://ui.adsabs.harvard.edu/abs/2013ApJS..205...12K} {205, 2, 12}

\bibitem[\protect\citeauthoryear{Kohonen}{Kohonen}{1982}]{Kohonen:1982:SOM}
Kohonen T.,  1982, Biological Cybernetics, 43, 1, 59

\bibitem[\protect\citeauthoryear{{Kohonen}}{{Kohonen}}{2001}]{Kohonen:2001:SOM}
{Kohonen} T.,  2001, {Self-Organizing Maps}

\bibitem[\protect\citeauthoryear{{Kong} et~al.,}{{Kong} et~al.}{2024}]{Kong:2024:ObiWan}
{Kong} H.,  et~al., 2024, \mn@doi [arXiv e-prints] {10.48550/arXiv.2405.16299}, \href {https://ui.adsabs.harvard.edu/abs/2024arXiv240516299K} {p. arXiv:2405.16299}

\bibitem[\protect\citeauthoryear{{Laigle} et~al.,}{{Laigle} et~al.}{2016}]{Laigle2016}
{Laigle} C.,  et~al., 2016, \mn@doi [\apjs] {10.3847/0067-0049/224/2/24}, \href {https://ui.adsabs.harvard.edu/abs/2016ApJS..224...24L} {224, 2, 24}

\bibitem[\protect\citeauthoryear{{Le F{\`e}vre} et~al.,}{{Le F{\`e}vre} et~al.}{2005}]{LeFevre:2005:VVDS}
{Le F{\`e}vre} O.,  et~al., 2005, \mn@doi [\aap] {10.1051/0004-6361:20041960}, \href {https://ui.adsabs.harvard.edu/abs/2005A&A...439..845L} {439, 3, 845}

\bibitem[\protect\citeauthoryear{{Li} et~al.,}{{Li} et~al.}{2023}]{Li2023b}
{Li} X.,  et~al., 2023, \mn@doi [\prd] {10.1103/PhysRevD.108.123518}, \href {https://ui.adsabs.harvard.edu/abs/2023PhRvD.108l3518L} {108, 12, 123518}

\bibitem[\protect\citeauthoryear{{Lilly} et~al.,}{{Lilly} et~al.}{2007}]{Lilly:2007:zCOSMOS}
{Lilly} S.~J.,  et~al., 2007, \mn@doi [\apjs] {10.1086/516589}, \href {https://ui.adsabs.harvard.edu/abs/2007ApJS..172...70L} {172, 1, 70}

\bibitem[\protect\citeauthoryear{{MacCrann \& Becker} et~al.,}{{MacCrann \& Becker} et~al.}{2022}]{Maccrann2022ImSim}
{MacCrann \& Becker} et~al., 2022, \mn@doi [\mnras] {10.1093/mnras/stab2870}, \href {https://ui.adsabs.harvard.edu/abs/2022MNRAS.509.3371M} {509, 3, 3371}

\bibitem[\protect\citeauthoryear{{Mandelbaum} et~al.,}{{Mandelbaum} et~al.}{2015}]{Mandelbaum2015}
{Mandelbaum} R.,  et~al., 2015, \mn@doi [\mnras] {10.1093/mnras/stv781}, \href {https://ui.adsabs.harvard.edu/abs/2015MNRAS.450.2963M} {450, 3, 2963}

\bibitem[\protect\citeauthoryear{{Masters} et~al.,}{{Masters} et~al.}{2015}]{Masters:2015:SOM}
{Masters} D.,  et~al., 2015, \mn@doi [\apj] {10.1088/0004-637X/813/1/53}, \href {https://ui.adsabs.harvard.edu/abs/2015ApJ...813...53M} {813, 1, 53}

\bibitem[\protect\citeauthoryear{{Masters}, {Stern}, {Cohen}, {Capak}, {Rhodes}, {Castander}  \& {Paltani}}{{Masters} et~al.}{2017}]{Masters:2017:C3R2_DR1}
{Masters} D.~C.,  {Stern} D.~K.,  {Cohen} J.~G.,  {Capak} P.~L.,  {Rhodes} J.~D.,  {Castander} F.~J.,   {Paltani} S.,  2017, \mn@doi [\apj] {10.3847/1538-4357/aa6f08}, \href {https://ui.adsabs.harvard.edu/abs/2017ApJ...841..111M} {841, 2, 111}

\bibitem[\protect\citeauthoryear{{Masters} et~al.,}{{Masters} et~al.}{2019}]{Masters:2019:C3R2_DR2}
{Masters} D.~C.,  et~al., 2019, \mn@doi [\apj] {10.3847/1538-4357/ab184d}, \href {https://ui.adsabs.harvard.edu/abs/2019ApJ...877...81M} {877, 2, 81}

\bibitem[\protect\citeauthoryear{{McCracken} et~al.,}{{McCracken} et~al.}{2012}]{McCracken2012}
{McCracken} H.~J.,  et~al., 2012, \mn@doi [\aap] {10.1051/0004-6361/201219507}, \href {https://ui.adsabs.harvard.edu/abs/2012A&A...544A.156M} {544, A156}

\bibitem[\protect\citeauthoryear{McKinney}{McKinney}{2011}]{Mckinney2011pandas}
McKinney W.,  2011, Python for High Performance and Scientific Computing, 14

\bibitem[\protect\citeauthoryear{{Mead}, {Brieden}, {Tr{\"o}ster}  \& {Heymans}}{{Mead} et~al.}{2021}]{Mead2021b}
{Mead} A.~J.,  {Brieden} S.,  {Tr{\"o}ster} T.,   {Heymans} C.,  2021, \mn@doi [\mnras] {10.1093/mnras/stab082}, \href {https://ui.adsabs.harvard.edu/abs/2021MNRAS.502.1401M} {502, 1, 1401}

\bibitem[\protect\citeauthoryear{{M{\'e}nard}, {Scranton}, {Schmidt}, {Morrison}, {Jeong}, {Budavari}  \& {Rahman}}{{M{\'e}nard} et~al.}{2013}]{Menard2013}
{M{\'e}nard} B.,  {Scranton} R.,  {Schmidt} S.,  {Morrison} C.,  {Jeong} D.,  {Budavari} T.,   {Rahman} M.,  2013, \mn@doi [arXiv e-prints] {10.48550/arXiv.1303.4722}, \href {https://ui.adsabs.harvard.edu/abs/2013arXiv1303.4722M} {p. arXiv:1303.4722}

\bibitem[\protect\citeauthoryear{{Miyazaki} et~al.,}{{Miyazaki} et~al.}{2018}]{Miyazaki:2018:HSC}
{Miyazaki} S.,  et~al., 2018, \mn@doi [\pasj] {10.1093/pasj/psx063}, \href {https://ui.adsabs.harvard.edu/abs/2018PASJ...70S...1M} {70, S1}

\bibitem[\protect\citeauthoryear{{Myles \& Alarcon} et~al.,}{{Myles \& Alarcon} et~al.}{2021}]{Myles:2021:DESY3}
{Myles \& Alarcon} et~al., 2021, \mn@doi [\mnras] {10.1093/mnras/stab1515}, \href {https://ui.adsabs.harvard.edu/abs/2021MNRAS.505.4249M} {505, 3, 4249}

\bibitem[\protect\citeauthoryear{{Newman}}{{Newman}}{2008}]{Newman:2008:WZ}
{Newman} J.~A.,  2008, \mn@doi [\apj] {10.1086/589982}, \href {https://ui.adsabs.harvard.edu/abs/2008ApJ...684...88N} {684, 1, 88}

\bibitem[\protect\citeauthoryear{{Nikutta}, {Fitzpatrick}, {Scott}  \& {Weaver}}{{Nikutta} et~al.}{2020}]{Nikutta:2020:DataLab}
{Nikutta} R.,  {Fitzpatrick} M.,  {Scott} A.,   {Weaver} B.~A.,  2020, \mn@doi [Astronomy and Computing] {10.1016/j.ascom.2020.100411}, \href {https://ui.adsabs.harvard.edu/abs/2020A&C....3300411N} {33, 100411}

\bibitem[\protect\citeauthoryear{Pedregosa et~al.,}{Pedregosa et~al.}{2011}]{Pedregosa2012Sklearn}
Pedregosa F.,  et~al., 2011, Journal of machine learning research, 12, Oct, 2825

\bibitem[\protect\citeauthoryear{{Rau} et~al.,}{{Rau} et~al.}{2023}]{Rau2023}
{Rau} M.~M.,  et~al., 2023, \mn@doi [\mnras] {10.1093/mnras/stad1962}, \href {https://ui.adsabs.harvard.edu/abs/2023MNRAS.524.5109R} {524, 4, 5109}

\bibitem[\protect\citeauthoryear{{Rowe} et~al.,}{{Rowe} et~al.}{2015}]{Rowe:2015:galsim}
{Rowe} B.~T.~P.,  et~al., 2015, \mn@doi [Astronomy and Computing] {10.1016/j.ascom.2015.02.002}, \href {https://ui.adsabs.harvard.edu/abs/2015A&C....10..121R} {10, 121}

\bibitem[\protect\citeauthoryear{{S{\'a}nchez \& Prat} et~al.,}{{S{\'a}nchez \& Prat} et~al.}{2022}]{Sanchez2022}
{S{\'a}nchez \& Prat} et~al., 2022, \mn@doi [\prd] {10.1103/PhysRevD.105.083529}, \href {https://ui.adsabs.harvard.edu/abs/2022PhRvD.105h3529S} {105, 8, 083529}

\bibitem[\protect\citeauthoryear{{S{\'a}nchez} et~al.,}{{S{\'a}nchez} et~al.}{2020a}]{Sanchez:2020:CosmoDC2}
{S{\'a}nchez} J.,  et~al., 2020a, \mn@doi [\mnras] {10.1093/mnras/staa1957}, \href {https://ui.adsabs.harvard.edu/abs/2020MNRAS.497..210S} {497, 1, 210}

\bibitem[\protect\citeauthoryear{{S{\'a}nchez}, {Raveri}, {Alarcon}  \& {Bernstein}}{{S{\'a}nchez} et~al.}{2020b}]{Sanchez:2020:NoiseSOM}
{S{\'a}nchez} C.,  {Raveri} M.,  {Alarcon} A.,   {Bernstein} G.~M.,  2020b, \mn@doi [\mnras] {10.1093/mnras/staa2542}, \href {https://ui.adsabs.harvard.edu/abs/2020MNRAS.498.2984S} {498, 2, 2984}

\bibitem[\protect\citeauthoryear{{S{\'a}nchez} et~al.,}{{S{\'a}nchez} et~al.}{2023}]{Sanchez:2023:highzY3}
{S{\'a}nchez} C.,  et~al., 2023, \mn@doi [\mnras] {10.1093/mnras/stad2402}, \href {https://ui.adsabs.harvard.edu/abs/2023MNRAS.525.3896S} {525, 3, 3896}

\bibitem[\protect\citeauthoryear{{Schlegel}, {Finkbeiner}  \& {Davis}}{{Schlegel} et~al.}{1998}]{Schlegel:1998:Dust}
{Schlegel} D.~J.,  {Finkbeiner} D.~P.,   {Davis} M.,  1998, \mn@doi [\apj] {10.1086/305772}, \href {https://ui.adsabs.harvard.edu/abs/1998ApJ...500..525S} {500, 2, 525}

\bibitem[\protect\citeauthoryear{{Schmidt}}{{Schmidt}}{2008}]{Schmidt:2008:MG_WL}
{Schmidt} F.,  2008, \mn@doi [\prd] {10.1103/PhysRevD.78.043002}, \href {https://ui.adsabs.harvard.edu/abs/2008PhRvD..78d3002S} {78, 4, 043002}

\bibitem[\protect\citeauthoryear{{Schneider}, {Knox}, {Zhan}  \& {Connolly}}{{Schneider} et~al.}{2006}]{Schneider:2006:WZ}
{Schneider} M.,  {Knox} L.,  {Zhan} H.,   {Connolly} A.,  2006, \mn@doi [\apj] {10.1086/507675}, \href {https://ui.adsabs.harvard.edu/abs/2006ApJ...651...14S} {651, 1, 14}

\bibitem[\protect\citeauthoryear{{Schneider}, {Teyssier}, {Stadel}, {Chisari}, {Le Brun}, {Amara}  \& {Refregier}}{{Schneider} et~al.}{2019}]{Schneider2019Baryonification}
{Schneider} A.,  {Teyssier} R.,  {Stadel} J.,  {Chisari} N.~E.,  {Le Brun} A. M.~C.,  {Amara} A.,   {Refregier} A.,  2019, \mn@doi [\jcap] {10.1088/1475-7516/2019/03/020}, \href {https://ui.adsabs.harvard.edu/abs/2019JCAP...03..020S} {2019, 3, 020}

\bibitem[\protect\citeauthoryear{{Secco \& Samuroff} et~al.,}{{Secco \& Samuroff} et~al.}{2022}]{Secco2022}
{Secco \& Samuroff} et~al., 2022, \mn@doi [\prd] {10.1103/PhysRevD.105.023515}, \href {https://ui.adsabs.harvard.edu/abs/2022PhRvD.105b3515S} {105, 2, 023515}

\bibitem[\protect\citeauthoryear{{Sheldon}}{{Sheldon}}{2015}]{Sheldon:2015:ngmix}
{Sheldon} E.,  2015, {NGMIX: Gaussian mixture models for 2D images}, Astrophysics Source Code Library, record ascl:1508.008

\bibitem[\protect\citeauthoryear{{Sheldon} \& {Huff}}{{Sheldon} \& {Huff}}{2017}]{Sheldon2017}
{Sheldon} E.~S.,  {Huff} E.~M.,  2017, \mn@doi [\apj] {10.3847/1538-4357/aa704b}, \href {https://ui.adsabs.harvard.edu/abs/2017ApJ...841...24S} {841, 1, 24}

\bibitem[\protect\citeauthoryear{{Sheldon}, {Becker}, {MacCrann}  \& {Jarvis}}{{Sheldon} et~al.}{2020}]{Sheldon2020}
{Sheldon} E.~S.,  {Becker} M.~R.,  {MacCrann} N.,   {Jarvis} M.,  2020, \mn@doi [\apj] {10.3847/1538-4357/abb595}, \href {https://ui.adsabs.harvard.edu/abs/2020ApJ...902..138S} {902, 2, 138}

\bibitem[\protect\citeauthoryear{{Stanford} et~al.,}{{Stanford} et~al.}{2021}]{Stanford:2021:C3R2_DR3}
{Stanford} S.~A.,  et~al., 2021, \mn@doi [\apjs] {10.3847/1538-4365/ac0833}, \href {https://ui.adsabs.harvard.edu/abs/2021ApJS..256....9S} {256, 1, 9}

\bibitem[\protect\citeauthoryear{{Suchyta} et~al.,}{{Suchyta} et~al.}{2016}]{Suchyta:2016}
{Suchyta} E.,  et~al., 2016, \mn@doi [\mnras] {10.1093/mnras/stv2953}, \href {https://ui.adsabs.harvard.edu/abs/2016MNRAS.457..786S} {457, 1, 786}

\bibitem[\protect\citeauthoryear{{Takahashi}, {Sato}, {Nishimichi}, {Taruya}  \& {Oguri}}{{Takahashi} et~al.}{2012}]{Takahashi2012}
{Takahashi} R.,  {Sato} M.,  {Nishimichi} T.,  {Taruya} A.,   {Oguri} M.,  2012, \mn@doi [\apj] {10.1088/0004-637X/761/2/152}, \href {https://ui.adsabs.harvard.edu/abs/2012ApJ...761..152T} {761, 2, 152}

\bibitem[\protect\citeauthoryear{{Tanaka} et~al.,}{{Tanaka} et~al.}{2018}]{Tanaka2018}
{Tanaka} M.,  et~al., 2018, \mn@doi [\pasj] {10.1093/pasj/psx077}, \href {https://ui.adsabs.harvard.edu/abs/2018PASJ...70S...9T} {70, S9}

\bibitem[\protect\citeauthoryear{{The LSST Dark Energy Science Collaboration} et~al.,}{{The LSST Dark Energy Science Collaboration} et~al.}{2018}]{DESC:2018:SRD}
{The LSST Dark Energy Science Collaboration} et~al., 2018, \mn@doi [arXiv e-prints] {10.48550/arXiv.1809.01669}, \href {https://ui.adsabs.harvard.edu/abs/2018arXiv180901669T} {p. arXiv:1809.01669}

\bibitem[\protect\citeauthoryear{{Van der Walt}, {Colbert}  \& {Varoquaux}}{{Van der Walt} et~al.}{2011}]{vanderWalt2011Numpy}
{Van der Walt} S.,  {Colbert} S.~C.,   {Varoquaux} G.,  2011, \mn@doi [Computing in Science and Engineering] {10.1109/MCSE.2011.37}, \href {https://ui.adsabs.harvard.edu/abs/2011CSE....13b..22V} {13, 2, 22}

\bibitem[\protect\citeauthoryear{{Virtanen} et~al.,}{{Virtanen} et~al.}{2020}]{Virtanen2020Scipy}
{Virtanen} P.,  et~al., 2020, \mn@doi [Nature Methods] {https://doi.org/10.1038/s41592-019-0686-2}, \href {https://rdcu.be/b08Wh} {17, 261}

\bibitem[\protect\citeauthoryear{{Wang}, {Bahcall}  \& {Turner}}{{Wang} et~al.}{1998}]{Wang:1998:redshiftcolor}
{Wang} Y.,  {Bahcall} N.,   {Turner} E.~L.,  1998, \mn@doi [\aj] {10.1086/300592}, \href {https://ui.adsabs.harvard.edu/abs/1998AJ....116.2081W} {116, 5, 2081}

\bibitem[\protect\citeauthoryear{{Weaver} et~al.,}{{Weaver} et~al.}{2022}]{Weaver:2022:Cosmos}
{Weaver} J.~R.,  et~al., 2022, \mn@doi [\apjs] {10.3847/1538-4365/ac3078}, \href {https://ui.adsabs.harvard.edu/abs/2022ApJS..258...11W} {258, 1, 11}

\bibitem[\protect\citeauthoryear{{Wright}, {Hildebrandt}, {van den Busch}  \& {Heymans}}{{Wright} et~al.}{2020a}]{Wright:2020:KidsSOMS}
{Wright} A.~H.,  {Hildebrandt} H.,  {van den Busch} J.~L.,   {Heymans} C.,  2020a, \mn@doi [\aap] {10.1051/0004-6361/201936782}, \href {https://ui.adsabs.harvard.edu/abs/2020A&A...637A.100W} {637, A100}

\bibitem[\protect\citeauthoryear{{Wright}, {Hildebrandt}, {van den Busch}, {Heymans}, {Joachimi}, {Kannawadi}  \& {Kuijken}}{{Wright} et~al.}{2020b}]{Wright2020}
{Wright} A.~H.,  {Hildebrandt} H.,  {van den Busch} J.~L.,  {Heymans} C.,  {Joachimi} B.,  {Kannawadi} A.,   {Kuijken} K.,  2020b, \mn@doi [\aap] {10.1051/0004-6361/202038389}, \href {https://ui.adsabs.harvard.edu/abs/2020A&A...640L..14W} {640, L14}

\bibitem[\protect\citeauthoryear{{de Jong} et~al.,}{{de Jong} et~al.}{2015}]{deJong2015}
{de Jong} J.~T.~A.,  et~al., 2015, \mn@doi [\aap] {10.1051/0004-6361/201526601}, \href {http://adsabs.harvard.edu/abs/2015A%26A...582A..62D} {582, A62}

\bibitem[\protect\citeauthoryear{{van den Busch} et~al.,}{{van den Busch} et~al.}{2022}]{vandenBusch2022}
{van den Busch} J.~L.,  et~al., 2022, \mn@doi [\aap] {10.1051/0004-6361/202142083}, \href {https://ui.adsabs.harvard.edu/abs/2022A&A...664A.170V} {664, A170}

\makeatother
\end{thebibliography}



\appendix

\section{Combination of SOMPZ and WZ}\label{appx:SOMPZ_WZ}

\begin{figure*}
    \centering
    \includegraphics[width = 2\columnwidth]{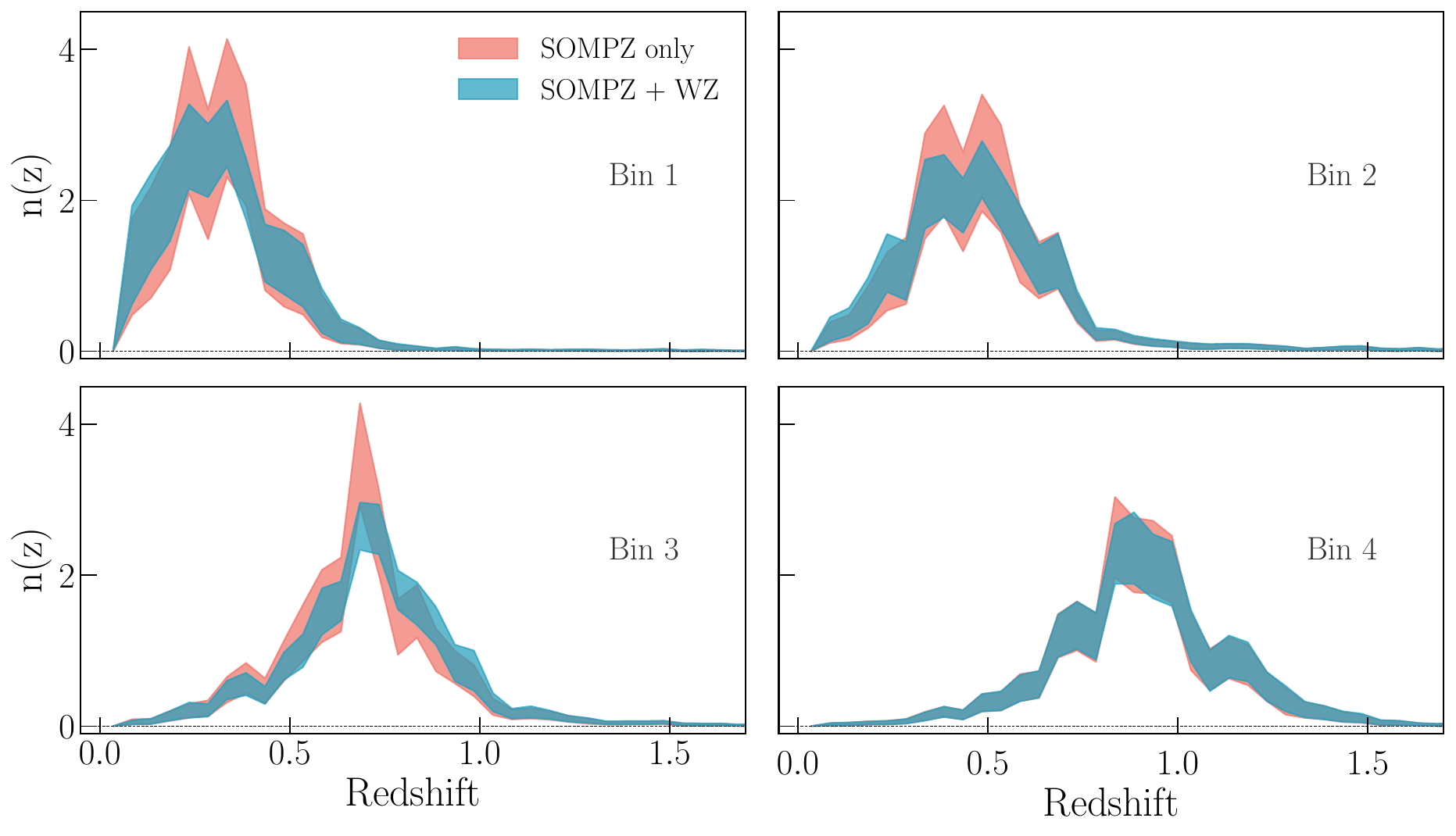}
    \caption{The fiducial redshift distributions from the SOMPZ method, compared with the results from combining the SOMPZ and clustering redshift methods. See Section~\ref{sec:sec:SOMPZ} for details on the SOMPZ only estimate, and then the text for details on the combination being shown here.}
    \label{fig:final_n_of_z}
\end{figure*}

In Section~\ref{sec:sec:SOMPZWZ} we describe the method for forward modelling the SOMPZ $n(z)$ into a WZ prediction. This was then used in Section~\ref{sec:Results_Nz:Crosscheck} to cross-check the fiducial SOMPZ $n(z)$ estimates.

In this Appendix, we now combine the SOMPZ and WZ methods to generate a new set of $n(z)$.
\citetalias{Gatti:2022:WzY3} perform a similar task by importance sampling the SOMPZ $n(z)$ using the WZ measurement/likelihood. The $f^X$ terms in Equation \eqref{eqn:3sdir} --- which are sampled by assuming a set of nested Dirichlet distributions; see Equation \eqref{eqn:Probf_Dir} --- are now sampled while including the likelihood from Equation \eqref{eqn:SOMPZWZ:likelihood}. This approach is computationally efficient as, under this modified procedure, the generated SOMPZ samples would have a higher likelihood of agreeing with the WZ measurements. 

The SOMPZ algorithm, on its own, is agnostic to the concept of galaxies and structure formation. The uncertainty on the redshift distribution in weak lensing surveys is limited by the shot noise and sample variance of the deep-field sample. As a consequence, the SOMPZ realizations can have somewhat rapid fluctuations across redshift. The WZ measurements, on the other hand, are smoother given they originate from cosmological correlations, which are smooth functions of redshift and length scale. The likelihood in Equation \eqref{eqn:SOMPZWZ:likelihood} is therefore maximized when the SOMPZ realization is smooth, but the probability of the 3sDir algorithm generating such a smooth realization is low. By modifying the SOMPZ algorithm to include the WZ information/likelihood during sampling, this probability can be significantly increased.

In this work, however, we take an alternative approach by splitting the sampling steps. We first generate a large sample of SOMPZ-based $n(z)$ distributions, of $\mathcal{O}(10^7)$, using the uncertainty quantification methods presented above. Each SOMPZ sample is then evaluated against the WZ measurement to obtain a likelihood of that sample. We only retain samples with a likelihood greater than a threshold that we define below. Thus, in this approach, we start from our Monte Carlo samples of $n(z)$ and subselect that population using the WZ information. Compared to the method above, this is a computationally expensive method as only a small subset of the generated SOMPZ samples will have a high likelihood given the WZ measurement: we find acceptance rates of $\approx 0.002\%$. However, this method has the advantage of being simplistic in nature, and requires no modifications to the likelihoods we define; see Section D5 in \citetalias{Myles:2021:DESY3} for details on the modified 3sDir algorithm needed for combining the SOMPZ and clustering redshift methods. Importantly, this simpler method still adds valuable WZ information to the $n(z)$ samples by discarding all SOMPZ samples that have significant disagreement with the WZ measurements.

For the subselection, we use the condition
\begin{equation}\label{eqn:WZ_criteria}
    \ln\mathcal{L_{\rm max}} - \ln\mathcal{L} = \Delta \ln \mathcal{L} < 22.78,
\end{equation}
where the choice of $22.78$ ensures the selected samples are at most $5\sigma$ from the maximum-likelihood point. We estimate this threshold by computing the cumulative probability within $5\sigma$ for a Gaussian distribution, and then calculating the $\chi^2$ value corresponding to that probability (from a standard $\chi^2$ distribution with $N_{\rm dof} = 8$, which is the size of the vector $\boldsymbol{q}$). We then use $0.5\chi^2$ as our threshold, since $\ln \mathcal{L} \propto -\chi^2/2$. We apply this threshold to all $10^7$ samples from the SOMPZ technique. Thus, after this step, the remaining $n(z)$ samples include uncertainties from shot noise, sample variance, redshift bias, and photometric zeropoint uncertainties, and are consistent with the clustering redshift measurements.

The $n(z)$ estimates from combining both SOMPZ and WZ is shown in Figure~\ref{fig:final_n_of_z}. We also show the the SOMPZ-only estimate for comparison. In all cases, the combined $n(z)$ is smoother than the SOMPZ-only estimate. Thus, the primary impact of including clustering redshift information is to smooth the SOMPZ-only estimates, and Figure~\ref{fig:final_n_of_z} shows exactly this result. The smoothing effect is best seen in the first tomographic bin, where the double peak at $z \sim 0.3$ is found in the SOMPZ-only estimate but is more smoothed out in the combined one. For the fourth tomographic bin, the SOMPZ-only and SOMPZ plus WZ estimates are quite similar to each other. The WZ measurements have higher uncertainties in the redshift range probed by this bin and so the WZ-based selection criteria in Equation \eqref{eqn:WZ_criteria} is passed by more SOMPZ-only samples of the $n(z)$. 

This change in the smoothness of the distribution has minimal impact on the final cosmic shear cosmology results (see \citetalias{paper4}), similar to what was found in DES Y3 \citep{Amon2022}. Note that we generally expect the WZ results to be less impactful in our work than was the case in \citetalias{Gatti:2022:WzY3}. Their work used a second, more informative galaxy sample --- a luminous red galaxy sample (\textsc{RedMaGiC}), constructed using DES data --- which had precise (photometric) redshift estimates and an order of magnitude more galaxies, whereas in this work we are limited to the sparser \Boss and \eBoss datasets. However, \citetalias{Gatti:2022:WzY3}, with their higher-precision data, find the addition of WZ information does not change their redshift uncertainty (see their Table 4). Thus, given the lower-precision data in our work, the combination of SOMPZ and WZ above is done as a slightly different test of the consistency between the results of the two methods, in addition to the cross-check of Section~\ref{sec:Results_Nz:Crosscheck}). The method does not sufficiently improve the uncertainties of the actual $n(z)$ estimates; see Table~\ref{tab:z_calibration}.


\label{lastpage}
\end{document}